\documentclass{jfm}
\usepackage{graphicx}
\usepackage{amsmath}
\usepackage{xcolor}
\usepackage{mathrsfs}

\shorttitle{Hemorheology in dilute, semi-dilute and dense suspensions of RBCs}

\shortauthor{N. Takeishi, M. E. Rosti, Y. Imai, S. Wada and L. Brandt}

\title{Hemorheology in dilute, semi-dilute and dense suspensions of red blood cells}

\author{
  Naoki Takeishi\aff{1} \corresp{\email{ntakeishi@me.es.osaka-u.ac.jp}},
  Marco E. Rosti\aff{2},
  Yohsuke Imai\aff{3},
  Shigeo Wada\aff{1} \and
  Luca Brandt\aff{2}  
}
 
\affiliation{
  \aff{1}Graduate School of Engineering Science, Osaka University, 1-3 Machikaneyama, Toyonaka, Osaka 560-8531, Japan
  \aff{2}Linn\'e Flow Centre and SeRC (Swedish e-Science Research Centre), KTH Mechanics, SE 100 44 Stockholm, Sweden
  \aff{3}Graduate School of Engineering, Kobe University, 1-1 Rokkodai, Nada, Kobe 657-8501, Japan\\ 	
}

\begin{document}

\maketitle

\begin{abstract}
We present a numerical analysis of the rheology of a suspension of red blood cells (RBCs) in a wall-bounded shear flow. The flow is assumed as almost inertialess. The suspension of RBCs, modeled as biconcave capsules whose membrane follows the Skalak constitutive law, is simulated for a wide range of viscosity ratios between the cytoplasm and plasma: $\lambda$ = 0.1--10, for volume fractions up to $\phi$ = 0.41 and for different capillary numbers ($Ca$). Our numerical results show that an RBC at low $Ca$ tends to orient to the shear plane and exhibits the so-called rolling motion, a stable mode with higher intrinsic viscosity than the so-called tumbling motion. As $Ca$ increases, the mode shifts from the rolling to the swinging motion. Hydrodynamic interactions (higher volume fraction) also allows RBCs to exhibit both tumbling or swinging motions resulting in a drop of the intrinsic viscosity for dilute and semi-dilute suspensions. Because of this mode change, conventional ways of modeling the relative viscosity as a polynomial function of $\phi$ cannot be simply applied in suspensions of RBCs at low volume fractions. The relative viscosity for high volume fractions, however, can be well described as a function of an effective volume fraction, defined by the volume of spheres of radius equal to the semi-middle axis of the deformed RBC. We find that the relative viscosity successfully collapses on a single non-linear curve independently of $\lambda$ except for the case with $Ca \geq$ 0.4, where the fit works only in the case of low/moderate volume fraction, and fails in the case of a fully dense suspension.
\end{abstract}


\section{Introduction}
The blood viscosity is a basic biological parameter affecting the blood flow both in large arteries and in microcirculations, and hence studies of hemorheology from single cell level to macro scale blood flow have been intensively conducted for many decades \citep{Pedley1980, Mohandas2008, Secomb2017}. Since human blood is a dense suspension consisting of 55\% fluid (plasma) and 45\% blood cells, with over 98\% of the blood cells being red blood cells (RBCs), hydrodynamic interactions of individual RBCs are of fundamental importance for hemorheology. Despite a number of studies about hemorheology, much is still unknown, in particular how the single cell behavior relates to the behavior in suspensions and then rheology. Therefore, the objective of this study is to clarify the behavior of individual RBCs from dilute to dense suspensions, and to elucidate the relationship between behaviors of individual RBCs and hemorheology.

Clarifying the cellular scale dynamics allows us to build precise continuum models of suspension \citep{Ishikawa2012, Rivera2015}, and potentially leads us to a novel diagnosis about patients with blood diseases \citep{Ito2017}. Therefore, researchers have made every effort to reveal the dynamics of single RBC as well as the rheological description of blood flow. By means of experimental observations, the dynamics of single RBC have been well investigated, e.g., RBCs subjected to low shear rate exhibit rigid-body-like flipping, the so-called tumbling motion \citep{Schmid-Schonbein1969, Fischer2004, Dupire2010} and wheel-like rotation, the so-called rolling motion \citep{Dupire2012, Lanotte2016}, while RBCs subjected to high shear rates exhibit the so-called tank-treading motion \citep{Schmid-Schonbein1969, Fischer1978, Fischer2004}.
The swinging motion was introduced by \cite{Abkarian2007} as an oscillating orientation of a tank-treading motion in the case of relatively low viscosity ratio $\lambda \sim$ 0.5. By means of numerical simulations, it has been observed that the dynamics transitions from rolling/tumbling to kayaking (or oscillating-swinging) and swinging, following tank-treading motions for a wide range of viscosity ratios (0.1 $< \lambda <$ 10) \citep{Cordasco2013, Sinha2015}. Moreover, recent experimental and numerical studies demonstrated that a rolling or tumbling RBC can shift to stomatocyte first and finally attaining a polylobed shapes (or multilobes) as the shear rate increases with relatively high viscosity ratios ($\lambda \sim$ 3 to 5) \citep{Lanotte2016, Mauer2018}.
Despite those insights, it is not known how those different motions of individual RBCs affect the bulk suspension rheology.

On the other hands, the rheological description of suspensions, especially of rigid particles, was addressed in the pioneering work by \cite{Batchelor1970}, showing that stress due to the presence of particles is evaluated using a particle stress tensor, which can be expressed as a summation of stresslet in a domain. \cite{Pozrikidis1992} analytically derived the effective stresslet of a deformable capsule consisting of an internal fluid enclosed by a thin elastic membrane. It is known that the usual distribution of hemoglobin concentration in individual RBCs ranges from 27 to 37 g/dL corresponding to the internal fluid viscosity being $\mu_1$ = 5--15 cP (= 5--15$\times$10$^{-3}$ Pa$\cdot$s) \citep{Mohandas2008}, while the normal plasma (external fluid) viscosity is $\mu_0$ = 1.1--1.3 cP (= 1.1--1.3$\times$10$^{-3}$ Pa$\cdot$s) for plasma at 37 $^{\circ}\mathrm{C}$ \citep{Harkness1970}. If the plasma viscosity is set to be $\mu_0$ = 1.2 cP, the physiological relevant viscosity ratio can be taken as $\lambda$ (= $\mu_1/\mu_0$) = 4.2--12.5. At the single cell level, the effect of viscosity ratio $\lambda$ on steady motions has been well investigated \citep{Cordasco2014, Mauer2018}. In suspensions, numerical studies of the behaviors of deformable particles modeled as neo-Hookean spherical capsules \citep{Clausen2011, Kumar2014, Matsunaga2016} or as viscoelastic materials \citep{Rosti2018a, Rosti2018b} have been conducted, while numerical studies of the behaviors of RBCs modeled as deformable biconcave capsules are still limited \citep{Fedosov2011, ReasorJr2013, Gross2014, Lanotte2016}, where the shear-thinning behaviors of a suspension of RBCs was systematically investigated. However, it remains unclear how the viscosity ratio $\lambda$ affects the bulk suspension rheology of RBCs. 

Nowadays, a rheological description of blood is important considering the fast increasing of diabetes mellitus worldwide. \cite{Skovborg1966} measured the viscosity of blood from diabetic patients, and found that it was approximately 20\% higher than in the controls. Elevated blood viscosity was also found in other hematologic disorders, e.g., multiple myeloma \citep{Dintenfass1975, Somer1987} and sickle cell disease \citep{Evans1984, Embury1984}. An experimental study using coaxial cylinder viscometer revealed that the blood from patients with sickle cell anemia, which is an inherited blood disorder exhibiting heterogeneous cell morphology, has higher hemoglobin concentration resulting in abnormal rheology \citep{Chien1970b, Usami1975, Kaul1991}. Numerical study about two body interactions of RBCs also concluded that the viscosity ratio is one of the most important parameters in hemorheology for the dilute and the semi-dilute regimes \citep{Omori2014}. As a conventional rheological description, the relative viscosity $\mu_{re}$ is often modeled as a polynomial function of the volume fraction $\phi$. For example, \cite{Einstein1911} proposed the viscosity law for the dilute suspension of rigid particles: $\mu_{re} = 1 + 2.5 \phi$, while \cite{Taylor1932} proposed a modified law for particles including internal fluid: $\mu_{re} = 1 + 2.5 \tilde{\lambda} \phi$, where $\tilde{\lambda}$ is Taylor's factor defined as $\tilde{\lambda} = \left( \lambda + 0.4 \right)/\left( \lambda + 1 \right)$. More recently, such polynomial approach has been applied to dense suspensions of deformable particles, with high order terms of $\phi$ \citep{Matsunaga2016, Rosti2018a}. However, a polynomial law in dense suspensions of non-spherical deformable particles such as RBCs is still missing due to complexity of the phenomenon. 

To reveal the rheological description of RBC suspensions, we investigate the effect of a wide range of viscosity ratios $\lambda$ = 0.1--10, non-dimensional shear rates (capillary number; $Ca$), and volume fractions $\phi$. We performed numerical simulations to study the behavior of RBCs subjected to various $Ca$ in wall-bounded shear flow from dilute ($\phi$ = 6$\times$10$^{-4}$; single RBC level) to dense suspensions ($\phi$ = 0.41). The contribution of the individual deformed RBC to the bulk suspension rheology is quantified by the stresslet tensor \citep{Batchelor1970}. The RBC is modeled as a biconcave capsule, whose membrane follows the Skalak constitutive law \citep{Skalak1973}. Since this problem require heavy computational resources, we resort to GPU computing, using the lattice-Boltzmann method for the inner and outer fluid and the finite element method to follow the deformation of the RBC membrane. The models have been successfully applied to the analysis of hydrodynamic interactions between RBCs-leukocyte \citep{Takeishi2014}, -circulating tumor cell \citep{Takeishi2015} and -microparticles/platelets \citep{Takeishi2017, Takeishi2019} in channel flows.

The remainder of this paper is organized as follows. Section 2 gives the problem statement and numerical methods. Section 3 presents the numerical results about single RBC and the semi-dilute/dense suspensions, and Section 4 a discussion and comparison between our numerical results and previous experimental/numerical results, followed by a summary of the main conclusions in Section 5. The validation of our numerical model is described in the Appendix.

\section{Problem statement}
\subsection{Flow and cell models} \label{setup}
We consider a cellular flow consisting of plasma and RBCs with radius $a$ in a rectangular box of size 16$a\times$10$a\times$16$a$ along the span-wise $x$, wall-normal $y$, and stream-wise $z$ directions, with a resolution of 8 fluid lattices per radius of RBC. Although the domain size used here has been shown to be adequate to investigate suspensions of rigid and deformable spherical particles in previous studies \citep{Picano2013, Rosti2018a}, we have preliminary checked its effect for RBCs as well as the effect of wall in the appendix, \S\ref{appA2} and \S\ref{appA3}. An RBC is modeled as a biconcave capsule, or a Newtonian fluid enclosed by a thin elastic membrane, with a major diameter 8 $\mu$m (= 2$a$), and maximum thickness 2 $\mu$m (= $a$/2). Although some recent numerical studies argued about the stress-free shape of RBCs \citep{Peng2014, Tsubota2014, Sinha2015}, we define the initial shape of RBC as a biconcave shape.

The shear flow is generated by moving the top and bottom walls located at $y = \pm H/2$ with constant velocity $U_{wall} = \pm \dot{\gamma}H/2$, where $H$ (= 10$a$) is the domain height and $\dot{\gamma}$ (= $U_c/a$) the shear rate defined using the characteristic velocity $U_c$. Periodic boundary conditions are imposed on the two homogeneous directions ($x$ and $z$ directions). The cytoplasmic viscosity is taken to be $\mu_1$ = 6.0$\times$10$^{-3}$ Pa$\cdot$s, which is five times higher than the plasma viscosity; $\mu_0$ = 1.2$\times$10$^{-3}$ Pa$\cdot$s \citep{Harkness1970}. Hence, in our study, the physiological relevant viscosity ratio is set to be $\lambda$ (= $\mu_1/\mu_0$) = 5, and the range of viscosity ratios $\lambda$ = 0.1--10 are considered. The problem is characterized by the capillary number ($Ca$),
\begin{equation}
  Ca = \frac{\mu_0 U_c}{G_s} = \frac{\mu_0 \dot{\gamma} a}{G_s}.
  \label{Ca}
\end{equation}
where $G_s$ is the surface shear elastic modulus. To counter the computational costs, we set $Re = \rho U_c a/\mu_0$ = 0.2, where $\rho$ is the plasma density. This value well represents capsule dynamics in unbounded shear flows solved by the boundary integral method in Stokes flow \citep{Omori2012, Matsunaga2016} (see also Appendix \S\ref{appA1} and \S\ref{appA3}). In this study, the range of $Ca$ = 0.05--1.2 is considered covering typical venule wall-shear rates of 333 s$^{-1}$ \citep{Koutsiaris2013}, corresponding to $Ca$ = 0.4, and arteriole wall shear rate of 670 s$^{-1}$ \citep{Koutsiaris2007} corresponding to $Ca$ = 0.8.

The membrane is modeled as an isotropic and hyperelastic material. The surface deformation gradient tensor $\boldsymbol{F}_s$ is given by
\begin{equation}
  d\boldsymbol{x}_m = \boldsymbol{F}_s \boldsymbol{\cdot} d\boldsymbol{X}_m,
  \label{Fs}
\end{equation}
where $\boldsymbol{X}_m$ and $\boldsymbol{x}_m$ are the membrane material points of the reference and deformed states, respectively. The local deformation of the membrane can be measured by the Green-Lagrange strain tensor
\begin{equation}
  \boldsymbol{E} = \frac{1}{2} \left( \boldsymbol{C} - \boldsymbol{I}_s \right),
  \label{e}
\end{equation}
where $\boldsymbol{I}_s$ is the tangential projection operator. The two invariants of the in-plane strain tensor $\boldsymbol{E}$ can be given by 
\begin{equation}
  I_1 = \lambda_1^2 + \lambda_2^2 - 2, \quad I_2 = \lambda_1^2 \lambda_2^2 - 1 = J_s^2 - 1,
  \label{I1andI2}
\end{equation}
where $\lambda_1$ and $\lambda_2$ are the principal extension ratios. The Jacobian $J_s = \lambda_1\lambda_2$ expresses the ratio of the deformed to the reference surface areas. The elastic stresses in an infinitely thin membrane are replaced by elastic tensions. The Cauchy tension $\boldsymbol{T}$ can be related to an elastic strain energy per unit area, $w_s \left( I_1, I_2 \right)$:
\begin{equation}
  \boldsymbol{T} = \frac{1}{J_s} \boldsymbol{F}_s \boldsymbol{\cdot} \frac{\partial w_s \left( I_1, I_2 \right)}{\partial \boldsymbol{E}} \boldsymbol{\cdot} \boldsymbol{F}_s^T,
  \label{T}
\end{equation}
where the strain energy function $w_s$ satisfies the Skalak (SK) constitutive law \citep{Skalak1973}
\begin{equation}
  w_s^{SK} = \frac{G_s}{4} \left( I_1^2 + 2 I_1 - 2I_2 + C I_2^2\right),
  \label{SK}
\end{equation}
being $C$ a coefficient representing the area incompressibility. In this study, we set $G_s$ = 4 $\mu$N/m and $C$ = 10$^2$. Bending resistance is also considered \citep{Li2005}, with a bending modulus $k_b$ = 5.0$\times$10$^{-19}$ J \citep{Puig-de-Morales-Marinkovic2007}: these values successfully reproduce the deformation of RBCs in shear flow and also the thickness of the cell-depleted peripheral layer, see appendix \S\ref{appA1}.\citep{Takeishi2014}.

\subsection{Numerical method}
The in-plane elastic tensions $\boldsymbol{T}$ are obtained from the Skalak constitutive law (\ref{SK}). Neglecting inertial effects on the membrane deformation, the static local equilibrium equation of the membrane is given by
\begin{equation}
  \nabla_s \boldsymbol{\cdot} \boldsymbol{T} + \boldsymbol{q} = \boldsymbol{0},
  \label{StrongForm}
\end{equation}
where $\nabla_s$ is the surface gradient operator. Based on the virtual work principle, the above strong form equation (\ref{StrongForm}) can be rewritten in weak form as 
\begin{equation}
  \int_S \boldsymbol{\hat{u}} \boldsymbol{\cdot} \boldsymbol{q} dS = \int_S \boldsymbol{\hat{\epsilon}} \boldsymbol{:} \boldsymbol{T} dS,
  \label{WeakForm}
\end{equation}
where $\boldsymbol{\hat{u}}$ and $\boldsymbol{\hat{\epsilon}} = ( \nabla_s \boldsymbol{\hat{u}} + \nabla_s \boldsymbol{\hat{u}}^T )\big/2$ are the virtual displacement and virtual strain, respectively. The FEM is used to solve equation (\ref{WeakForm}) and obtain the load $\boldsymbol{q}$ acting on the membrane \citep[see also][]{Walter2010}.

The LBM based on the D3Q19 model \citep{Chen1998, Dupin2007} is used to solve the fluid velocity field in the plasma and cytoplasm within the RBC membrane. In the  LBM, the macroscopic flow is obtained by collision and streaming of hypothetical particles described by the lattice-Boltzmann-Gross-Krook (LBGK) equation \citep{Bhatnagar1954}, which is given as
\begin{equation}
    f_i \left( \boldsymbol{x}_f + \boldsymbol{c}_i \Delta t, t + \Delta t \right) - f \left( \boldsymbol{x}_f, t \right) =
  - \frac{1}{\tau} \left[ f_i \left( \boldsymbol{x}_f, t \right) - f_i^{eq} \left( \boldsymbol{x}_f, t \right) \right] + F_i \Delta t,
  \label{LB}
\end{equation}
where $f_i$ is the particle distribution function for particles with velocity $\boldsymbol{c}_i$ ($i$ = 0--18) at the fluid node $\boldsymbol{x}_f$, $\Delta t$ is the time step size, $f_i^{eq}$ is the equilibrium distribution function, and $\tau$ is the nondimensional relaxation time. The external force term $F_i$ can be written as
\begin{equation}
    F_i =  \left( 1 - \frac{1}{2\tau} \right) w_i \left[ \frac{\boldsymbol{c}_i - \boldsymbol{v}}{c_s^2} + \frac{\left( \boldsymbol{c}_i \cdot \boldsymbol{v} \right)}{c_s^4} \right] \cdot \boldsymbol{F} \left( \boldsymbol{x}_f \right).
  \label{F}
\end{equation}
where $c_s = \Delta x_f/(\sqrt{3} \Delta t)$ is the speed of sound. The external force $\boldsymbol{F} \left( \boldsymbol{x}_f \right)$ is a distributed force applied from the membrane material points with the immersed boundary method (IBM)\citep{Peskin2002}. The particle velocity $\boldsymbol{c}_i$ is written by using the time-step size $\Delta t$ and the lattice size $\Delta x_f$ as
\begin{equation}
  \boldsymbol{c}_i = c_i^I \frac{\Delta x_f}{\Delta t} \boldsymbol{e}_I,
  \label{ci}
\end{equation}
where $\boldsymbol{e}_I$ is the Cartesian basis. The equilibrium distribution function is given by
\begin{equation}
     f_i^{eq} \left( \boldsymbol{x}_f, t \right) = \rho w_i \left[ 1 + \frac{\boldsymbol{v} \cdot \boldsymbol{c}_i}{c_s^2} + \frac{\left( \boldsymbol{v} \otimes \boldsymbol{v} \right) : \left( \boldsymbol{c}_i \otimes \boldsymbol{c}_i - c_s^2 \boldsymbol{I} \right) }{2c_s^4} \right],
  \label{f_eq}
\end{equation}
where $w_i$ is the weight ($w_i$ = 0 for $i = 0$, $w_i$ = 1/18 for the non-diagonal directions, and $w_i$ = 1/36 for the diagonal directioins) and $\boldsymbol{I}$ is the identity tensor. The macroscopic variables $\rho$ and $\boldsymbol{v}$ are defined as
\begin{eqnarray}
  &&\rho = \sum_i f_i, \\
  &&\rho \boldsymbol{v} = \sum_i c_i f_i + \cfrac{1}{2} \boldsymbol{F} \left( \boldsymbol{x}_f \right) \Delta t.
  \label{rho_u}
\end{eqnarray}
In the IBM \citep{Peskin2002}, the membrane force $\boldsymbol{f} \left( \boldsymbol{x}_m \right)$ at the membrane node $\boldsymbol{x}_m$ is distributed to the neighboring fluid nodes $\boldsymbol{x}_f$, and the external force $\boldsymbol{F} \left( \boldsymbol{x}_f \right)$ in (\ref{F}) is computed as
\begin{equation}
  \boldsymbol{F} \left( \boldsymbol{x}_f \right) = \sum_m D \left( \boldsymbol{x}_f - \boldsymbol{x}_m \right) \boldsymbol{f} \left( \boldsymbol{x}_m \right),
\end{equation}
where $D \left( \boldsymbol{x} \right)$ is a smoothed delta function approximating the Dirac delta function, given by
\begin{equation}
  D \left( \boldsymbol{x} \right) =
  \begin{cases}
  \dfrac{1}{64 \Delta x_f^3} \prod_{k = 1}^3 \Bigl( 1 + \cos \dfrac{\pi x_k}{2 \Delta x_f} \Bigr) & \text{if} \ \left| x_k \right| \leq 2 \Delta x_f, \ x_1 = x, x_2 = y, x_3 = z, \\
  0 & \text{otherwise}.
  \end{cases}
  \label{delta_f}
\end{equation}
The velocity at the membrane node $\boldsymbol{v} \left( \boldsymbol{x}_m \right)$ is obtained by interpolating the velocities at the fluid nodes as
\begin{equation}
  \boldsymbol{v} \left( \boldsymbol{x}_m \right) = \sum_f D \left( \boldsymbol{x}_f - \boldsymbol{x}_m \right) \boldsymbol{v} \left( \boldsymbol{x}_f \right).
\end{equation}
The membrane node $\boldsymbol{x}_m$ is updated by Lagrangian tracking with the no-slip condition, i.e.
\begin{equation}
\frac{d \boldsymbol{x}_m}{dt} = \boldsymbol{v} \left( \boldsymbol{x}_m \right).
\end{equation}
The explicit forth-order Runge-Kutta method is used for the time-integration. 
Note that, by using our coupling method of LBM and IBM, the hydrodynamic interaction of individual RBCs is solved without modeling a non-hydrodynamic inter-membrane repulsive force in the case of vanishing inertia, as also shown in appendix \S\ref{appA3}.

The viscosity of a LB node $x_f$ is found using the volume-of-fluid (VOF) $\psi (x_f)$ (0 $\leq \psi \leq$ 1) of the internal fluid of the RBCs:
\begin{equation}
  \mu = (1 - \psi) \mu_0 + \psi \mu_1 = \bigl\{ 1 + (\lambda - 1) \psi \bigr\} \mu_0,
\end{equation}
and the kinematic viscosity as
\begin{equation}
  \nu = \frac{\mu}{\rho} = \bigl\{ 1 + (\lambda - 1) \psi \bigr\} \frac{\mu_0}{\rho} = \frac{2 \tau - 1}{2} c_s^2 \Delta t = \frac{2 \tau - 1}{6} \frac{\Delta r^2}{\Delta t}.
\end{equation}
To update the viscosity on the fluid lattice, we consider the VOF function $\psi$, which is governed by an advection equation:
\begin{equation}
  \frac{\partial \psi}{\partial t} + \nabla \cdot  \left( \boldsymbol{v} \psi \right) - \psi \nabla \cdot \boldsymbol{v} = 0.
  \label{vof}
\end{equation}
Equation (\ref{vof}) is solved by the THINC/WLIC (Tangent of Hyperbola for Interface Capturing/Weighted Line Interface Calculation) method \citep{Yokoi2007},
which is a combination of the THINC scheme and the WLIC method. As a characteristic function of the THINC scheme, the piecewise modified hyperbolic tangent function \citep{Xiao2005} is used. In the WLIC formulation, the interface is reconstructed by taking an average of the interfaces along the $x$, $y$ and $z$ coordinates with weights calculated from the surface normal. To counter the divergence between the interface of $\psi$ and the membrane surface, we also solve the Poisson's equation of the indicator function $\mathcal{I} \left( \boldsymbol{x}_f \right)$ used in the front-tracking method \citep{Unverdi1992}:
\begin{equation}
\nabla^2 \mathcal{I} \left( \boldsymbol{x}_f \right) = \nabla \cdot \boldsymbol{G} \left( \boldsymbol{x}_f \right),
\end{equation}
where $\mathcal{I} = 1$ in the interior of a cell and $\mathcal{I} = 0$ outside a cell, and $\boldsymbol{G} \left( \boldsymbol{x}_f \right)$ is described by the smoothed delta function (\ref{delta_f}):
\begin{equation}
\boldsymbol{G} \left( \boldsymbol{x}_f \right) = \nabla \mathcal{I} \left( \boldsymbol{x}_f \right) = \sum_e D \left( \boldsymbol{x}_f - \boldsymbol{x}_e \right) \boldsymbol{n}_e \Delta s_e,
\end{equation}
where $\boldsymbol{n}_e$ is the outward unit normal vector to an element with the area $\Delta s_e$, whose centroid is $\boldsymbol{x}_e$.
To speed-up the numerical simulations, we only solve the Poisson equation of the indicator function every ten thousand steps. Our methods are validated for different viscosity ratios by comparing the values of the Taylor parameter of deformed spherical capsules with those reported in \cite{Foessel2011}, as detailed in the appendix \S\ref{appA1}. A volume constraint is implemented to counteract the accumulation of small errors in the volume of the individual cells \citep{Freund2007}: in our simulation, the volume error is always maintained lower than 1.0$\times$10$^{-3}$ \%, as tested and validated in our previous study on cell adhesion \citep{Takeishi2016}.   

All numerical procedures are fully implemented on graphics processing unit (GPU) to accelerate the numerical simulation \citep{Miki2012}. The mesh size of the Lattice-Boltzmann method (LBM) for the fluid solution is set to be 250 nm, and that of the finite elements describing the membrane is approximately 250 nm (an unstructured mesh with 5,120 elements is used for the FEM). This resolution has been shown to successfully represent single- and multi-cellular dynamics \citep{Takeishi2014}; also, the results of multi-cellular dynamics are not changing with twice the resolution of both the fluid- and membrane-mesh \citep[see also][]{Takeishi2014}.

\subsection{Analysis of capsules suspensions}
\begin{figure}
  \centering
  \includegraphics[height=6.0cm]{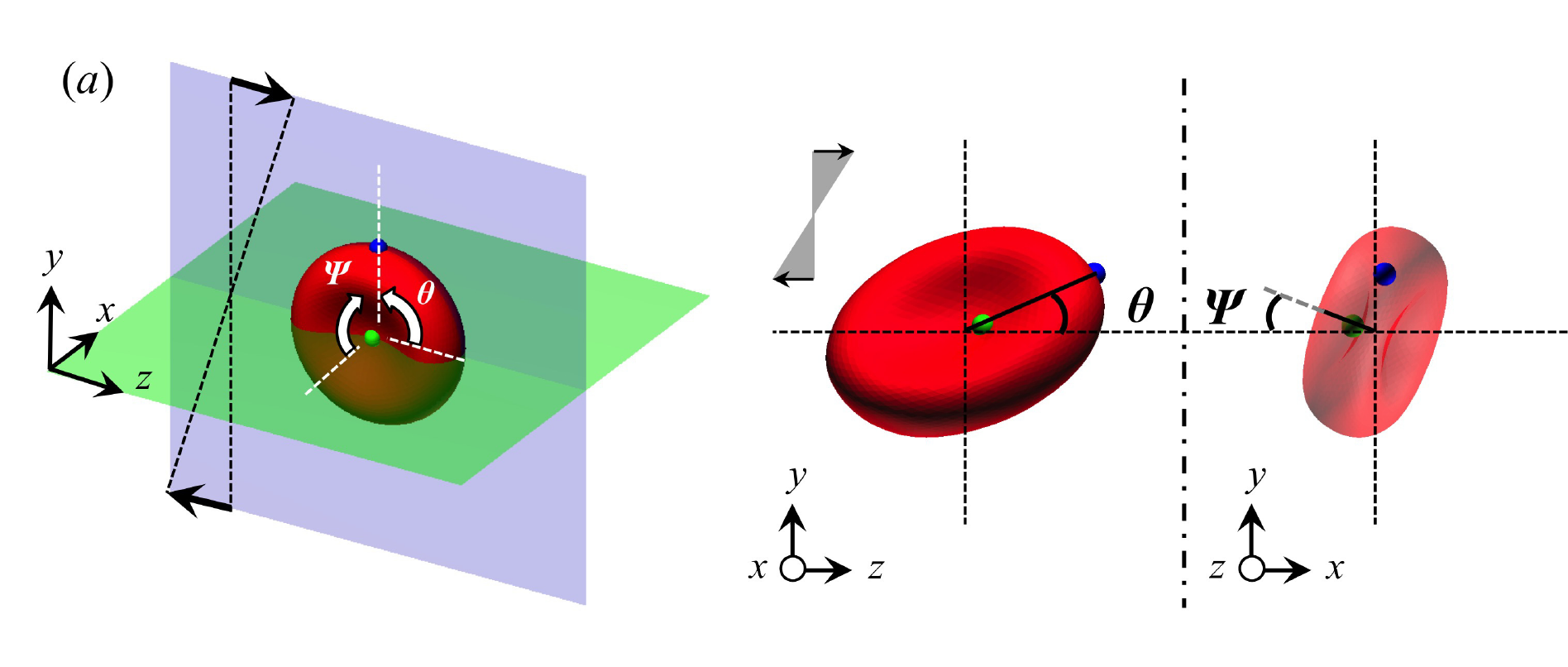} \\
  \includegraphics[height=5.5cm]{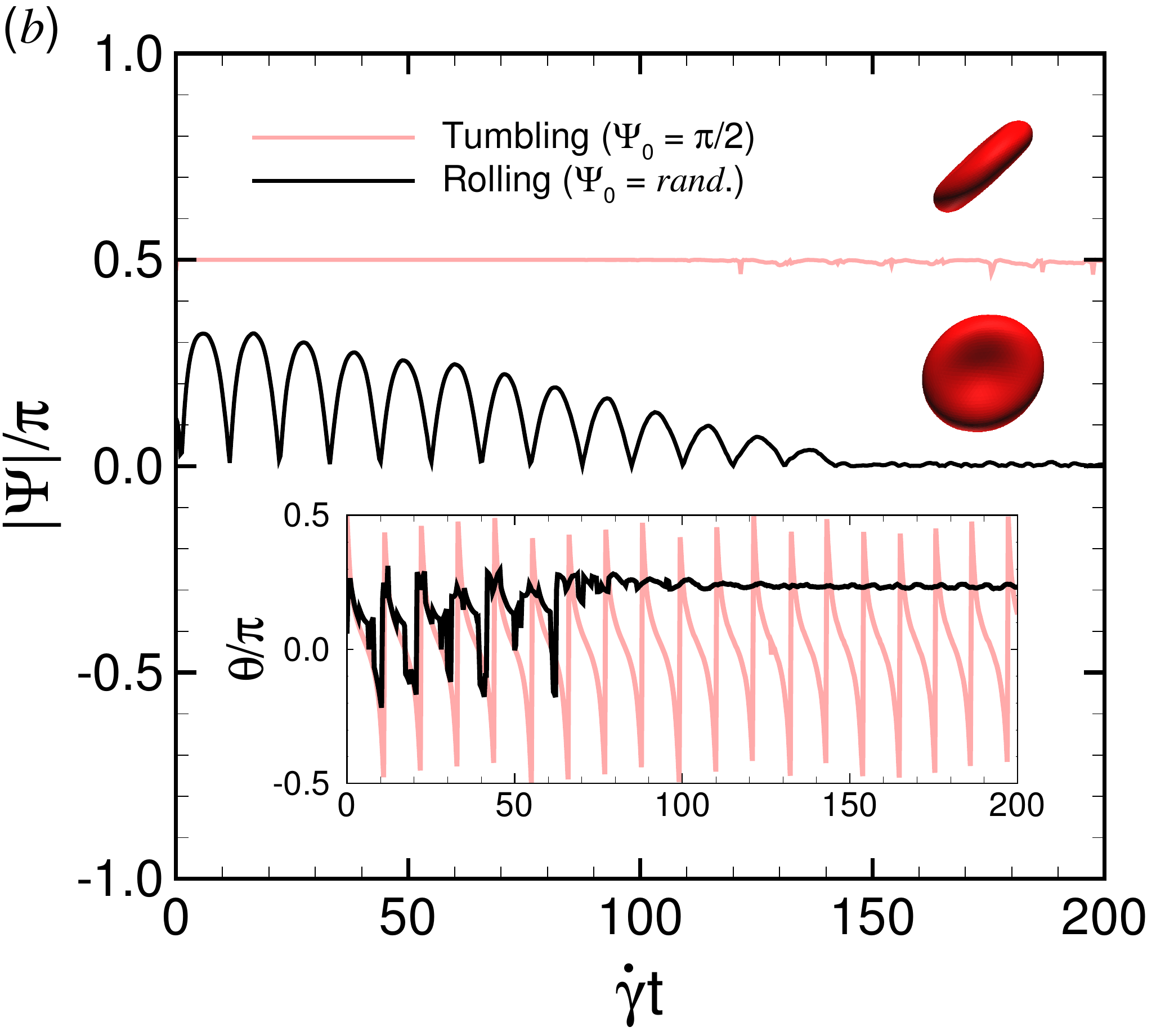}
  \includegraphics[height=5.5cm]{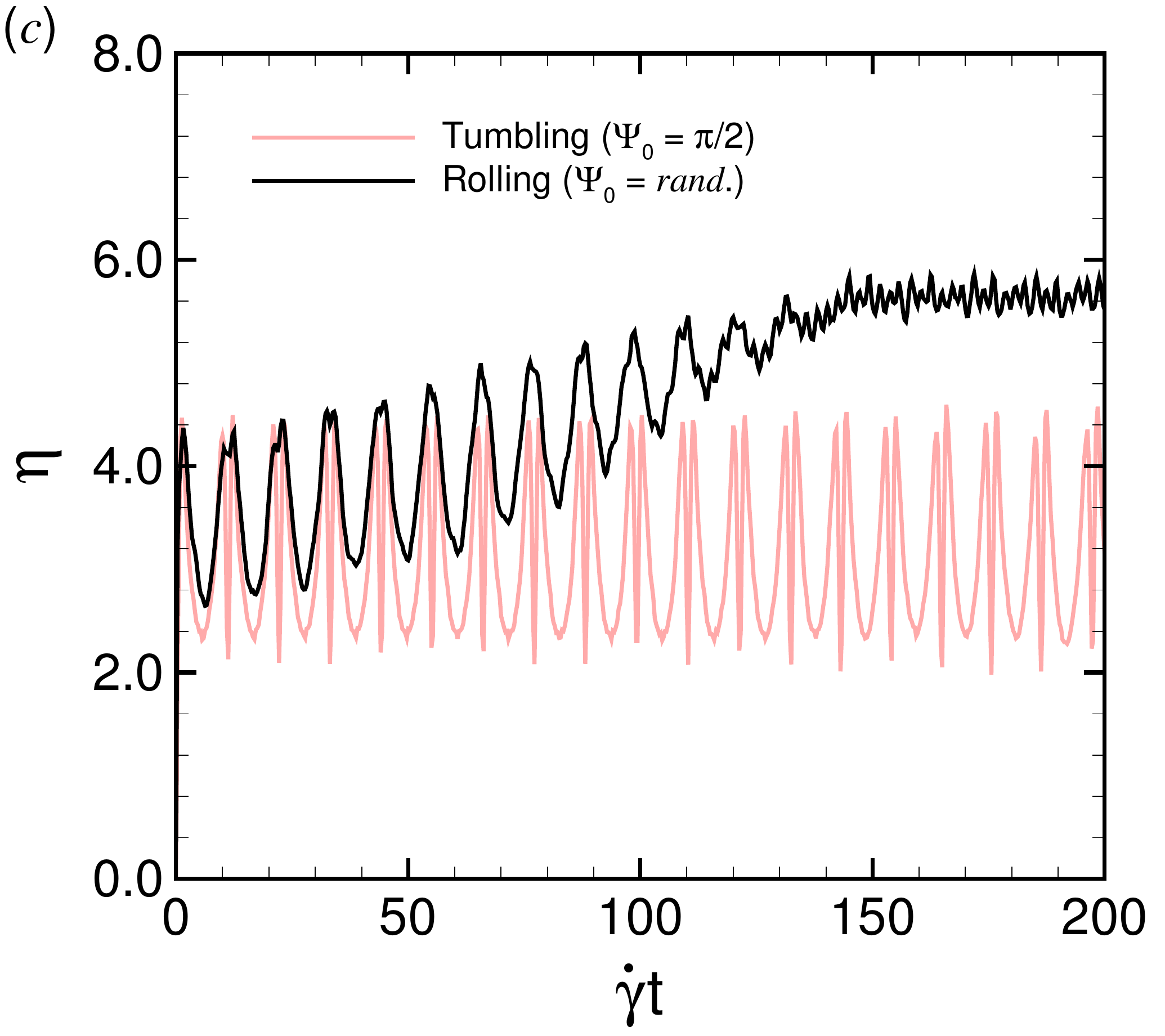}
  \caption{
  ($a$) Schematic of the 3D orientation of an RBC (\emph{top left}). The shear flow is driven along the $z$ direction by moving the top and bottom walls. Periodic boundary conditions are imposed on the flow ($z$ direction) and vortex ($x$ direction) directions. Color dots on the membrane denote material points to measure the RBC orientation, which is defined by the angle $\theta$ between the major axis of the deformed RBC and the shear direction, and the angle $\Psi$ between the vortex axis and the green dot (\emph{top right}).
  ($b$) Time history of the orientation angles $\Psi$ and $\theta$ for different motions; tumbling motion (light red line) for RBC initial orientation $\Psi_0$ = $\pi$/2, and rolling motion (black line) for random initial state, i.e., $\Psi_0$ = $rand$. The upper inset figure in ($b$) represents the tumbling motion (see the supplementary movie1), and the lower inset figure the rolling motion of RBC (see the supplementary movie2).
  ($c$) Time history of the intrinsic viscosity $\eta$ (= $\mu_{sp}/\phi$) for tumbling (light red line) and rolling (black line) motions. The results are obtained with $\lambda$ = 5 at $Ca$ = 0.05.
  }
  \label{fig:single_rbc_lam5ca005}
\end{figure}

For the following analysis, the behavior of RBCs subjected to shear flow is quantified by two different orientation angles $\theta$ and $\Psi$ as shown in Fig.\ref{fig:single_rbc_lam5ca005}$a$, where $\theta$ is the angle between the major axis of the deformed RBC and the shear direction, and $\Psi$ is the angle between the vortex axis and the normal vector at the initial concave node point (green-dot in Fig.\ref{fig:single_rbc_lam5ca005}$a$). In suspensions of RBCs, the ensemble average of these orientation angles are calculated,
\begin{eqnarray}
\langle \xi \rangle = \frac{1}{MN} \sum^M_m \sum^N_n \xi^{m, n} \quad (\xi = \theta \ \text{or} \ \Psi),
\end{eqnarray}
where $M$, $N$ are the number of time steps and capsules (RBCs), respectively. Time average starts after the non-dimensional time $\dot{\gamma}t$ = 40 to reduce the influence of the initial conditions, and continues to over $\dot{\gamma}t$ = 100.

For the analysis of the suspension rheology, we consider the contribution of the suspended particles to the bulk viscosity in terms of the particle stress tensor $\boldsymbol{\Sigma}^{(p)}$ \citep{Batchelor1970}:
\begin{align}
  \boldsymbol{\Sigma}^{(p)} = \frac{1}{V} \sum_i^N \boldsymbol{S}_i,
\end{align}
where $V$ is the volume of the domain and $\boldsymbol{S}_i$ the stresslet of the \emph{i}-th particle (capsule and RBC in the present study). \cite{Pozrikidis1992} analytically derived the stresslet of a deformable capsule for any viscosity ratio:
\begin{align}
  \boldsymbol{S}_i = \int \left[ \frac{1}{2} \left( \boldsymbol{x} \otimes \boldsymbol{\hat{q}} + \boldsymbol{\hat{q}} \otimes \boldsymbol{x} \right) - \mu_0 \left( 1 - \lambda \right) \left( \boldsymbol{v} \otimes \boldsymbol{n} + \boldsymbol{n} \otimes \boldsymbol{v} \right) \right] dA_i,
\end{align}
where $\boldsymbol{x}$ is the membrane position relative to the centre of the RBC, $\boldsymbol{\hat{q}}$ the load acting on the membrane including a contribution of bending rigidity, $\boldsymbol{n}$ the surface normal vector, $\mu_0$ the outer fluid (plasma) viscosity, $\boldsymbol{v}$ the interfacial velocity of membrane, and $A_i$ the membrane surface area of the \emph{i}-th RBC. The suspension shear viscosity $\mu^{\ast}$ is often expressed in terms of the viscosity $\mu_0$ of the carrier fluid and a perturbation $\delta \mu$ (i.e., $\mu^{\ast} = \mu_0 + \delta \mu$), sometimes analytically  obtained by truncating a perturbative approach at leading order (e.g., for small deformability or very dilute conditions). This leads to the introduction of the relative viscosity $\mu_{re}$ and specific viscosity $\mu_{sp}$ defined as:
\begin{align}
  & \mu_{re} = \frac{\mu^{\ast}}{\mu_0} = 1 + \mu_{sp} \\
  & \mu_{sp} = \frac{\delta \mu}{\mu_0} = \frac{\Sigma^{(p)}_{12}}{\mu_0 \dot{\gamma}}.
\end{align}
For example, in a dilute suspension of rigid spheres with a volume fraction $\phi$, it is well known that the specific viscosity $\mu_{sp}$ can be given by a polynomial equation of the first order of $\phi$, $\mu_{sp}$ = $\eta \phi$ (= 2.5$\phi$) \citep{Einstein1911}, where the coefficient $\eta$ (= 2.5) is the intrinsic viscosity which is defined as $\eta = \mu_{sp}/\phi$.

The first and second normal stress difference, typically used to quantify the suspension viscoelastic behavior, are defined as:
\begin{align}
  &\frac{N_1}{\mu_0 \dot{\gamma}} = \frac{\Sigma_{11}^{(p)} - \Sigma_{22}^{(p)}}{\mu_0 \dot{\gamma}}, \\
  &\frac{N_2}{\mu_0 \dot{\gamma}} = \frac{\Sigma_{22}^{(p)} - \Sigma_{33}^{(p)}}{\mu_0 \dot{\gamma}}.
  \label{eq:Ni}
\end{align}
The particle pressure \citep{Jeffrey1993}, which is the isotropic stress that exists in the particle phase,  is given by
\begin{align}
  \frac{\Pi_p}{\mu_0 \dot{\gamma}} = - \frac{\text{tr} \Sigma^{(p)} }{3 \mu_0 \dot{\gamma}},   
  \label{eq:p_pressure}
\end{align}
The particle pressure is analogous to the osmotic pressure in colloidal suspension caused by the hydrodynamic interactions among the suspended particles without Brownian motion, and has been previously quantified for suspensions of deformable capsules \citep{Clausen2011, ReasorJr2013, Gross2014}.
In the following section, we show the numerical results obtained with $\lambda$ = 5, and compare those at the other $\lambda$ to quantify its effect on the bulk suspension rheology.
 
\section{Results}
\subsection{Behavior of single RBC}
First, we investigate the behavior of a single RBC at small deformations. When the RBC is placed perpendicular to the shear direction ($\Psi_0 $ = $\pi$/2), it keeps flipping along the vortex angle for low $Ca$ (= 0.05), with $\Psi \sim$ $\pi$/2 and with -$\pi$/2 $< \theta < \pi$/2, the so-called tumbling motion (Fig.\ref{fig:single_rbc_lam5ca005}$b$; see the supplementary movie1). On the other hand, an RBC initially randomly placed, i.e., $\Psi_0$ = $rand.$ (at least $\Psi_0 \neq$ 0 or $\neq \pi$/2), tends to orient parallel to the shear plane, showing a wheel-like configuration with $\Psi$ = 0 and $\theta \sim \pi$/4, the so-called rolling motion (Fig.\ref{fig:single_rbc_lam5ca005}$b$; see the supplementary movie2). Our numerical results suggest that a free mode of RBC at low $Ca$ is the rolling motion. Our numerical results also show that the stable rolling RBC has higher intrinsic viscosity than the tumbling RBC, see Fig.\ref{fig:single_rbc_lam5ca005}($c$). Since the orientation angle $\theta$ of a single deformable spherical capsule converges to $\pi$/4 in shear flow as $Ca \to$ 0 \citep{Barthes-Biesel1980, Barthes-Biesel1985}, the orientation angle $\theta$ of the rolling RBC also converges to $\theta \to \pi$/4. \cite{Jeffery1922} investigated the motion of a single ellipsoid in simple shear flow in the Stokes flow regime, and hypothesized that ``The particle will tend to adopt that motion which, of all the motions possible under the approximated equations, corresponds to the least dissipation''. \cite{Taylor1923} experimentally confirmed Jeffery's hypothesis by investigating the orbit of a prolate or oblate spheroid in a Couette flow at a very low $Re$. However, our numerical results of intrinsic viscosity $\eta$ does not agree with Jeffery's hypothesis, i.e., maximum in $\eta$, while agree with previous numerical results of deformable biconcave capsules \citep{Gross2014}.

\begin{figure}
  \centering
  \includegraphics[height=5.5cm]{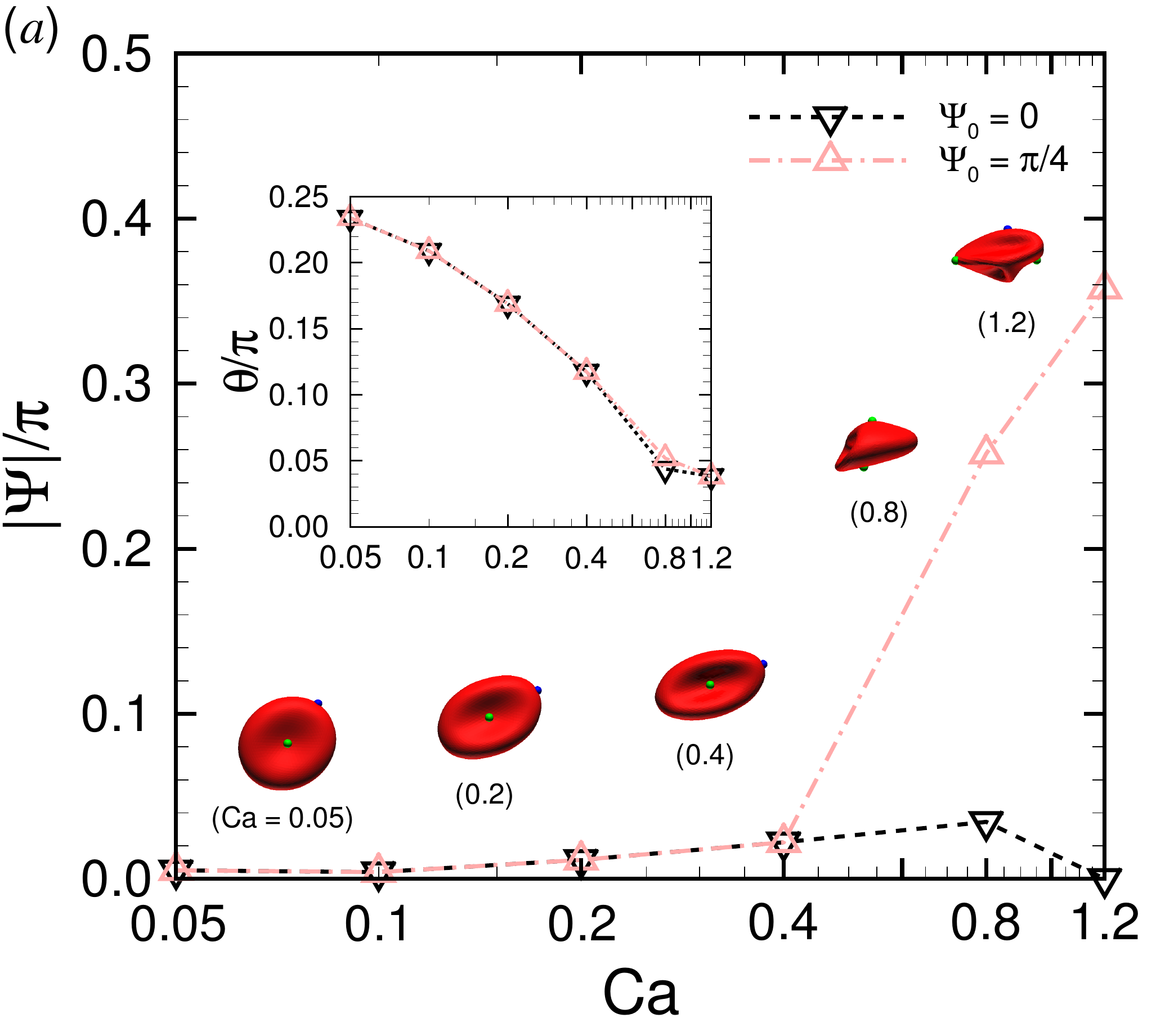}
  \includegraphics[height=5.5cm]{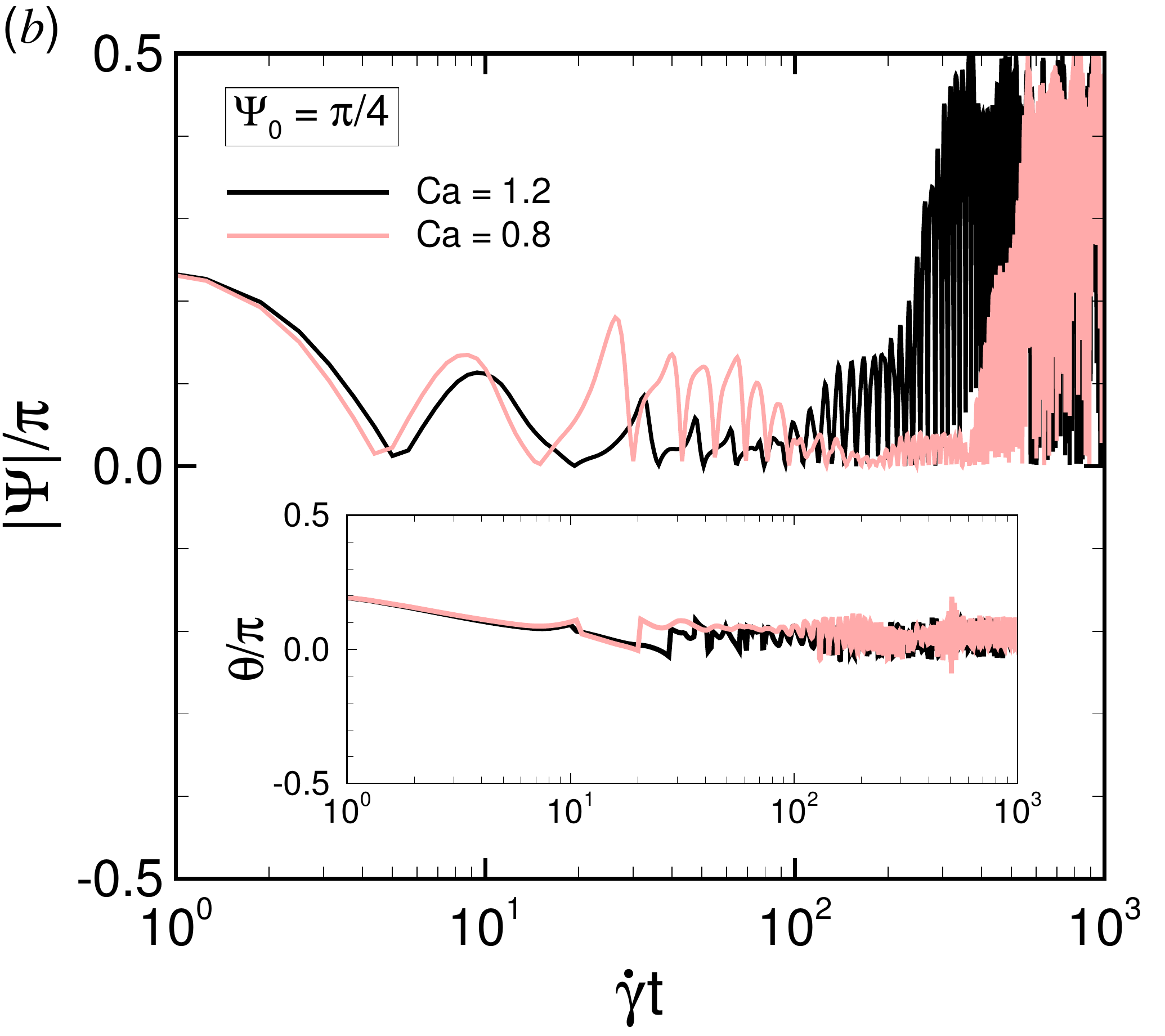}
  \includegraphics[height=5.5cm]{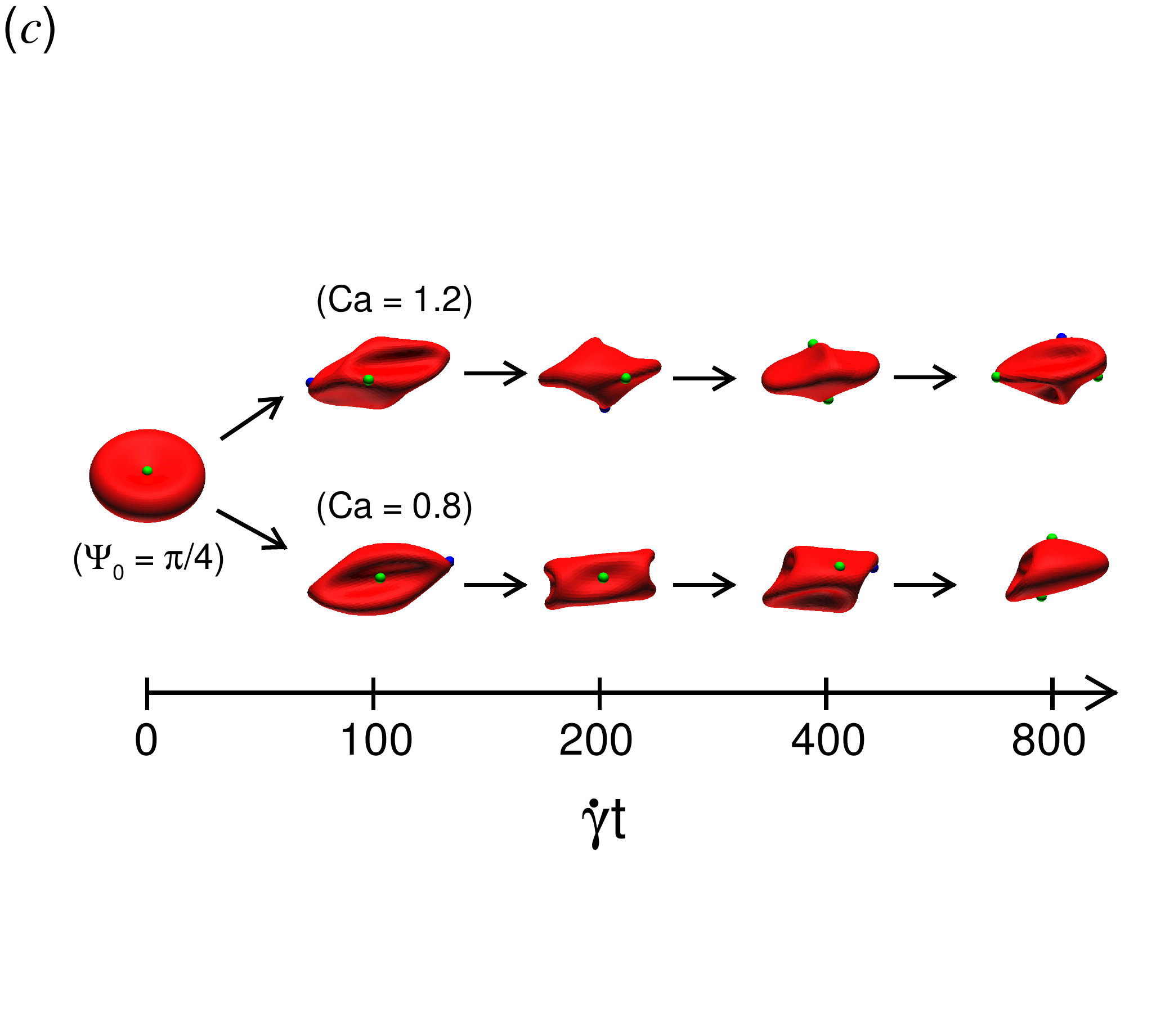}
  \includegraphics[height=5.5cm]{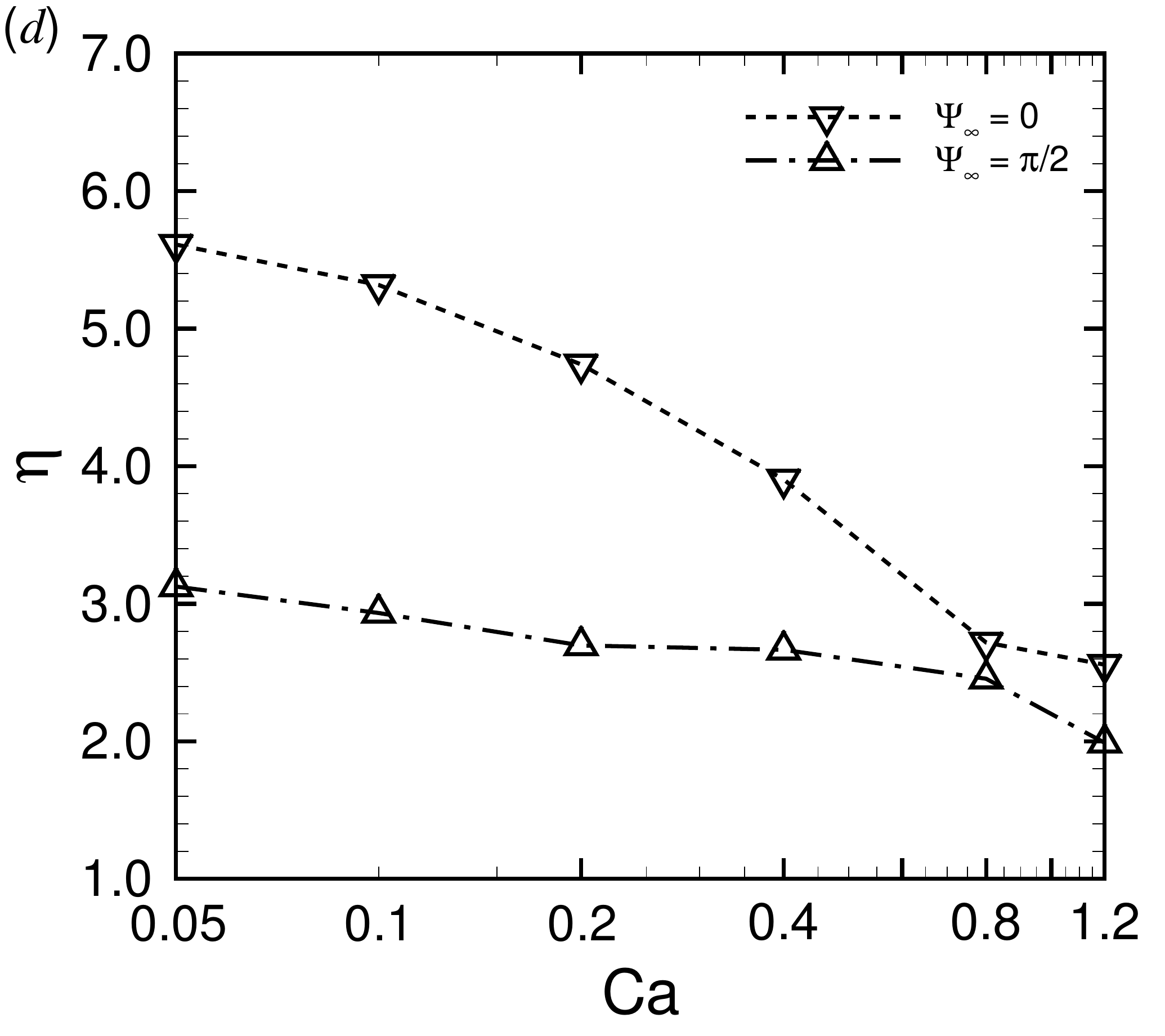} 
  \caption{
  ($a$) Time average of the orientation angle $\theta$ and $\Psi$ as a function of the logarithm of $Ca$. The simulations start from $\Psi_0$ = 0 (black inverse triangles) and $\Psi_0$ = $\pi$/4 (pale red triangles). Inset figures in ($a$) represent the stable configurations of the RBC with initial angle $\Psi_0$ = $\pi$/4 for each $Ca$.
  ($b$) Time history of the orientation angles $\theta$ and $\Psi$ for $Ca$ = 0.8 (pale red line) and $Ca$ = 1.2 (black line), for initial angle $\Psi_0$ = $\pi$/4.
  ($c$) Snapshots of deformed RBCs at different times for $Ca$ = 0.8 (down) and $Ca$ = 1.2 (top).
  ($d$) Time average of the intrinsic viscosity $\eta$ as a function of the logarithm of $Ca$ for different stable modes: $\Psi_{\infty}$ = 0 (inverse triangles) and $\Psi_{\infty}$ = $\pi$/2 (triangles), obtained with the simulations started from $\Psi_0$ = 0 and $\Psi_0$ = $\pi$/2, respectively.  These results are obtained with $\lambda$ = 5.
  }
  \label{fig:single_rbc_ca}
\end{figure}
The effects of $Ca$ on the stable mode and intrinsic viscosity are now investigated. At least for $Ca \leq$ 0.4, RBCs initially oriented with $\Psi_0 = \pi$/4 converge their orientation angles to the one obtained for RBCs with $\Psi_0$ = 0 as shown in Fig.\ref{fig:single_rbc_ca}($a$), where the inset photos represent the stable configurations at each $Ca$. Note that the final orientation is not changed if the RBCs are initially oriented with $\pi$/4 $< \Psi_0 < \pi$/2.
When $Ca$ increases ($\geq$ 0.8), the rolling motion becomes unstable. For instance, RBCs subjected to the highest $Ca$ that we investigated (i.e., $Ca$ = 1.2) fluctuates for 0 $\leq \Psi \leq \pi$/2,
showing a multilobes-like shape \citep{Lanotte2016}, while RBCs at $Ca$ = 0.8 for $\Psi \sim$ 0 show a tumbling stomatocyte-like shape \citep{Mauer2018} (Fig.\ref{fig:single_rbc_ca}$b$; see the supplementary movie3 for $Ca$ = 1.2 and movie4 for $Ca$ = 0.8). Such complex deformed shape of RBCs are qualitatively similar to the ones reported by \cite{Mauer2018}, where their smoothed dissipative particle dynamics model of RBCs with $\lambda$ = 5 shifts from rolling discocytes (similar to the inset in Fig.\ref{fig:single_rbc_ca}$a$ for $Ca$ = 0.05) to rolling/tumbling stomatocytes (similar to the inset in Fig.\ref{fig:single_rbc_ca}$a$ for $Ca$ = 0.2 or 0.4) and finally attains multilobes (similar to the inset in Fig.\ref{fig:single_rbc_ca}$a$ for $Ca$ = 0.8 or 1.2) as the shear rate increases.
Despite the different configurations, the intrinsic viscosity $\eta$ for high $Ca$ is similar as shown in Fig.\ref{fig:single_rbc_ca}($d$), and hence the effect of the stable modes on the intrinsic viscosity $\eta$ reduces for increasing $Ca$. According to Fig.\ref{fig:single_rbc_ca}($d$), the result for $\Psi_{\infty}$ = 0 demonstrates significant more shear-thinning \citep{Skovborg1966, Cokelet1968, Chien1970a}, than in the case of $\Psi_{\infty}$ = $\pi$/2.

\begin{figure}
  \centering
  \includegraphics[height=5.5cm]{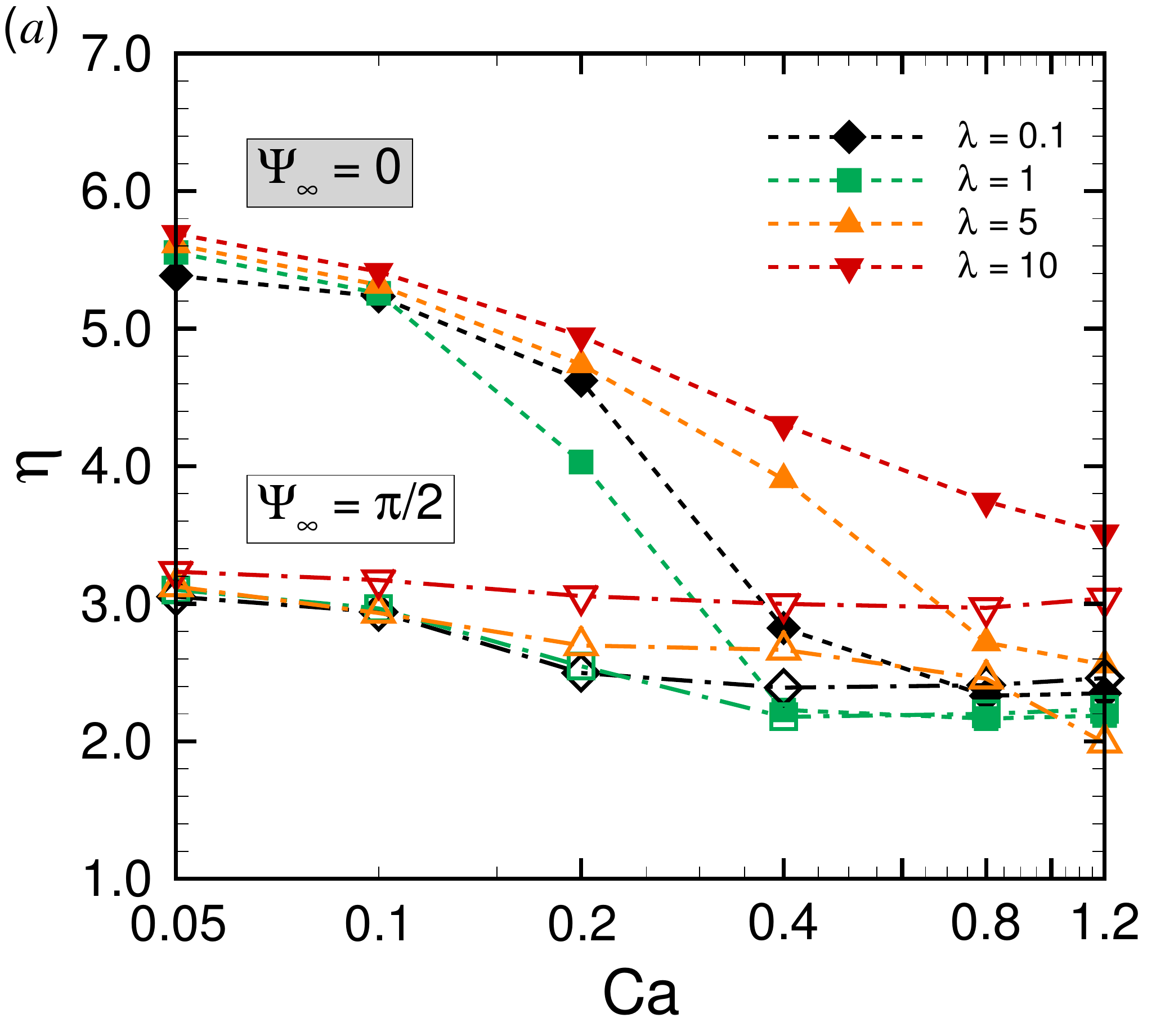}
  \includegraphics[height=5.5cm]{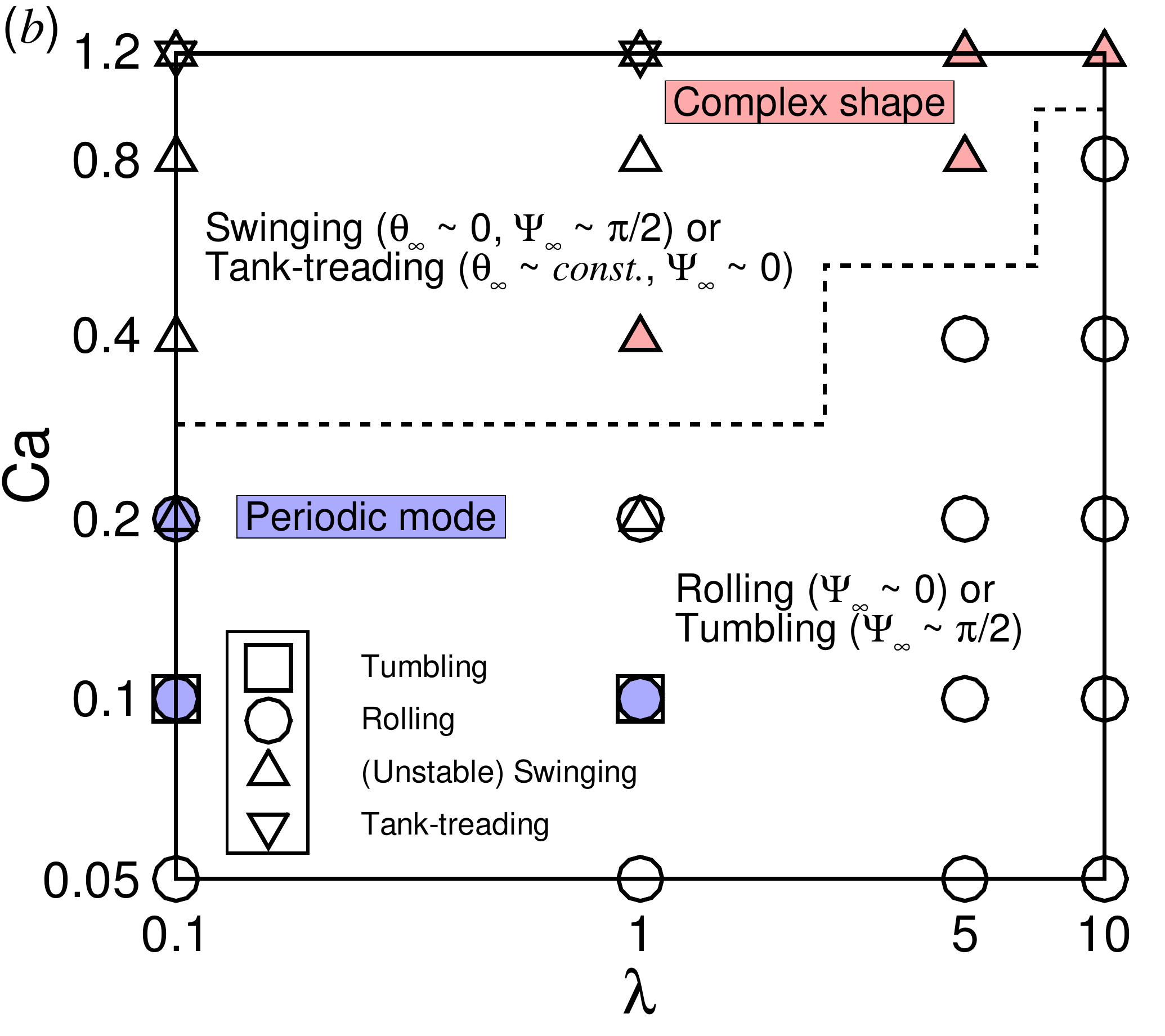}
  \caption{
  ($a$) Time average of the intrinsic viscosity $\eta$ obtained with different viscosity ratios $\lambda$ as a function of the logarithm of $Ca$ for each stable mode: $\Psi_{\infty}$ = 0 (dash lines) and $\Psi_{\infty}$ = $\pi/2$ (dash-dot lines). The results of $\Psi_{\infty}$ = 0 and $\Psi_{\infty} = \pi$/2 are obtained with initial orientation $\Psi_0$ = 0 and $\Psi_0$ = $\pi$/2, respectively.
  ($b$) Phase diagram of the stable modes of a single RBC as functions of the logarithm of $\lambda$ and $Ca$, where the squares ($\square$) denote the tumbling motion, circles ($\bigcirc$) the rolling motion, triangles ($\triangle$) the unstable or stable swinging motion, and inverse triangles ($\bigtriangledown$) the tank-treading motion. The solid blue dots represent the periodic motions (see the supplementary movie5), and the solid red dots the complex shapes which demonstrate an unstable swinging motion. The dash line separates the rolling from the (unstable) swinging motion. The results in ($b$) are obtained with random initial orientations.
  }
  \label{fig:single_rbc_mode}
\end{figure}
The effect of the viscosity ratio $\lambda$ on the intrinsic viscosity $\eta$ is quantified for each orientation (i.e., $\Psi_{\infty}$ = 0 or $\pi$/2), and shown in Fig.\ref{fig:single_rbc_mode}($a$), where the stable mode of $\Psi_{\infty}$ = 0 exhibits higher shear-thinning than the case of $\Psi_{\infty}$ = $\pi$/2 for all $\lambda$. To summarize, a phase diagram of stable modes of a single RBC based on the orientation angles is given in Fig.\ref{fig:single_rbc_mode}($b$) as functions of the viscosity ratio $\lambda$ and the logarithm of $Ca$, where all the results are obtained with the simulations started with random orientations $\Psi_0 = rand$. For low $Ca$, most of RBCs tend to show the rolling motion which corresponds to the rolling discocyte also reported by \citep{Mauer2018}, but some of them show an unstable periodic rolling motion even after a long period of time; in other words, the orientation angles do not converge during the simulation time (at least $\dot{\gamma}t \leq$ 1,000). As an example, the supplementary movie5 shows the result obtained with $\lambda$ = 0.1 at $Ca$ = 0.2. This periodic motion has been also called kayaking (oscillating-swinging) motion in previous numerical studies of biconcave capsules \citep{Cordasco2014, Sinha2015}, similarly to a classical Jeffery orbit \citep{Jeffery1922}. For increasing $Ca$, most of RBCs shift from the rolling to the
swinging motion almost independently of $\lambda$ (Fig.\ref{fig:single_rbc_mode}$b$). Based on several previous works on the dynamics of a single RBC, we define the swinging motion when $\theta$ is periodic and $\Psi \ne$ 0, while we define the tank-treading motion when $\theta$ is constant (\emph{const.}) and $\Psi \sim$ 0. In the swinging cases, especially for high viscosity ratios $\lambda \geq$ 5 (Fig.\ref{fig:single_rbc_mode}$b$), RBCs tend to show complex shapes (or multilobes) as $Ca$ increases. The stomatocyte, which can be assumed as one of the multilobe shapes, is found at $\lambda$ = 1 for $Ca$ = 0.4, which shifts to a stable swinging motion for higher $Ca$ (Fig.\ref{fig:single_rbc_mode}$b$). Hence, these complex shapes can be assumed as a transient to the stable swinging motion, and we thus call its mode ``\emph{unstable swinging motion}". \cite{Sinha2015} also showed that the biconcave capsule with $\lambda$ = 0.75, with membrane following the SK law ($G_s$ = 2.5 $\mu$N/m and $C$ = 10), transition from the tumbling to the oscillating-swinging motions for $\dot{\gamma} \sim$ 465 s$^{-1}$ and the to the tank-treading motion for $\dot{\gamma} \sim$ 930 s$^{-1}$ corresponding to $Ca \sim$ 1.12. \cite{Mauer2018} showed that tank-treading RBCs can only be found for low $\lambda$ ($\leq$ 3) and high $\dot{\gamma}$ ($\geq$ 820 s$^{-1}$), and that RBCs with $\lambda$ = 5 subject to high shear rates are tumbling stomatocytes or multilobes. The phase diagram that we obtained is indeed consistent with these literature results \citep{Mauer2018, Sinha2015}, since we also identify the tank-treading motion for $\lambda \leq$ 1 and $Ca$ = 1.2, and since RBCs with $\lambda \geq$ 5 subject to high $Ca$ exhibit multilobes shapes.
More precise descriptions of the stable modes of single RBCs are needed to investigate the effect of the initial orientation angle $\Psi_0$, which is however beyond the scope of present work. The dynamics of single RBC has been investigated in the past as well, e.g., in the study by \cite{Omori2012}, \cite{Cordasco2014}, \cite{Sinha2015} and \cite{Mauer2018}.

\subsection{Behavior of RBCs in semi-dilute and dense suspension}
\begin{figure}
  \centering
  \includegraphics[height=9cm]{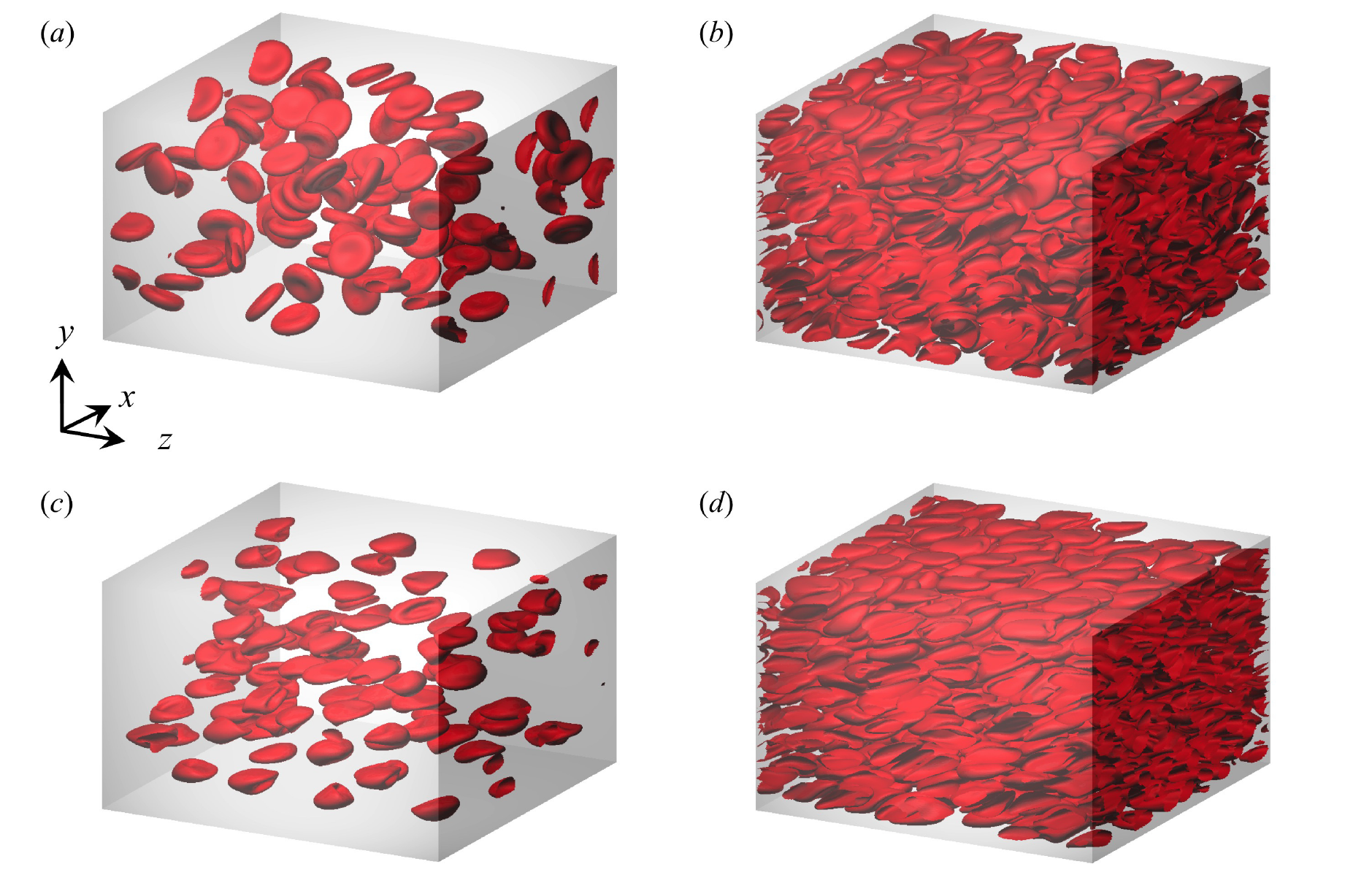}
  \caption{Snapshots of our numerical results for
  [($a$) and ($c$)] semi-dilute ($\phi$ = 0.05) and
  [($b$) and ($d$)] dense suspensions ($\phi$ = 0.41) with $\lambda$ = 5.
  The results for $Ca$ = 0.05 and 0.8 are reported in the top and bottom rows, respectively. The time evolution of these snapshots are shown in the supplementary movies: movie6 for ($a$), movie7 for ($b$), movie8 for ($c$) and movie9 for ($d$).
  }
  \label{fig:snapshots_suspension}
\end{figure}
\begin{figure}
  \centering 
  \includegraphics[height=5.5cm]{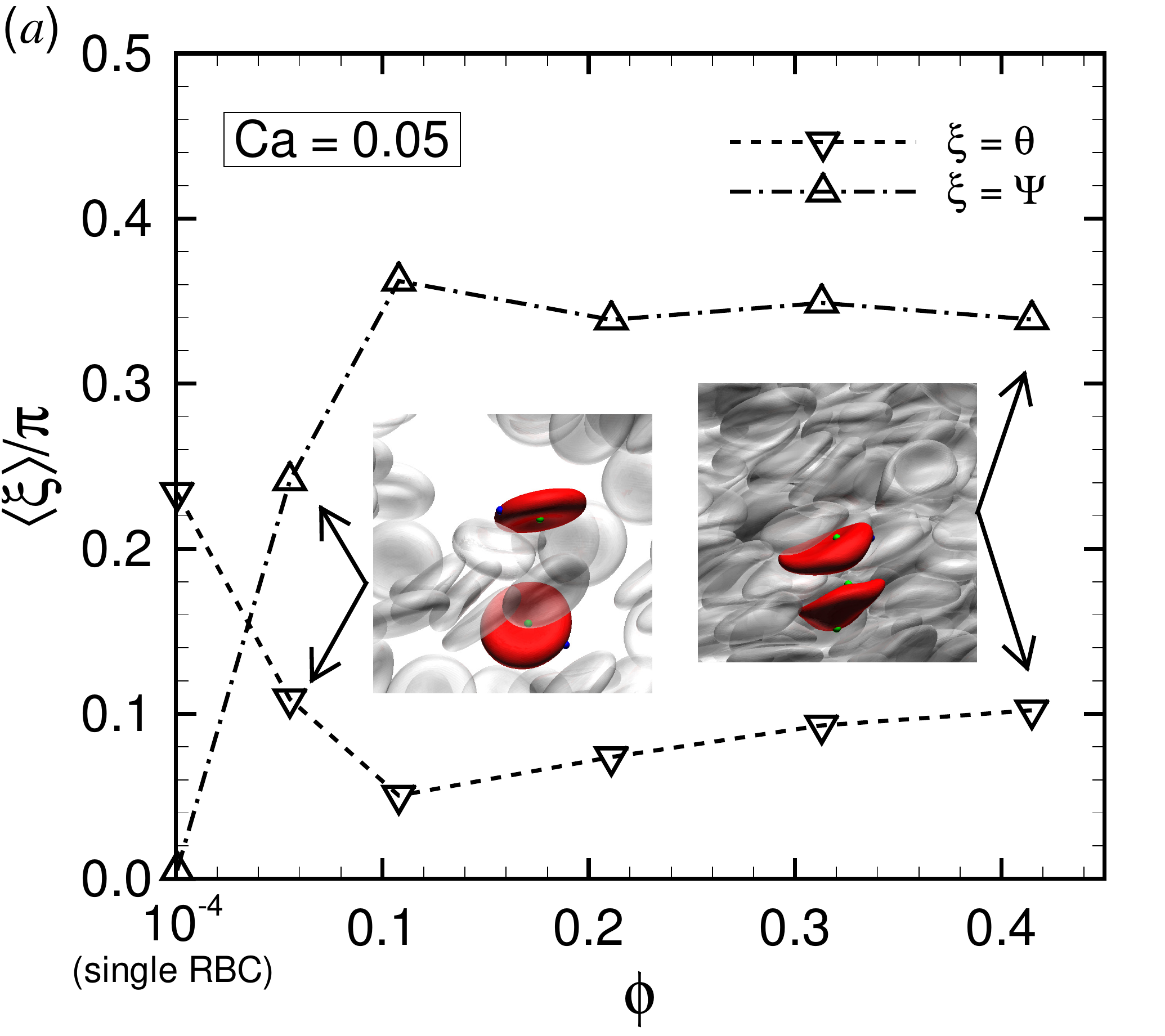}
  \includegraphics[height=5.5cm]{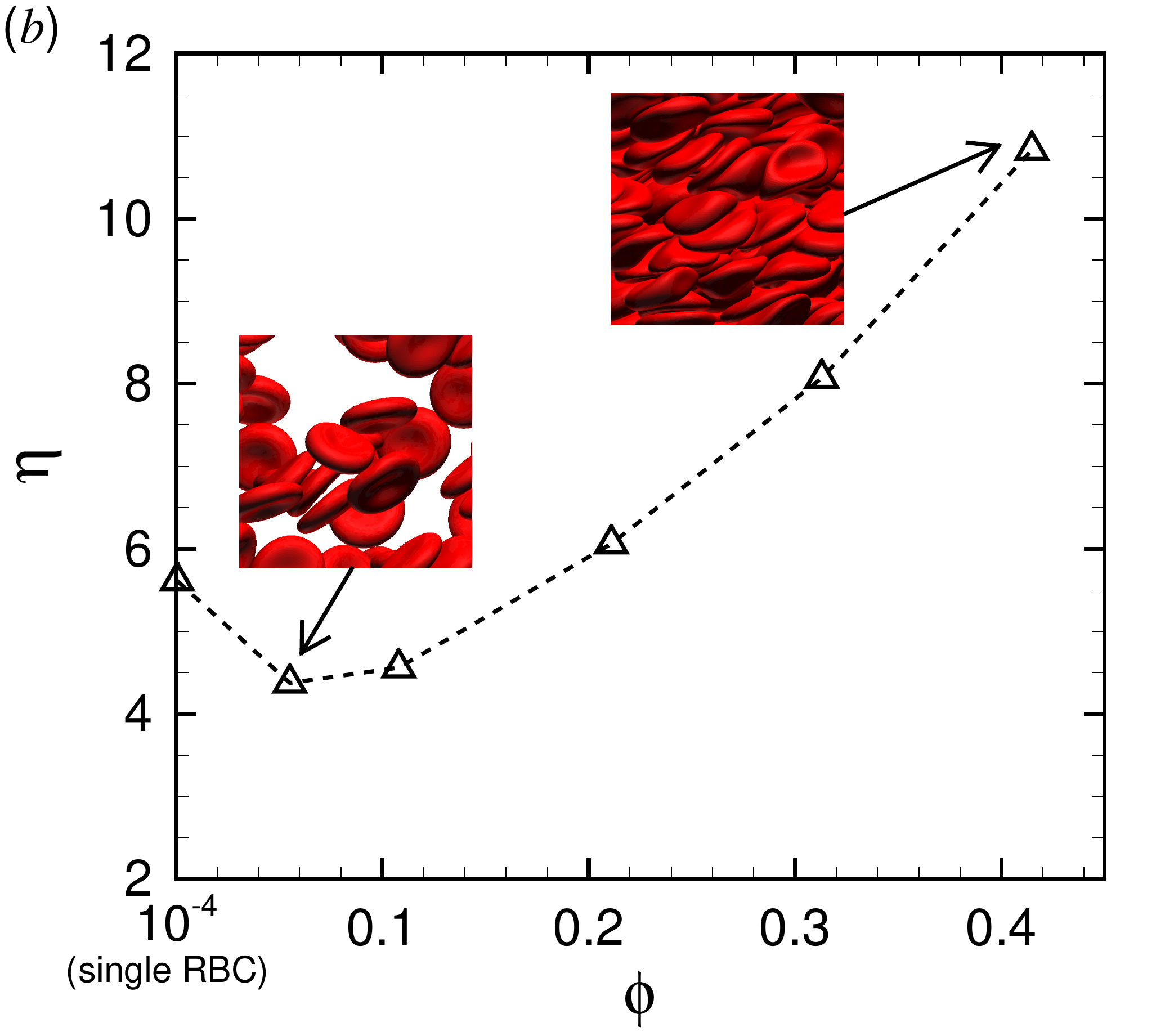}
  \includegraphics[height=5.5cm]{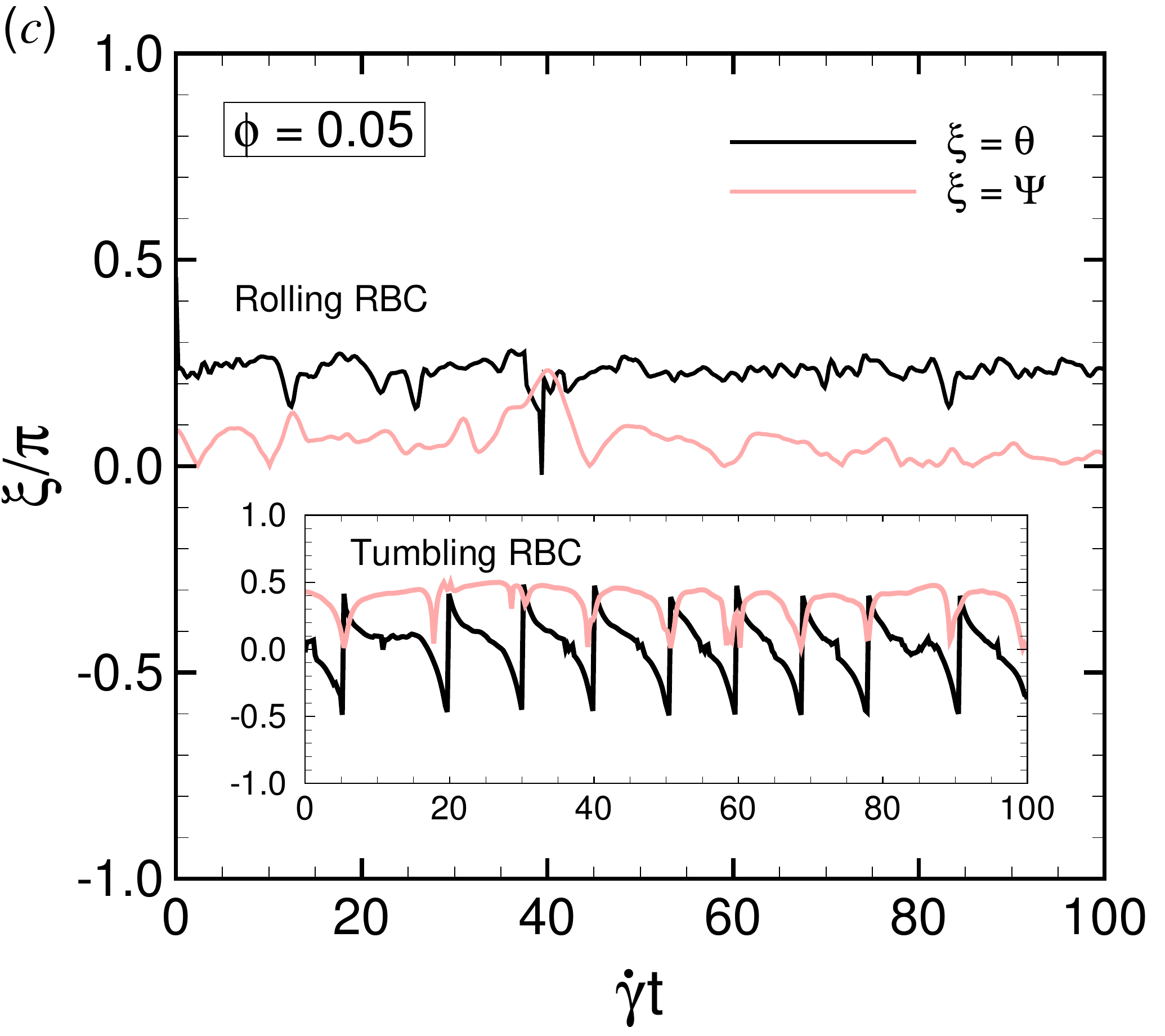}  
  \includegraphics[height=5.5cm]{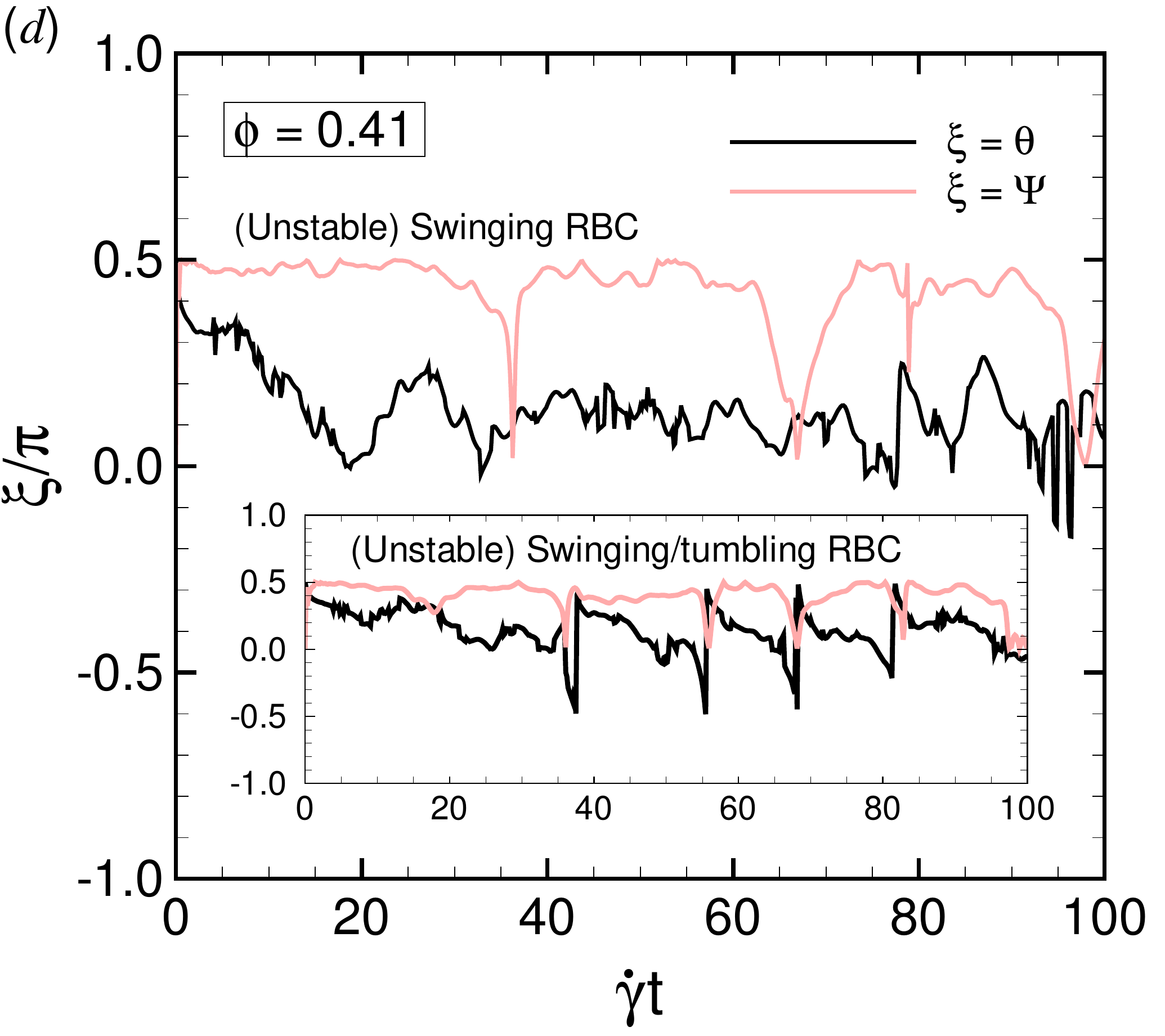}  
  \caption{
  ($a$) Ensemble average of the orientation angles of RBCs for $Ca$ = 0.05, and
  ($b$) the intrinsic viscosity $\eta$ as a function of the volume fraction $\phi$, where the inset figures are representative snapshots showing the coexistence of rolling and tumbling motions in a semi-dilute suspension and the unstable or stable swinging RBC in a dense suspension. These modes are identified by the time history of the orientation angles $\theta$ and $\Psi$ for
  ($c$) semi-dilute and
  ($d$) dense suspensions. These results are obtained with $\lambda$ = 5.
  }
  \label{fig:orientation_suspension_ca005lam5}
\end{figure}
Next, individual RBCs in semi-dilute and dense suspensions are investigated, and examples of snapshots of the numerical results are shown in Fig.\ref{fig:snapshots_suspension}, where, a semi-dilute suspension is defined for volume fraction $\phi$ = 0.05, and dense suspensions for the highest $\phi$ that we investigated, i.e., $\phi$ = 0.41. In semi-dilute suspensions, RBCs subjected to low $Ca$ = 0.05 show small deformations, where both the rolling and tumbling motions coexist (Fig.\ref{fig:snapshots_suspension}$a$) (see the supplementary movie6). In dense suspensions, however, due to the high packing, the RBCs are forced to exhibit only the swinging motion resulting in large elongations even for low $Ca$ = 0.05 (Fig.\ref{fig:snapshots_suspension}$b$; see the supplementary movie7). Indeed, as the volume fraction $\phi$ increases, the orientation angle $\Psi$ immediately increases and saturates around $\Psi \sim$ 0.34$\pi$, while the other orientation angle $\theta$ initially decreases and remains to be $\theta \leq$ 0.1$\pi$ as shown in Fig.\ref{fig:orientation_suspension_ca005lam5}($a$), where the insets represent enlarged views of semi-dilute suspensions showing the coexistence of the rolling and tumbling motions, and of dense suspensions dominated by swinging motions. The different motions are well characterized in the time history of both orientation angles as shown in Figs.\ref{fig:orientation_suspension_ca005lam5}($c$) and ($d$), respectively. These results clearly show that hydrodynamic interactions allow RBCs to shift from the rolling to the tumbling or swinging motions at different $Ca$ than for single RBCs. Since the tumbling and swinging motion is what allows low intrinsic viscosity $\eta$ as described above, hydrodynamic interactions decrease $\eta$ in semi-dilute suspensions but enhance it for high volume fractions as shown in Fig.\ref{fig:orientation_suspension_ca005lam5}($b$).

\begin{figure}
  \centering 
  \includegraphics[height=5.5cm]{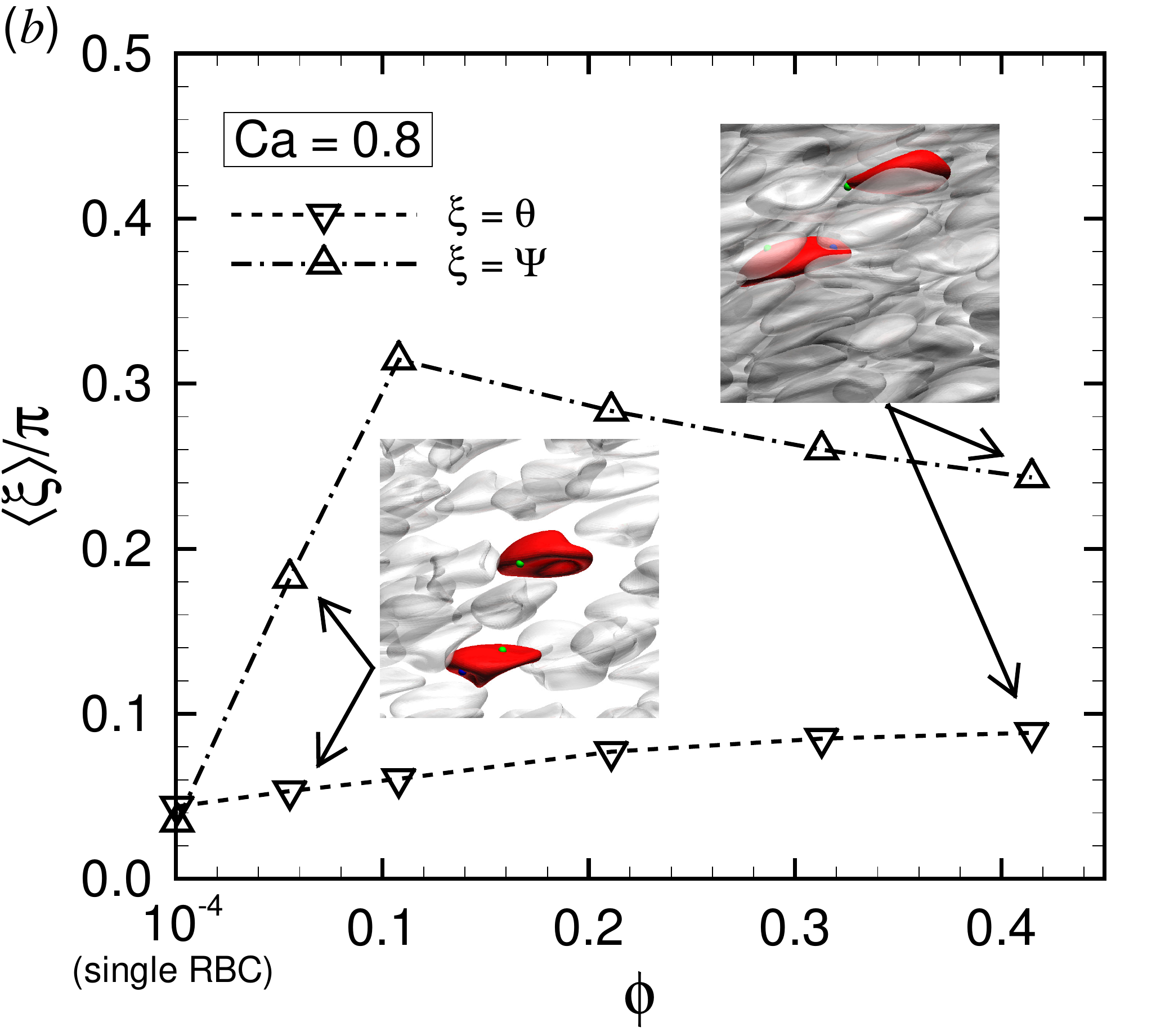}
  \includegraphics[height=5.5cm]{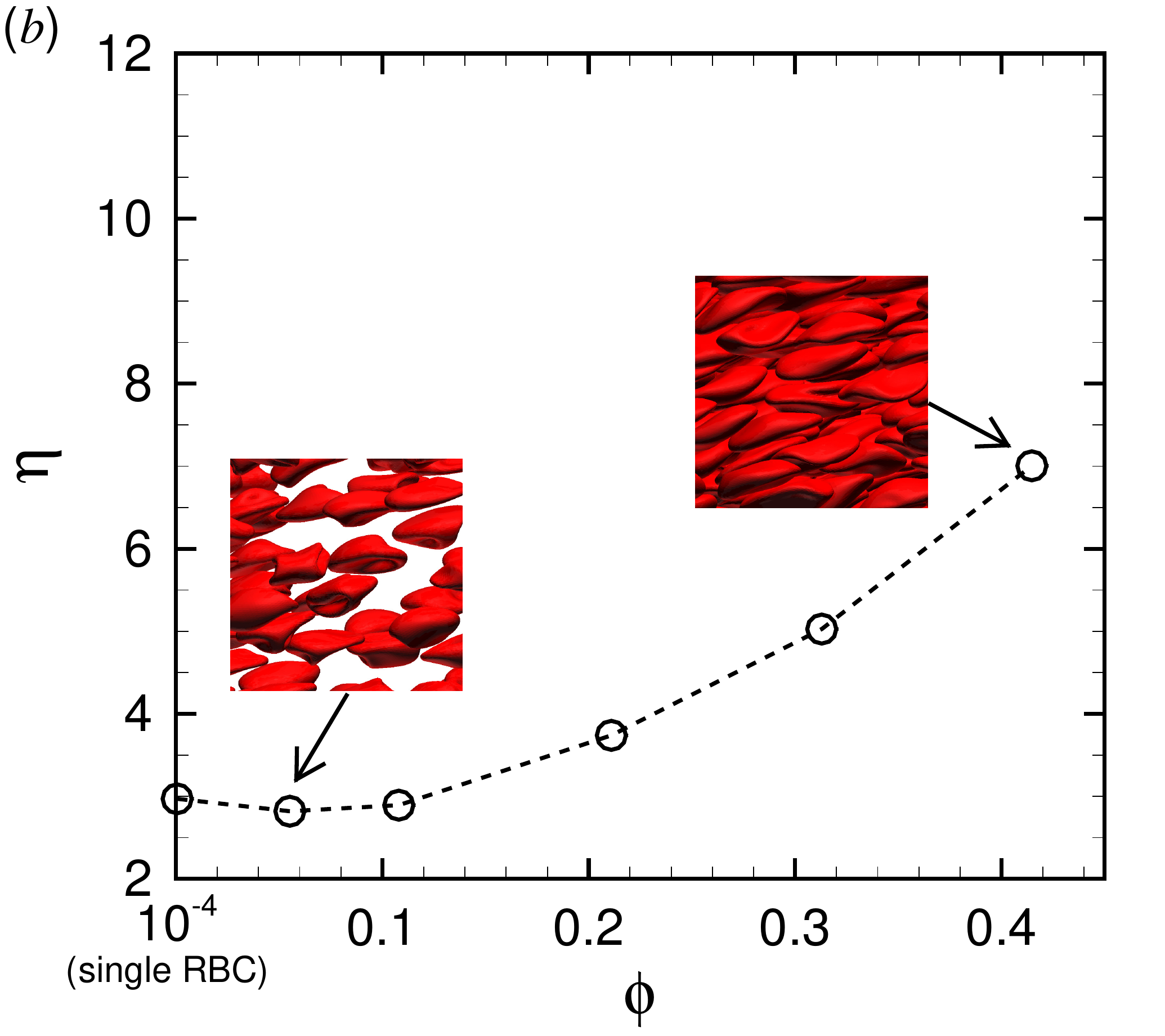}
  \includegraphics[height=5.5cm]{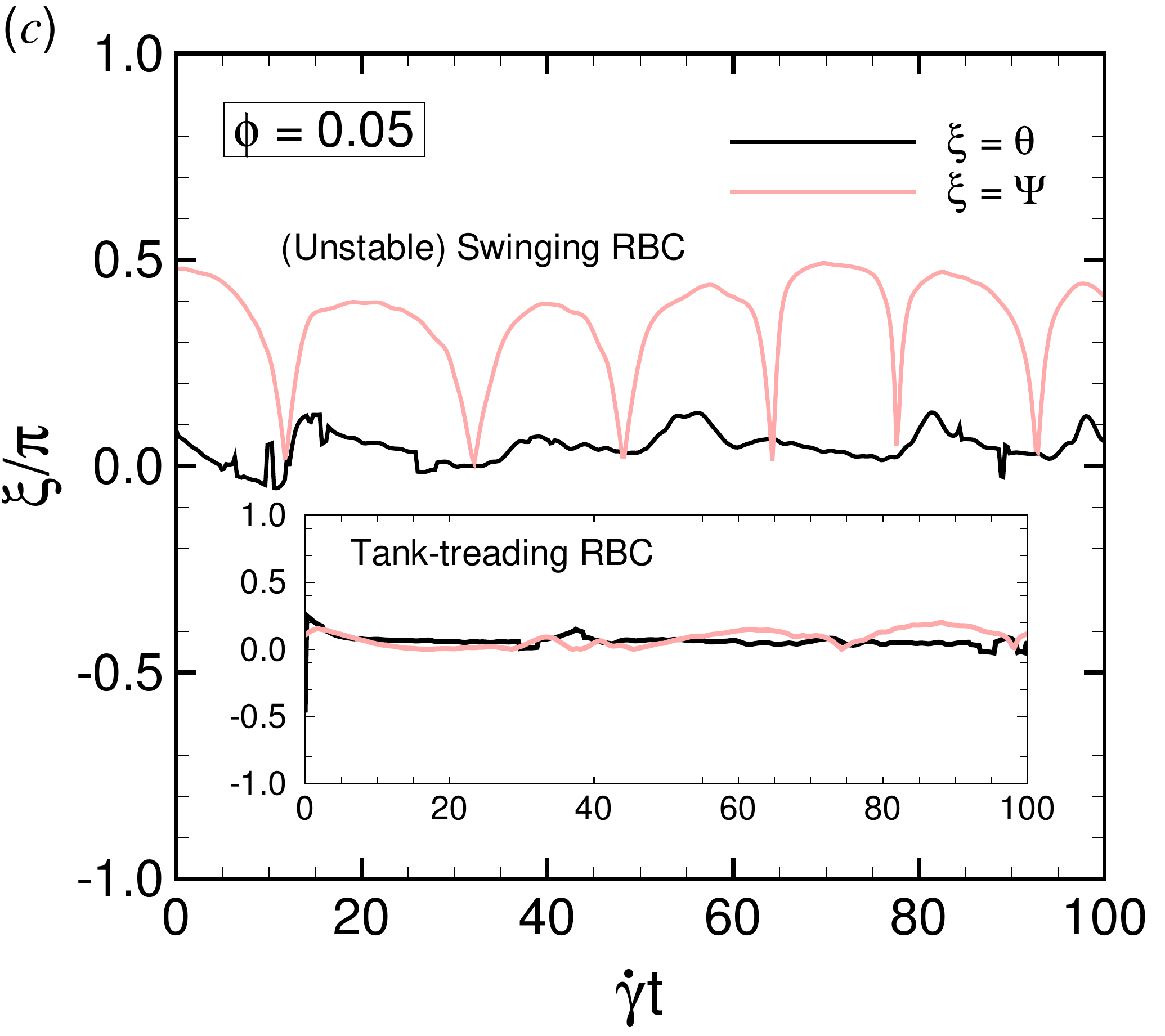}  
  \includegraphics[height=5.5cm]{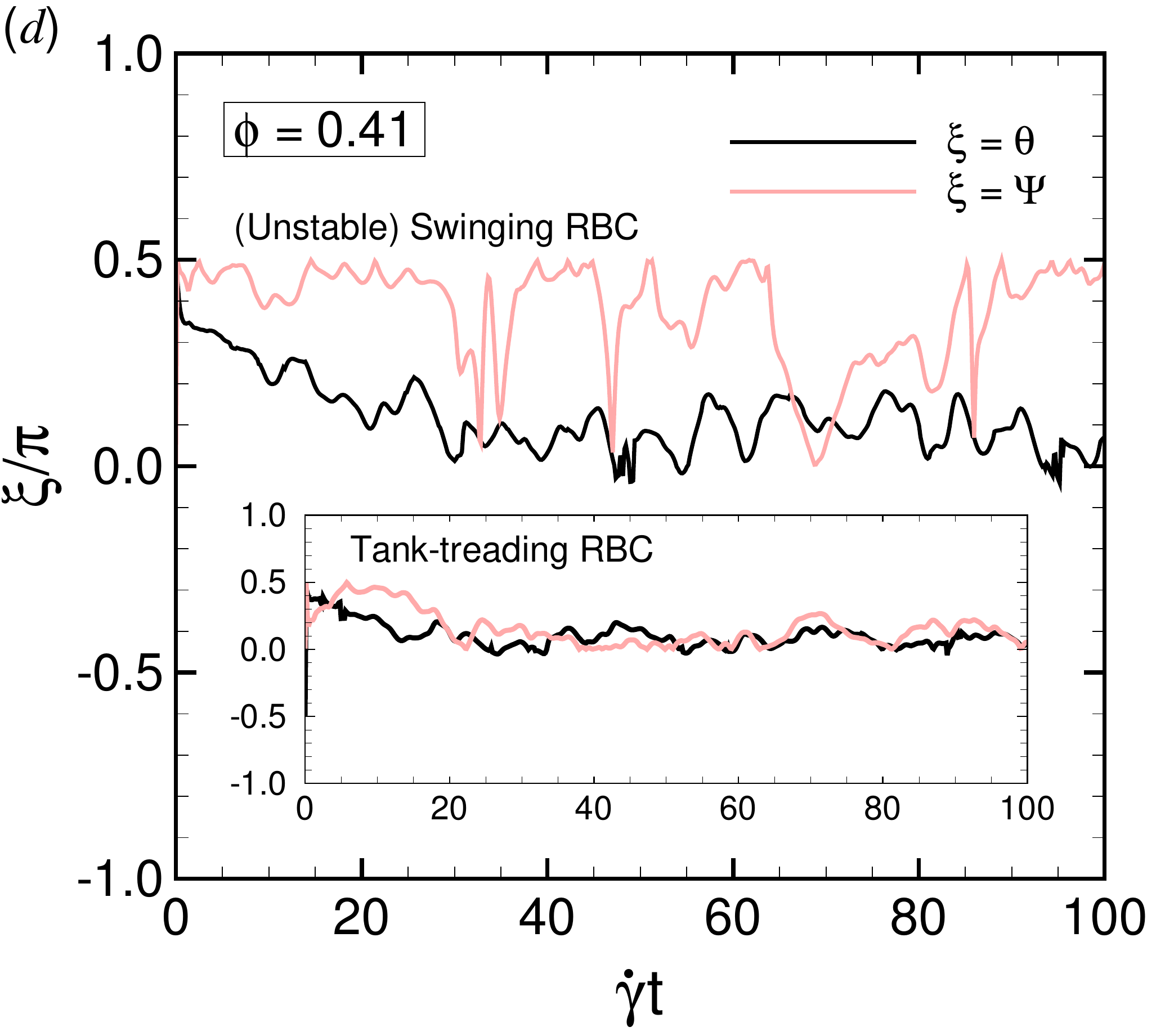}
  \caption{
  ($a$) Ensemble average of the orientation angles of RBCs for $Ca$ = 0.8, and
  ($b$) the intrinsic viscosity $\eta$ as a function of the volume fraction $\phi$.
  Time history of the orientation angles for a 
  ($c$) semi-dilute and
  ($d$) dense suspension. These results are obtained with $\lambda$ = 5.
  }
  \label{fig:orientation_suspension_ca08lam5}
\end{figure}
When $Ca$ increases to 0.8, individual RBCs in semi-dilute suspensions deform largely and show complex shapes (Fig.\ref{fig:snapshots_suspension}$c$; see the supplementary movie8). Since the deformation is induced by the hydrodynamic interactions, RBCs in dense suspensions elongate more than in the case with $Ca$ = 0.05 (Fig.\ref{fig:snapshots_suspension}$d$; see the supplementary movie9). Similarly to the case with $Ca$ = 0.05, the orientation angle $\Psi$ immediately increases as $\phi$ increases, but it slightly decreases from $\phi \sim$ 0.1 onwards. Since RBCs subject to high $Ca$ tend to show mostly swinging motions without multi-cellular interactions, the orientation angle $\theta$ is already small in the case of dilute suspensions (single RBC level), and slightly increases with $\phi$ as shown in Fig.\ref{fig:orientation_suspension_ca08lam5}($a$), where the insets represent enlarged views of semi-dilute and dense suspensions showing the coexistence of the swinging and tank-treading motions. Here, we define the tank-treading motion as $\Psi \sim$ 0 and $\theta \sim$ 0 as in the previous numerical study by \cite{Omori2012}. These are characterized in the time history by the two orientation angles reported in Figs.\ref{fig:orientation_suspension_ca08lam5}($c$) and \ref{fig:orientation_suspension_ca08lam5}($d$), where the swinging RBCs are the ones fluctuating in $\Psi$ (i.e., $\Psi \neq \pi$/2). Since rolling motions do not exist for $Ca$ = 0.8, the intrinsic viscosity does not much drop and it almost monotonically increases with $\phi$ as shown in Fig.\ref{fig:orientation_suspension_ca08lam5}($b$), where the insets display instantaneous configurations.

\begin{figure}
  \centering
  \includegraphics[height=5.5cm]{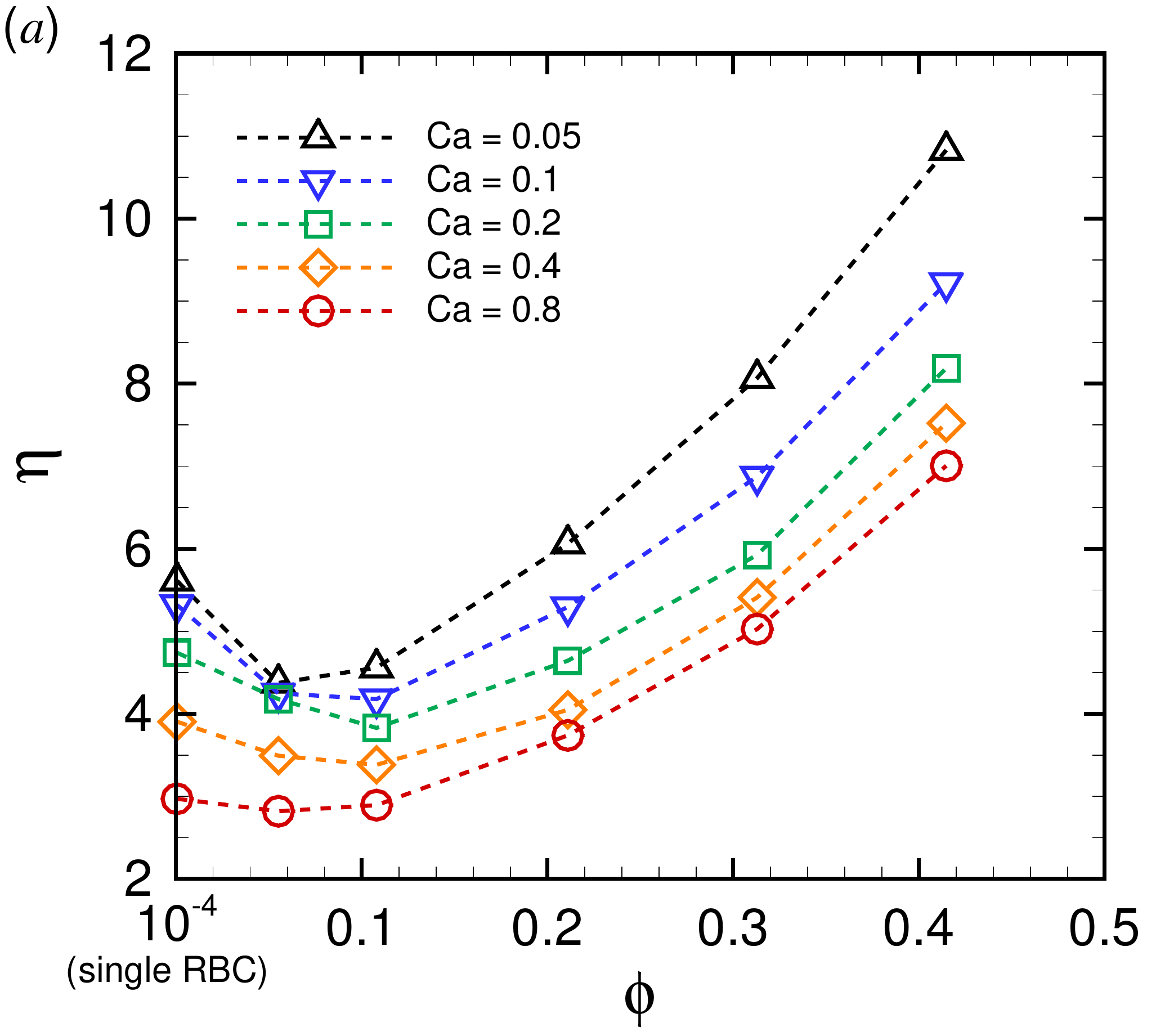}
  \includegraphics[height=5.5cm]{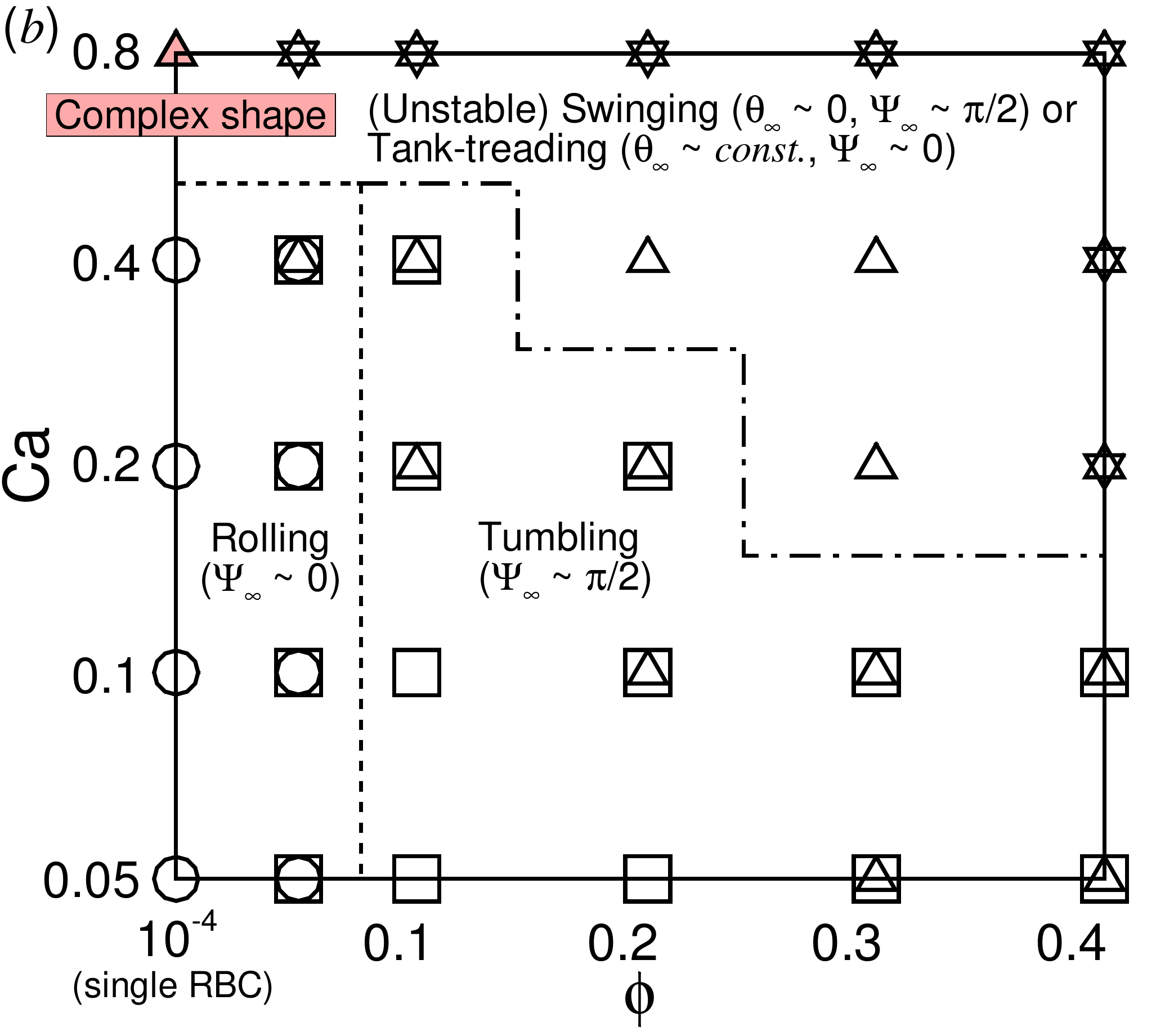} 
  \includegraphics[height=5.5cm]{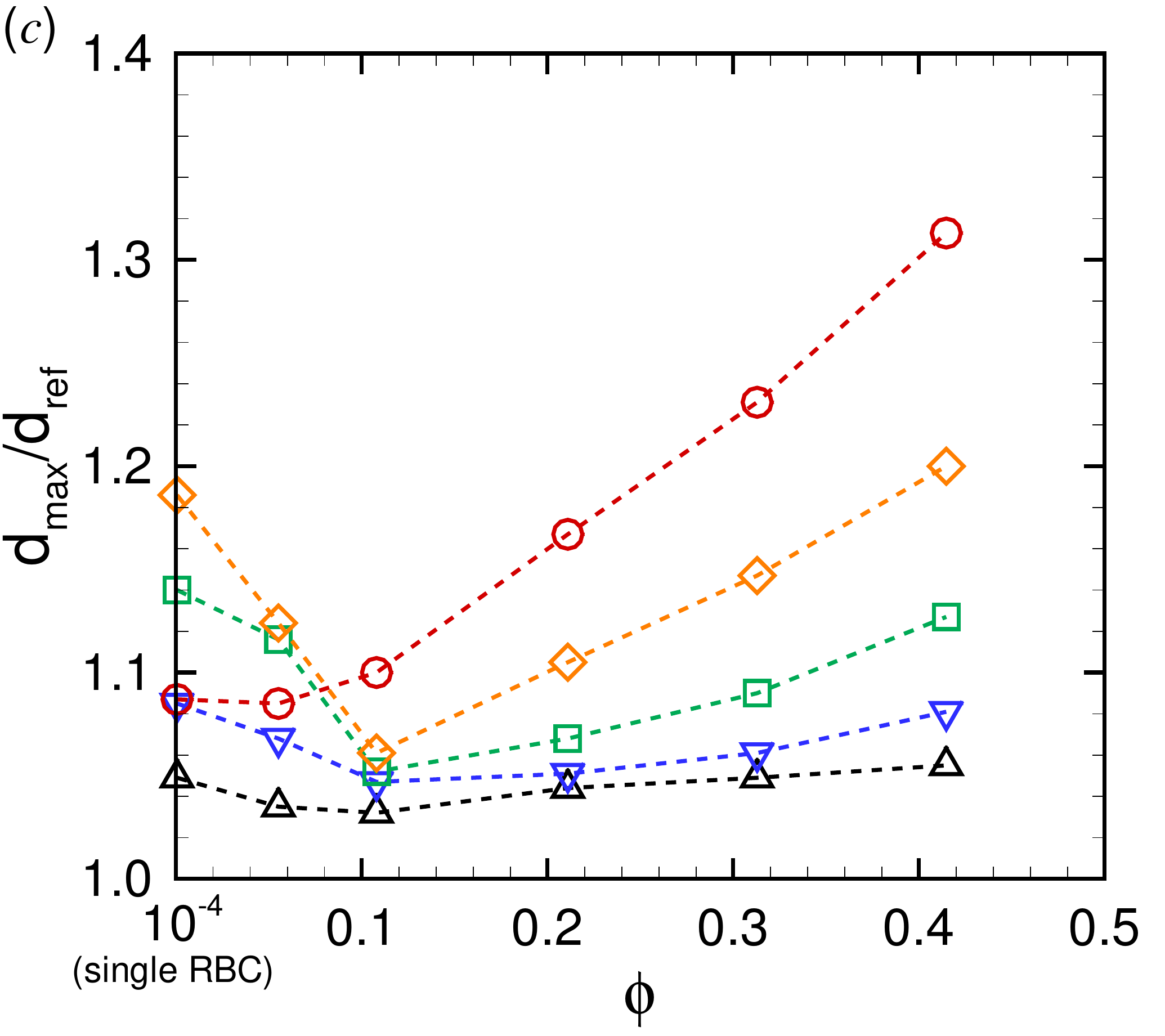}
  \includegraphics[height=5.5cm]{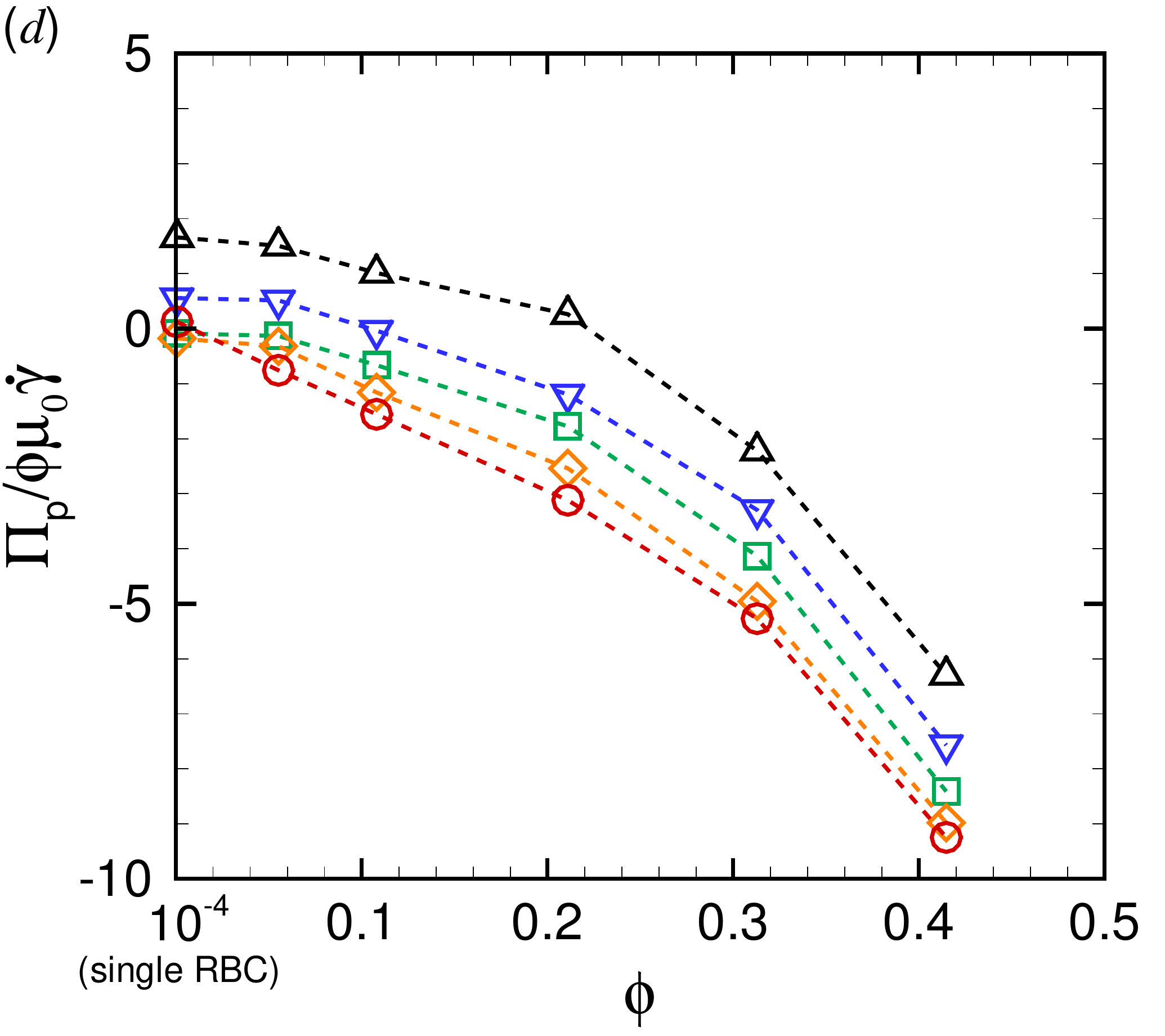} 
  \includegraphics[height=5.5cm]{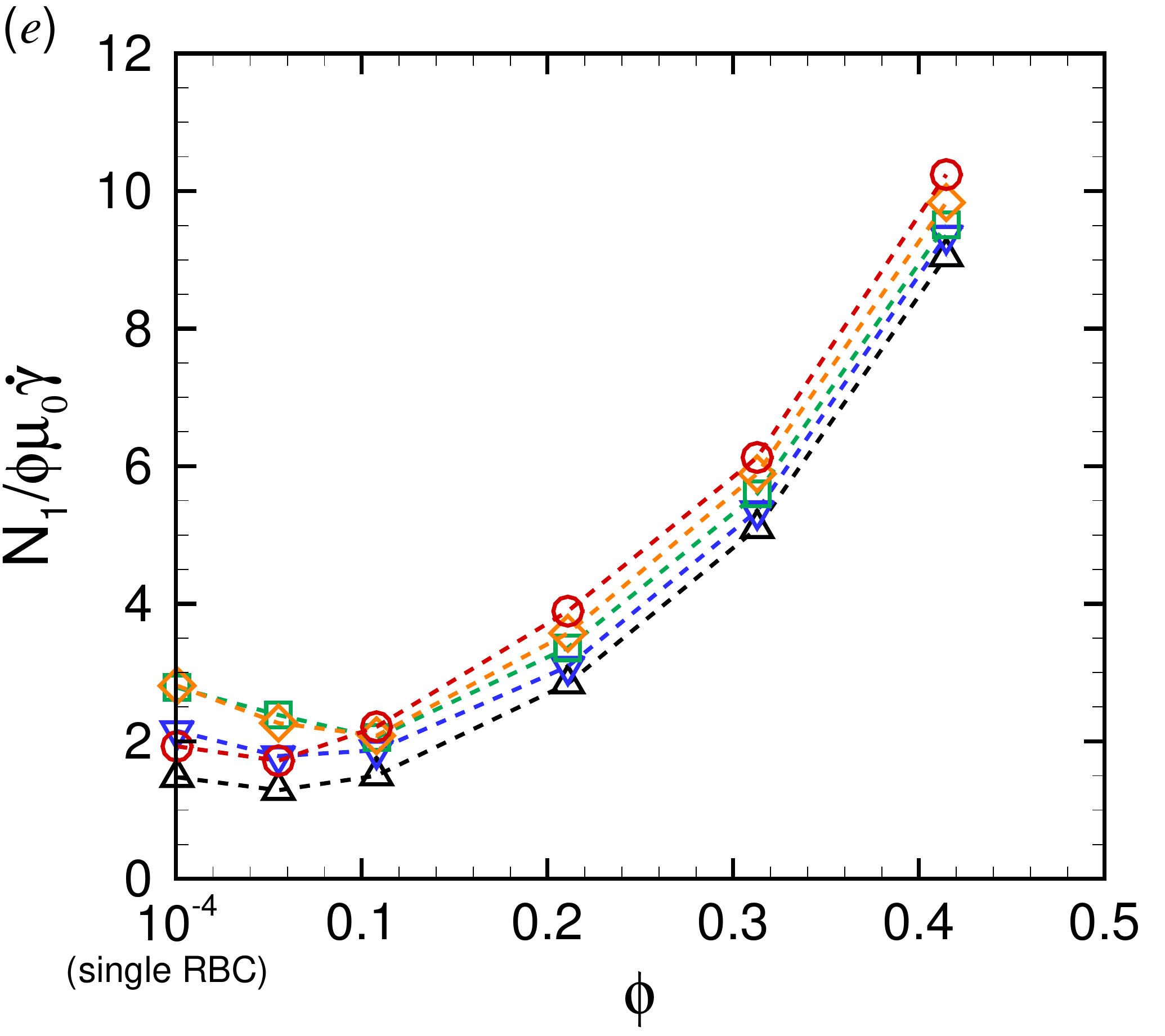}
  \includegraphics[height=5.5cm]{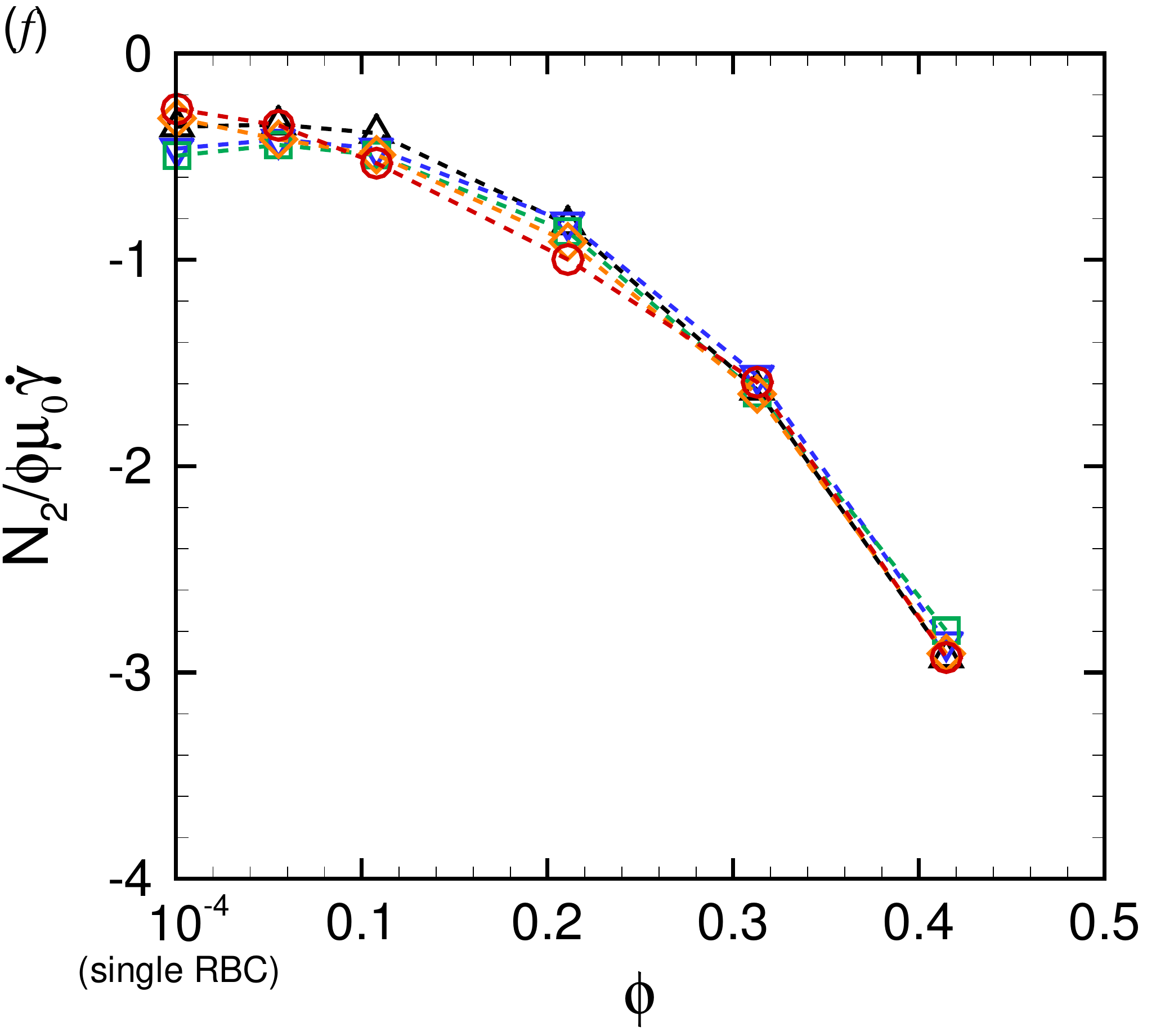}
   \caption{
  ($a$) The intrinsic viscosity $\eta$,
  ($b$) the phase diagram of the stable modes of RBCs as functions of $\phi$ and logarithm of $Ca$, where the squares ($\square$) denote the tumbling motion, circles ($\bigcirc$) the rolling motion, triangles ($\triangle$) the unstable or stable swinging motion and the inverse triangles ($\bigtriangledown$) the tank-treading motion. The red triangles denote the complex shape which demonstrates unstable swinging motions. The dash line separates the rolling from the other motions, and the dash-dot line the tumbling motions from the pure swinging/tank-treading motion.
  ($c$) The deformation index $d_{max}/d_{ref}$ , where $d_{max}$ and $d_{ref}$ are the maximum distances between two points on the deformed and reference membranes (i.e., no flow condition), respectively.
  ($d$) The particle pressure $\Pi_p/\phi\mu_0 \dot{\gamma}$ as a function of the volume fraction $\phi$ for different $Ca$.
  [($e$) and ($f$)] The first and  second normal stress difference $N_i/\phi\mu_0\dot{\gamma}$ ($i$ = 1 and 2), respectively.
  These results are obtained with $\lambda$ = 5.
  }
  \label{fig:multi-rbc_lam5}
\end{figure}
To summarize the results of stable modes of individual RBCs in dilute (single cell level), semi-dilute ($\phi \sim$ 0.05) and dense suspension ($\phi \sim$ 0.4), we show the intrinsic viscosity $\eta$ as a function of $\phi$ for different $Ca$ in Fig.\ref{fig:multi-rbc_lam5}($a$), where the drop of $\eta$ in semi-dilute suspensions is commonly found at each $Ca$. The drop of $\eta$ in semi-dilute suspensions is explained by the mode change from the rolling to tumbling motion, as reported in the phase diagram (Fig. \ref{fig:multi-rbc_lam5}$b$). Increasing $\phi$, the probability of swinging RBCs increases resulting in the increase of $\eta$ observed in the figures \ref{fig:multi-rbc_lam5}($a$) and \ref{fig:multi-rbc_lam5}($b$). To see the relationship between the intrinsic viscosity of the suspension and the RBCs deformation at high volume fractions, we show the deformation index $d_{max}/d_{ref}$ in Fig.\ref{fig:multi-rbc_lam5}($c$), where $d_{max}$ and $d_{ref}$ are the maximum distances between two points on the deformed and reference (i.e., without flow) membranes. We observe that $d_{max}/d_{ref}$ increases similarly to $\eta$ for relatively high volume fractions ($\phi >$ 0.05), and hence high $\eta$ can be associated with large deformation of swinging RBCs. The increase with $\phi$ is also found for the first normal stress difference $N_1/\phi\mu_0\dot{\gamma}$, while no significant differences are evident when varying $Ca$ (Fig.\ref{fig:multi-rbc_lam5}$e$). Instead, the second normal stress differences $N_2/\phi\mu_0\dot{\gamma}$ decreases with $\phi$, and again almost no difference is evident for different $Ca$ (Fig.\ref{fig:multi-rbc_lam5}$f$).
The particle pressure $\Pi_p/\phi \mu_0 \dot{\gamma}$ also decreases with $\phi$ (Fig.\ref{fig:multi-rbc_lam5}$d$), similarly to the second normal stress difference (Fig.\ref{fig:multi-rbc_lam5}$f$), although the effect of $Ca$ is more pronounced.

\begin{figure}
  \centering
  \includegraphics[height=5.5cm]{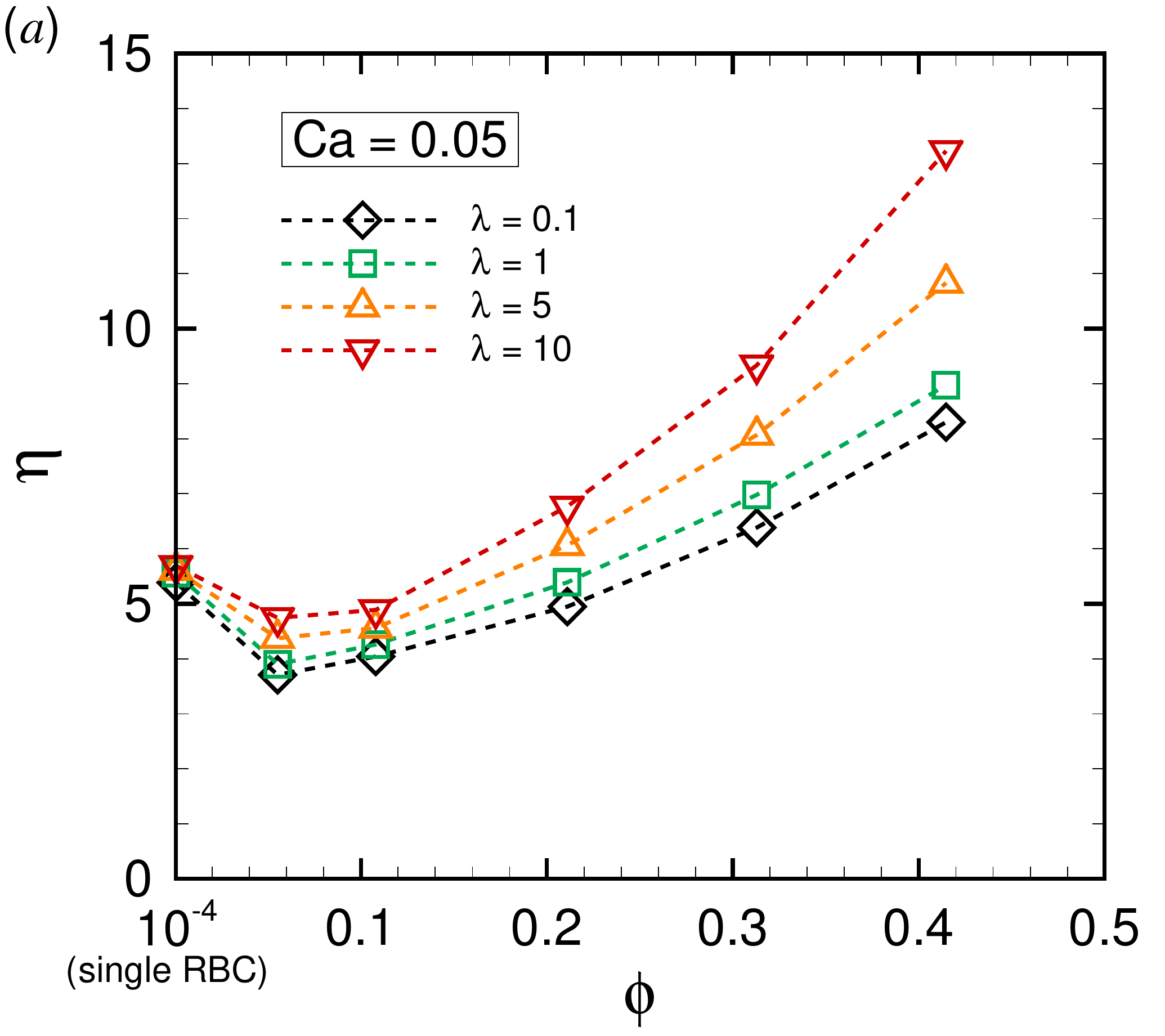}
  \includegraphics[height=5.5cm]{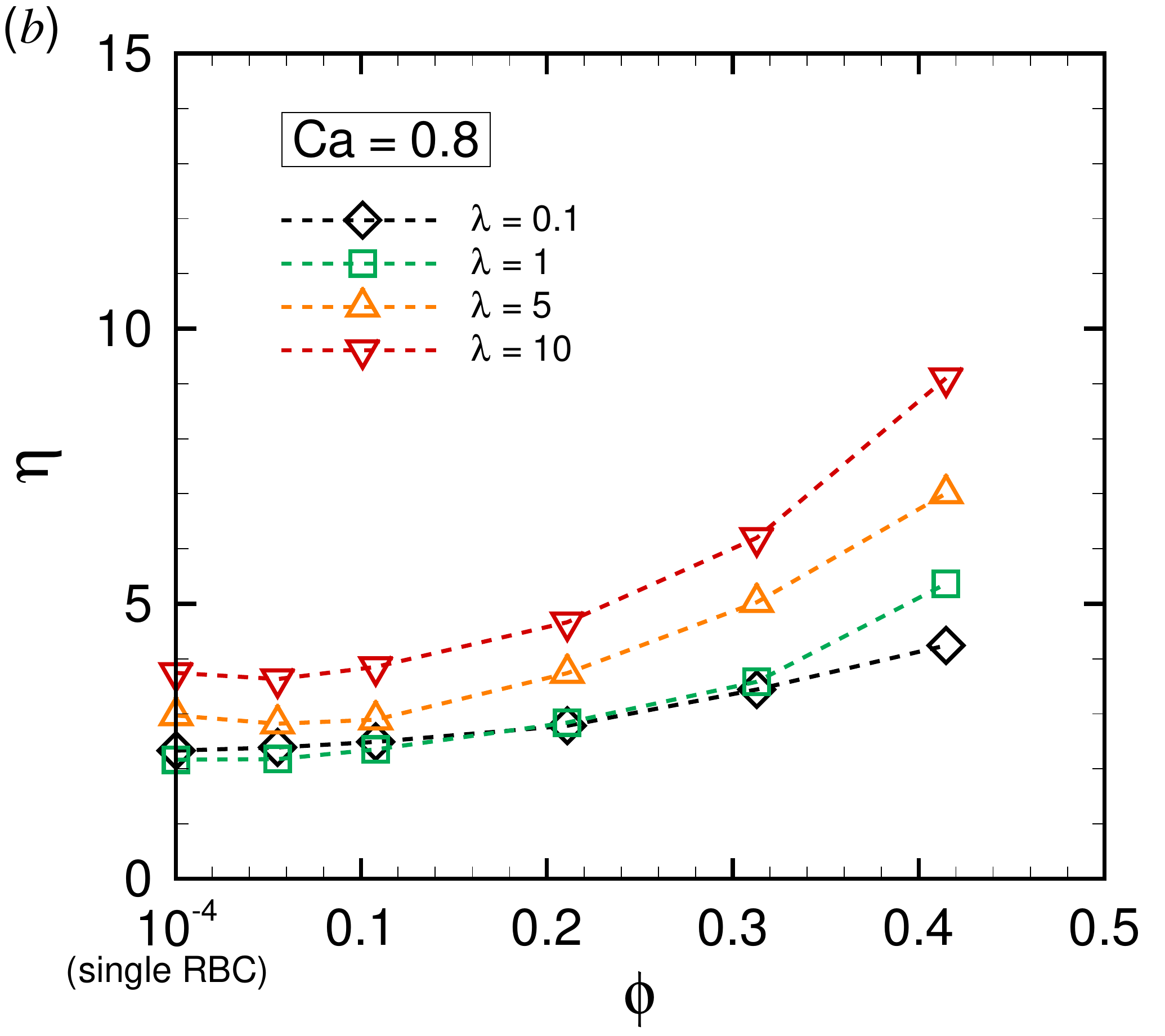}
  \includegraphics[height=5.5cm]{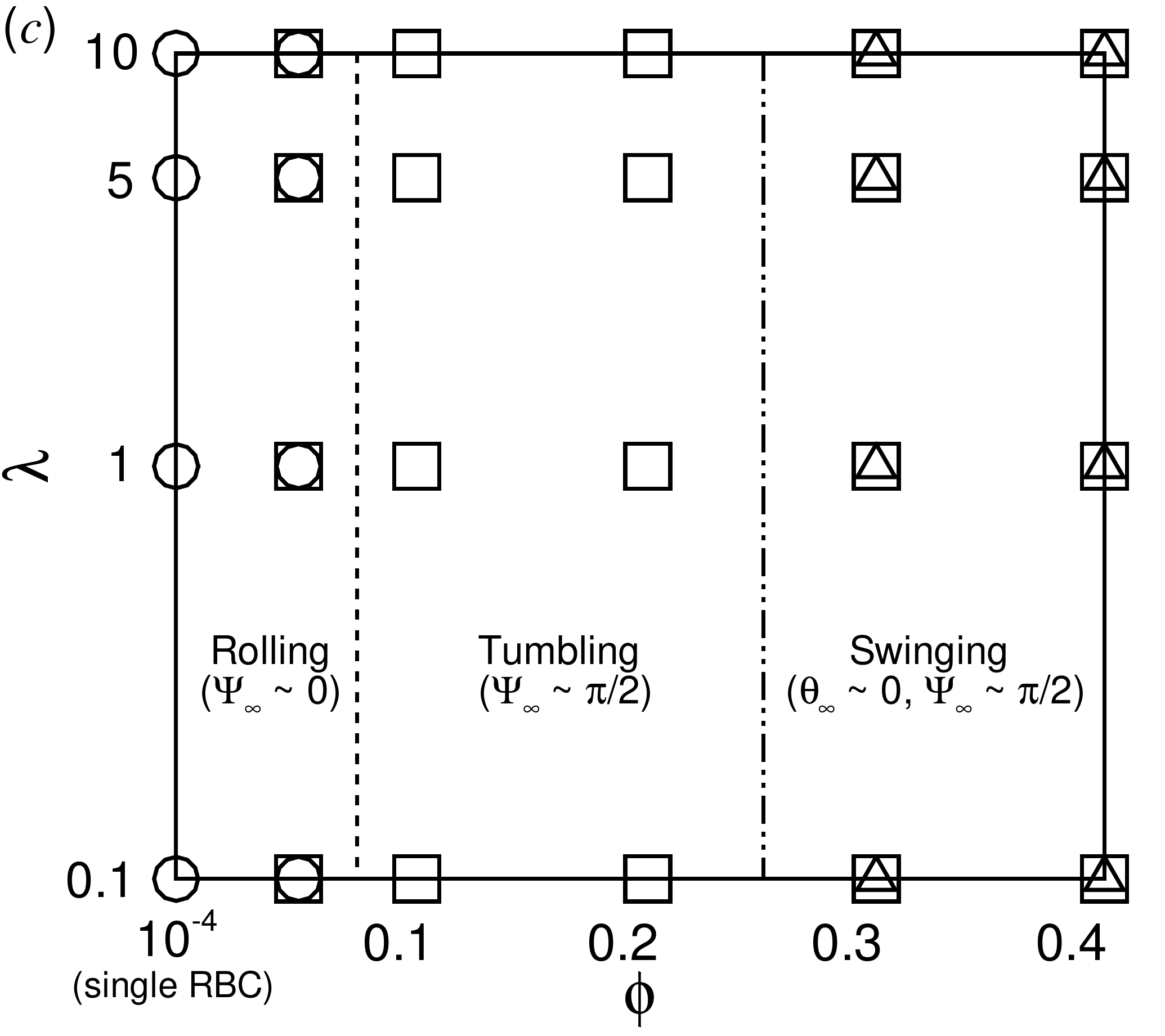}
  \includegraphics[height=5.5cm]{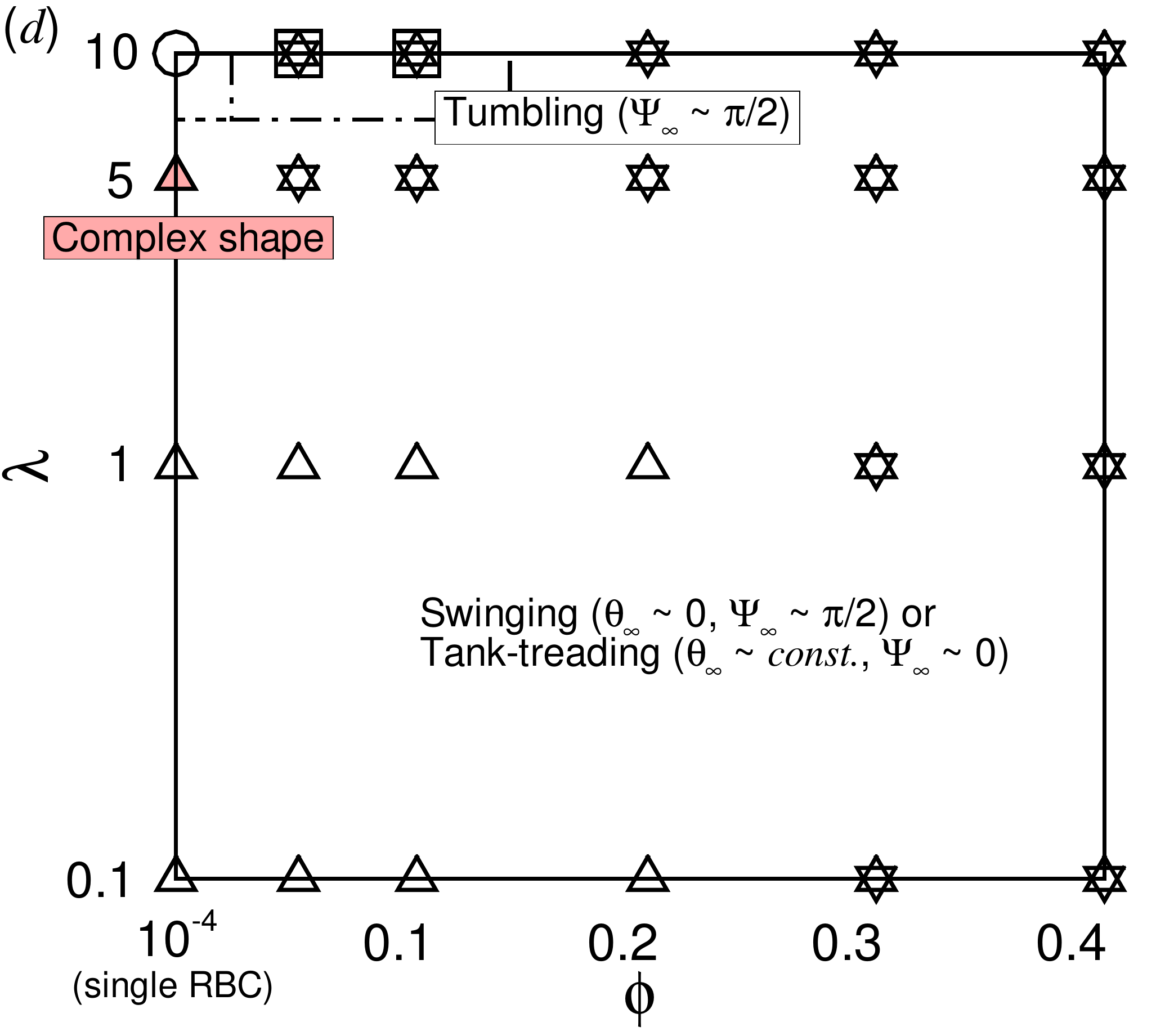}
  \includegraphics[height=5.5cm]{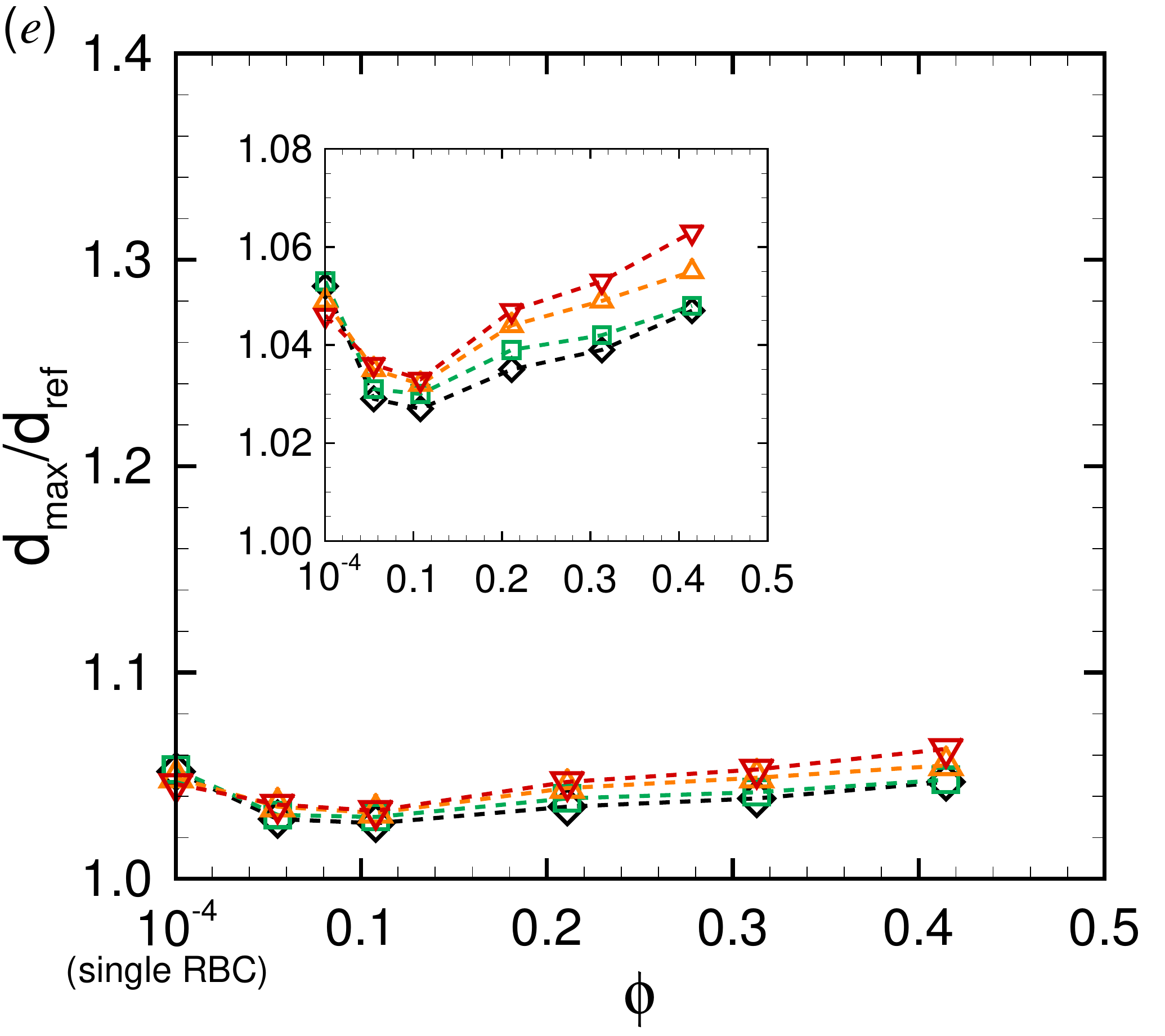}
  \includegraphics[height=5.5cm]{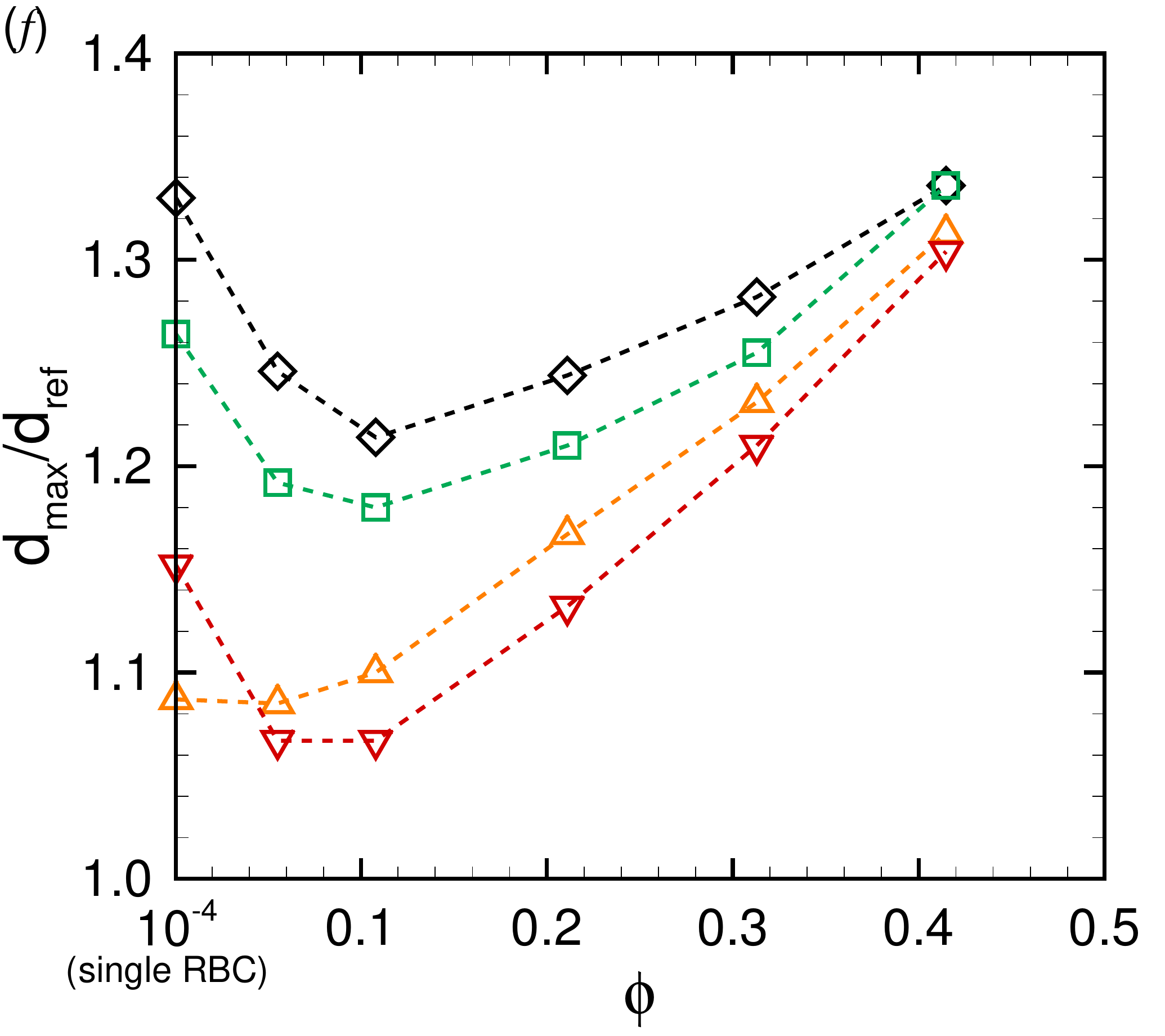}  
  \caption{
  [($a$) and ($b$)] The intrinsic viscosity $\eta$ as a function of the volume fraction $\phi$ for ($a$) low $Ca$ = 0.05 and ($b$) high $Ca$ = 0.8.
  [($c$) and ($d$)] The phase diagrams of the stable modes of RBCs as functions of $\phi$ and logarithm of $\lambda$ for ($e$) $Ca$ = 0.05 and ($f$) $Ca$ = 0.8, where the squares ($\square$) denote the tumbling motion, circles ($\bigcirc$) the rolling motion, triangles ($\triangle$) the unstable or stable swinging motion and the inverse triangles ($\bigtriangledown$) the tank-treading motion. The solid red triangles denote the complex shapes which demonstrate unstable swinging motions. The dash-line in ($e$) separates the rolling motion from the tumbling motion, the dash-two-dot line in ($e$) separates the pure tumbling motion from the swinging motion, and the dash-dot line in ($f$) separates the tumbling motion from the swinging/tank-treading motion.
  [($e$) and ($f$)] The deformation index $d_{max}/d_{ref}$ for ($e$) low $Ca$ = 0.05 and ($f$) high $Ca$ = 0.8.
  }
  \label{fig:multi-rbc_ca005ca08}
\end{figure}
The effects of $\lambda$ on the intrinsic viscosity $\eta$ and on the stable mode are investigated at a fixed $Ca$, and the results are shown in Fig.\ref{fig:multi-rbc_ca005ca08}. For low $Ca$ (= 0.05), the drop of $\eta$ in semi-dilute suspensions ($\phi \sim$ 0.05) is observed independently of the value of $\lambda$ (Fig.\ref{fig:multi-rbc_ca005ca08}$a$) because the mode changes from the rolling to the tumbling motion (Fig.\ref{fig:multi-rbc_ca005ca08}$c$). For high Ca (= 0.8), the mode change is not evident in semi-dilute suspensions (Fig.\ref{fig:multi-rbc_ca005ca08}$d$), and hence the drop of $\eta$ is not found for every $\lambda$ (Fig.\ref{fig:multi-rbc_ca005ca08}$b$). As shown in Figs.\ref{fig:multi-rbc_ca005ca08}($e$) and \ref{fig:multi-rbc_ca005ca08}($f$), the increase of $\eta$ for relatively high volume fractions can also be explained by large deformations of swinging RBC independently of $\lambda$.

\section{Discussion}
\subsection{Comparison with experiments}
\begin{figure}
  \centering 
  \includegraphics[height=5.5cm]{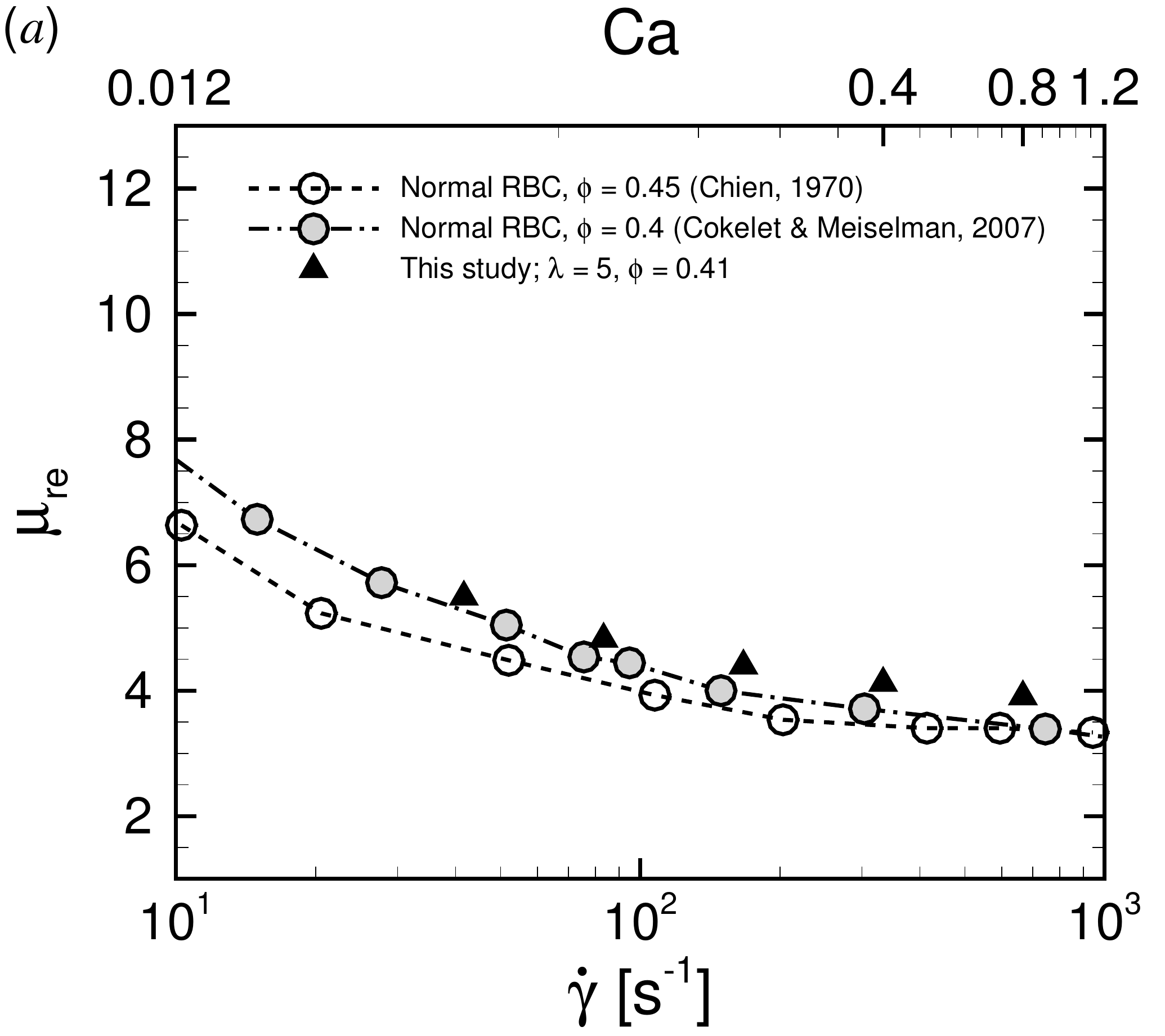} \\
  \includegraphics[height=5.5cm]{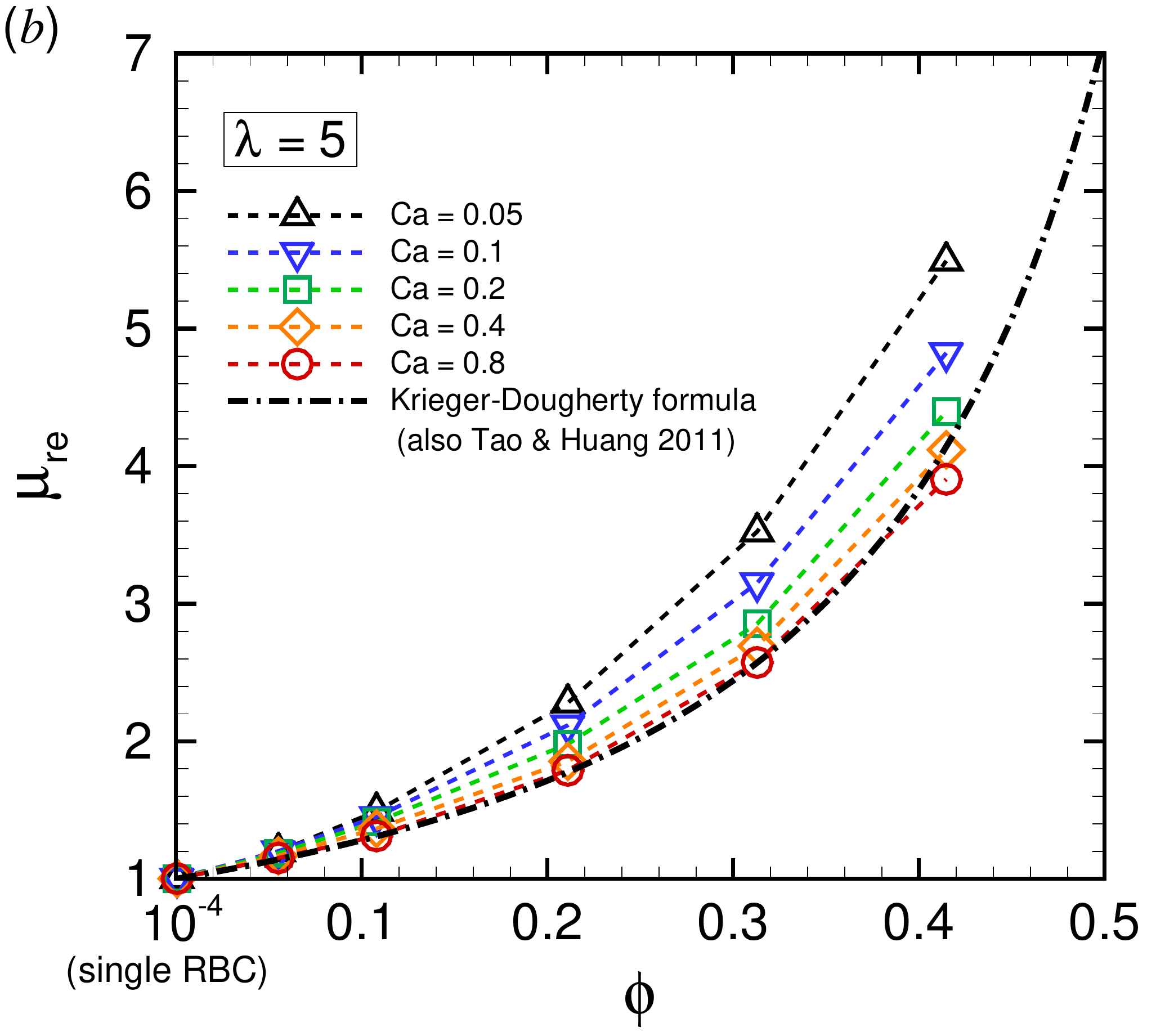}
  \includegraphics[height=5.5cm]{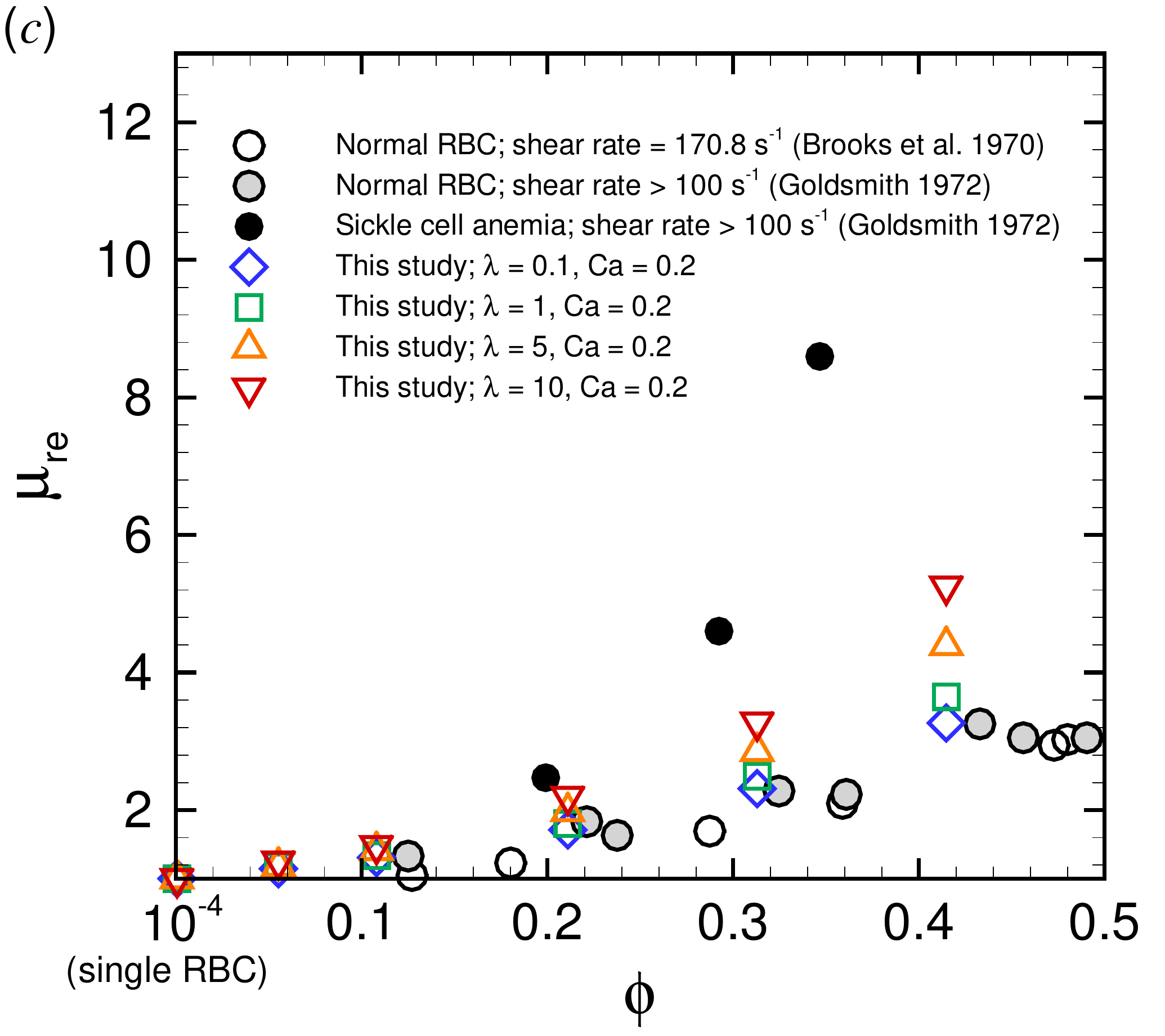}
  \caption{
  ($a$) Relative viscosity $\mu_{re}$ for $\phi$ = 0.41 as a function of the logarithm of the shear rate $\dot{\gamma}$ obtained with $\lambda$ = 5. The experimental results of normal human RBC suspension in plasma, 11\% albumin-Ringer solution for $\phi$ = 0.45 at 37$^{\circ}$C \citep{Chien1970a}, and in plasma at $\phi$ = 0.4 \citep{Cokelet2007} are also displayed as white and gray circles, respectively. The viscosity of 11\% albumin-Ringer solution is the same as plasma, i.e., $\mu_0$ = 1.2 cP.
  ($b$) Relative viscosity $\mu_{re}$ obtained with $\lambda$ = 5 as a function of the volume fraction $\phi$ for different $Ca$. The empirical expression reported in equation (\ref{Krieger1959}) with the parameters proposed by \cite{Tao2011} is also displayed as a dash-dot line.
  ($c$) Relative viscosity $\mu_{re}$ as a function of volume fraction $\phi$ for different viscosity ratios $\lambda$ at $Ca$ = 0.2 ($\dot{\gamma}$ = 167 s$^{-1}$). The experimental results of the suspension of acetaldehyde-fixed human RBCs in plasma for $\dot{\gamma}$ = 170.8 s$^{-1}$ are also displayed as white circles \citep{Brooks1970}. Moreover, the experimental data of a suspension of normal and sickle human RBCs for $\dot{\gamma} >$ 100 s$^{-1}$ are also displayed as black and gray circles, respectively \citep{Goldsmith1972}.
  }
  \label{fig:comparison_exp}
\end{figure}
Our numerical results of the relative viscosities $\mu_{re}$ in dense suspensions ($\phi$ = 0.41) are compared with the previous experimental results by \cite{Chien1970a} and \cite{Cokelet2007}. In the study by \cite{Chien1970a}, the viscosity of normal human RBC suspensions in heparinized plasma, or in 11\% albumin-Ringer solution at 45\%-volume fraction were measured in a coaxial cylinder viscometer at 37$^{\circ}$C, where the 11\% albumin-Ringer solution had the same viscosity as plasma (i.e., $\mu_0$ = 1.2 cP) but did not cause RBC aggregation. \cite{Cokelet2007} also measured the viscosity of normal human RBC suspension in plasma at 40\%-volume fraction. Therefore, these experimental conditions in RBC suspensions correspond to $\lambda$ = 5. Indeed, our numerical results obtained with $\lambda$ = 5 well agree with the experimental results, especially by \cite{Cokelet1968} as shown in Fig.\ref{fig:comparison_exp}($a$).

Figure \ref{fig:comparison_exp}($b$) shows the numerical results of $\mu_{re}$ as a function of $\phi$ for $\lambda$ = 5. The relative viscosity $\mu_{re}$ for each $Ca$ exponentially increases with $\phi$, and $\mu_{re}$ tends to decrease as $Ca$ increases. This behavior is the same when the viscosity ratio changes (data not shown). Our numerical results of $\mu_{re}$ are compared with the empirical expression proposed by \cite{Krieger1959}:
\begin{equation}
  \mu_{re} = \left( 1 - \frac{\phi}{\phi_m} \right)^{-\eta \phi_m},
  \label{Krieger1959}
\end{equation}
where $\phi_m$ is the maximum volume fraction. Although equation (\ref{Krieger1959}) was originally proposed for rigid sphere suspensions, it allows us to estimate the viscosity for particles of any shape by choosing a suitable $\phi_m$ and $\eta$, e.g., \cite{Tao2011} set $\phi_m$ = 0.72 and $\eta$ = 2.3 in order to estimate the relative viscosity $\mu_{re}$ experimentally obtained in blood with the plasma viscosity $\approx$ 1.0 cP at 37 $^{\circ}$C and the viscosity ratio $\lambda$ around 5. Our numerical results agree well also with this empirical expression (\ref{Krieger1959}) with the same parameters proposed by \cite{Tao2011}, especially for high $Ca$ (= 0.8).

Figure \ref{fig:comparison_exp}($c$) shows the numerical results of $\mu_{re}$ obtained with different $\lambda$ for $Ca$ = 0.2, which corresponds to a shear rate $\dot{\gamma}$ = 167 s$^{-1}$, as a function of the volume fraction $\phi$. These results are compared with previous measurements obtained with acetaldehyde-fixed human RBC suspension in plasma for $\dot{\gamma}$ = 170.8 s$^{-1}$ \citep{Brooks1970}, and also with normal human/sickle RBC suspension for $\dot{\gamma} >$ 100 s$^{-1}$ \citep{Goldsmith1972}. Again, we confirm that our numerical results are well within those of normal and sickle RBCs. It is also known that cytoplasmic viscosity nonlinearly increases with hemoglobin concentration, resulting in alteration of the cell deformability. The usual distribution of hemoglobin concentration in individual RBCs ranges from 27 to 37 g/dL corresponding to the internal fluid viscosity ($\mu_1$) being 5--15 cP \citep{Mohandas2008}. The physiological relevant viscosity ratio therefore can be taken as $\lambda$ = 4.2--12.5 if the plasma viscosity is set to $\mu_0$ = 1.2 cP. In the sickle cell anemia, on the other hands,
the hemoglobin concentration is abnormal, e.g., the mean corpuscular hemoglobin concentration of sickle cells is potentially elevated to 44.4-47.6 g/dl \citep{Evans1984}. Since previous studies shown that the viscosity of hemoglobin solution abruptly increases to 45 cP at 40 g/dL, up to 170 cP at 45 g/dL and 650 cP at 50 g/dL \citep{Cokelet1968, Mohandas2008}, sickle cells with 45 g/dL in hemoglobin concentration may have high cytoplasmic viscosity $\mu_1 \approx$ 170 cP, where the viscosity ratio is taken as $\lambda \approx$ 140 if physiological plasma viscosity ($\mu_0$ = 1.2 cP) is considered. Although the exact hemoglobin concentration is different depending on the type of sickle cell disease and on the person, the relative viscosity $\mu_{re}$ in the blood with sickle cell anemia should be high at any shear rates than that of normal blood \citep{Chien1970b, Usami1975, Kaul1991}.

\subsection{Comparison with other numerical models}
\begin{figure}
  \centering 
  \includegraphics[height=5.5cm]{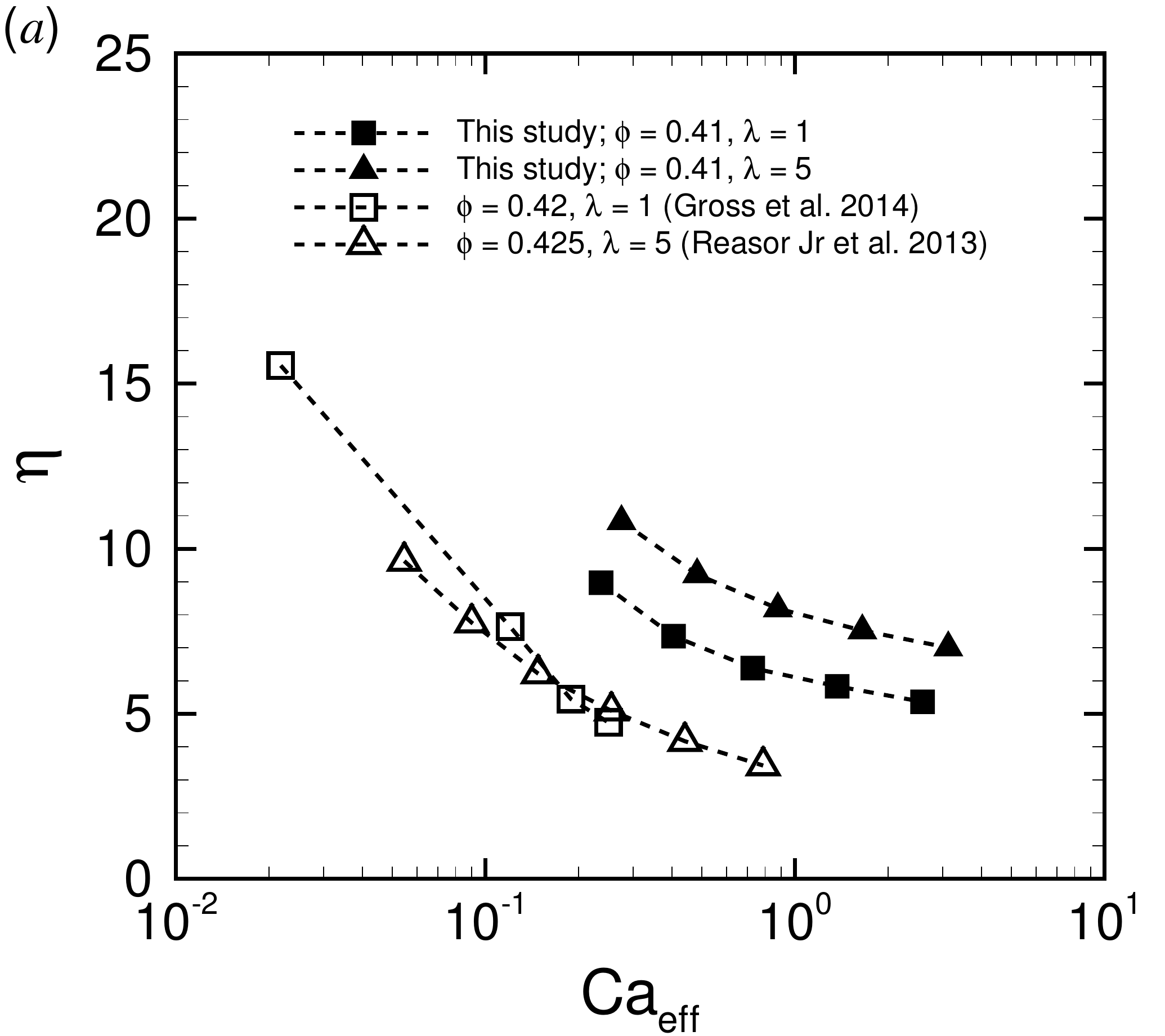}
  \includegraphics[height=5.5cm]{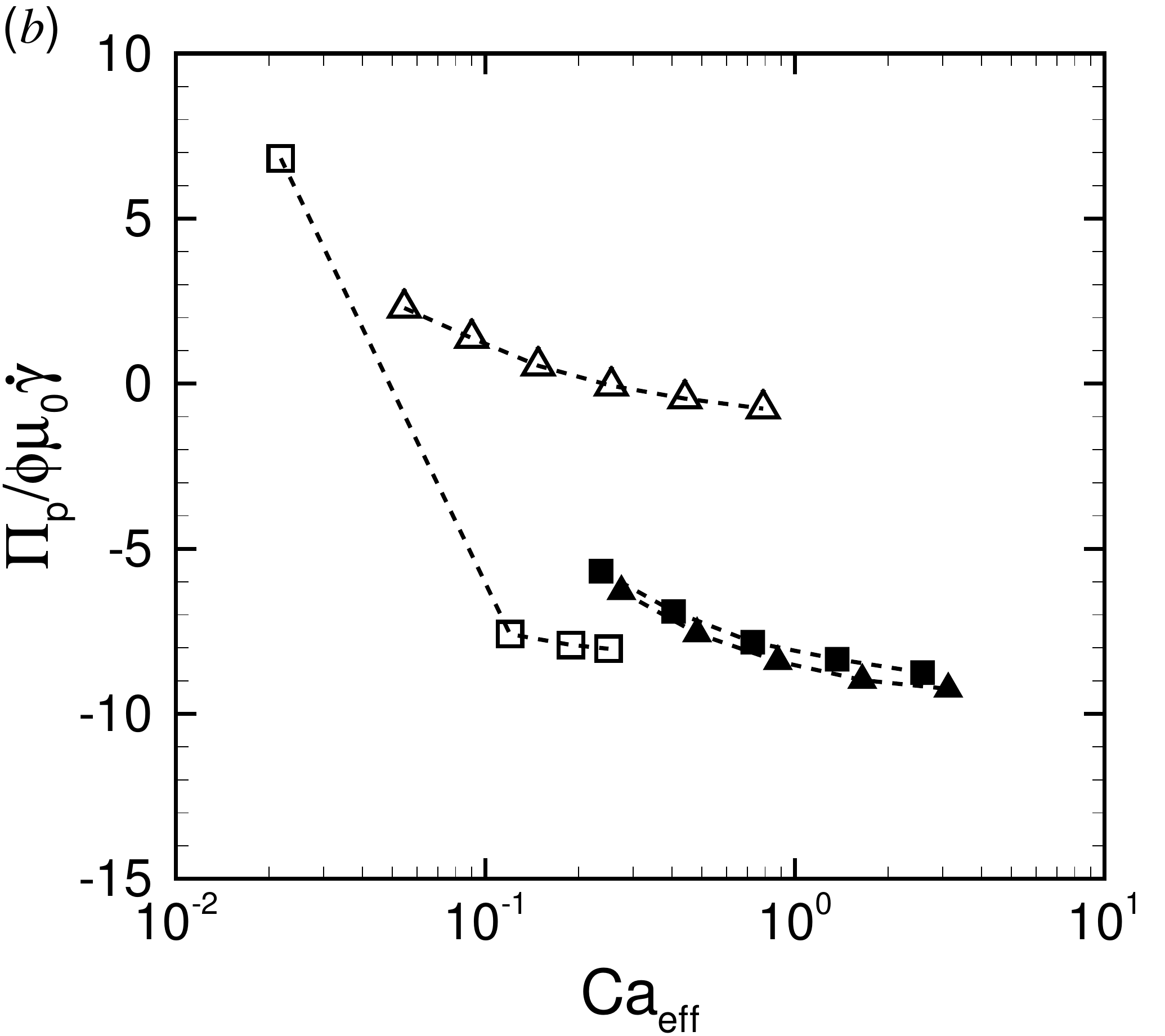}
  \includegraphics[height=5.5cm]{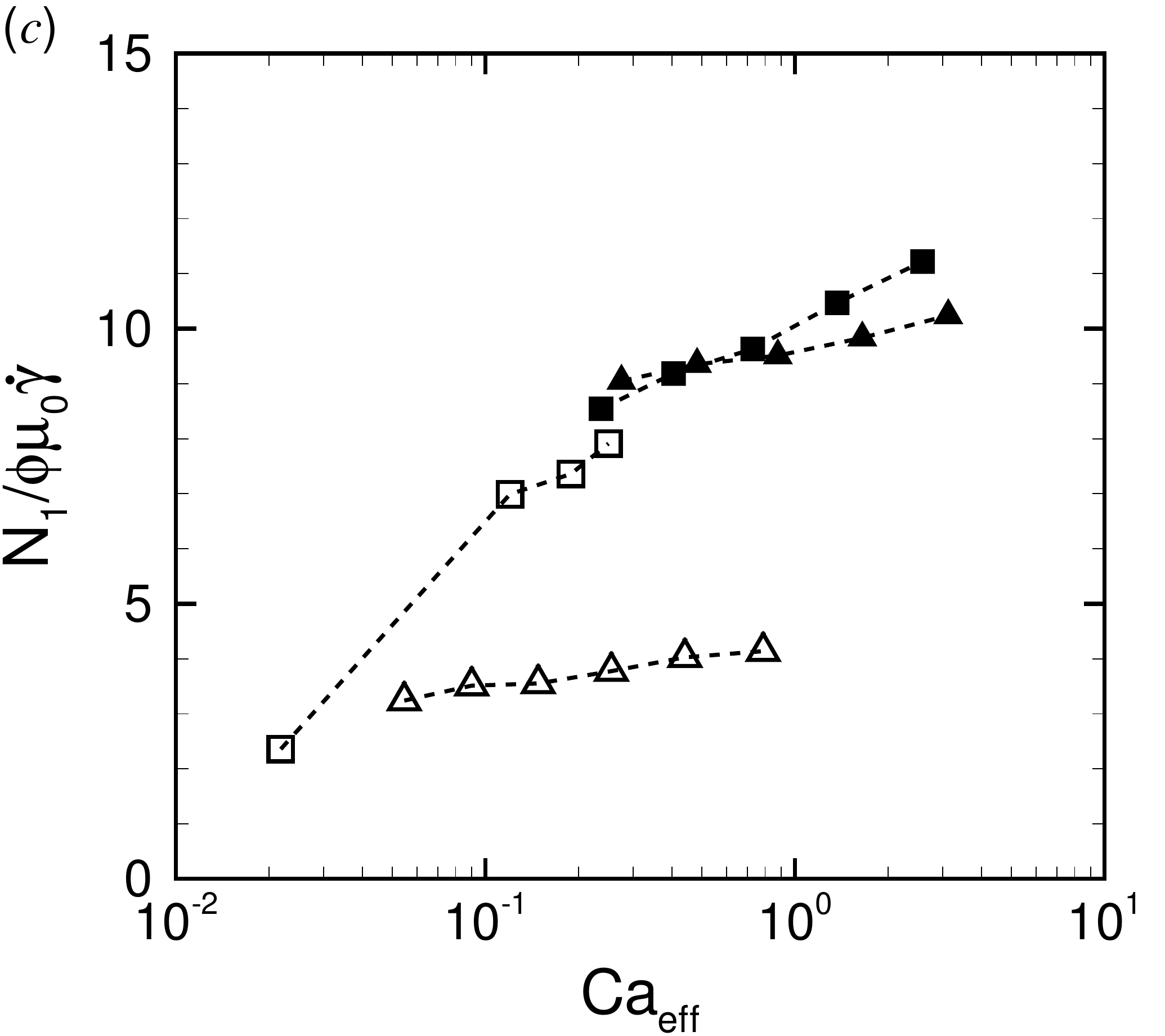}
  \includegraphics[height=5.5cm]{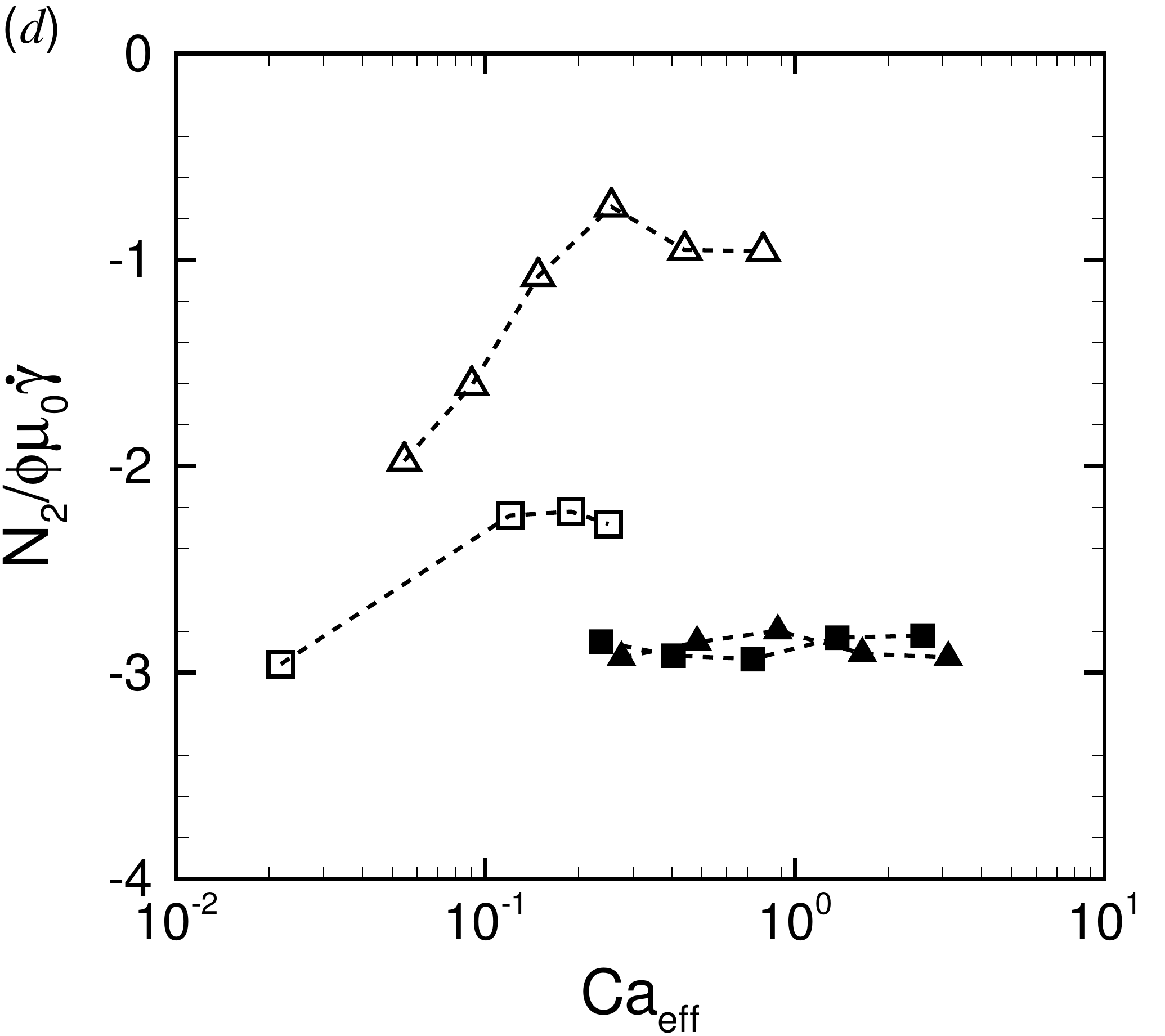}
  \caption{
  ($a$) The intrinsic viscosity $\eta$,
  ($b$) the particle pressure $\Pi_p/\mu_0 \dot{\gamma}$, and
  [($c$) and ($d$)] the first and second normal stress differences $N_i/\mu_0 \dot{\gamma}$ ($i$ = 1 and 2) for $\phi$ = 0.41 as a function of the effective capillary number $Ca_{eff}$. The previous numerical results for $\lambda$ = 1 ($\phi$ = 0.42) by \cite{Gross2014} and $\lambda$ = 5 ($\phi$ = 0.425, $k_b$ = 2.4$\times$10$^{-19}$ J) by \cite{ReasorJr2013} are also displayed for comparison.
  }
  \label{fig:comparison_num}
\end{figure}
Next, we compare our numerical results of stresslet with those obtained with other membrane constitutive models: the spectrin-link model \citep{ReasorJr2013} and the continuum-based capsule model \citep{Gross2014} whose membrane follows SK law but with repulsive forces between the RBCs. For reasonable comparison, we define the effective capillary number as $Ca_{eff}$ = $\mu_{re} Ca$. Note that the definition of $Ca$ in these previous works \citep{ReasorJr2013, Gross2014} is the same used here and reported in (\ref{Ca}). Figure \ref{fig:comparison_num}($a$) shows the intrinsic viscosity $\eta$ as a function of $Ca_{eff}$. Independently of the numerical model used, the RBC suspensions show a shear-thinning behavior; however, our results exhibit higher intrinsic viscosity than in the other two studies. While $\eta$ in our results decreases for all $Ca_{eff}$ as the viscosity ratio decreases from $\lambda$ = 5 to $\lambda$ = 1, in the previous studies $\eta$ was almost independent of the value of $\lambda$ (Fig.\ref{fig:comparison_num}$a$). Although the results by \cite{ReasorJr2013} and \cite{Gross2014} exhibit similar $\eta$, the particle pressure $\Pi_p/\mu_0 \dot{\gamma}$ (Fig.\ref{fig:comparison_num}$b$) and the two normal stress differences $N_i/\mu_0 \dot{\gamma}$ (Figs.\ref{fig:comparison_num}$c$ and \ref{fig:comparison_num}$d$) are quite different between these two previous studies.
We think that the discrepancies among three numerical studies of the stresslet values shown in Fig.\ref{fig:comparison_num} are mainly due to the difference in the choice of constitutive model for the RBC membrane, the contact model between RBCs, and the boundary conditions.
Comparing the results of the present study with the previous study by \cite{ReasorJr2013}, we conclude that the stresslet is sensitive to the membrane constitutive model.
Although the membrane model applied in \citep{Gross2014} and that of present study are the same, \cite{Gross2014} considered repulsive forces between the RBCs. Such contact model guarantees a certain amount of fluid between the RBCs, which is likely to decrease the relative viscosity ($S_{12}$) of the RBC suspension and also affect the other components of the particle stresslet tensor ($S_{ij}$).
The different boundary conditions are also likely to partially affect the stresslet. Reasor Jr et al. (2013) used the Lees-Edwards boundary condition \citep{Lees1972} to consider an unbounded shear flow, while \cite{Gross2014} and the present study consider a wall-bounded shear flow. However, we believe that the effect of a solid wall on the solution is limited because our numerical results (e.g., $\mu_{sp}$ and $N_i/\mu\dot{\gamma}$) for suspensions of spherical particles in a bounded shear flow well agree to the previous numerical studies by \cite{Matsunaga2016}, where an unbounded shear flow was solved by the boundary element method (see Fig.\ref{fig:nh-sphere_deformation}). These results suggest that the stresslet will be independent of numerical methods if the same membrane constitutive model and contact model between paticles are used, and also indicates that the domain size used here is adequate.
\cite{Gross2014} systematically investigated the stresslet for relatively low $Ca_{eff}$ ($10^{-4} \leq Ca_{eff} \leq 10^{-1}$) at $\lambda$ = 1. Our numerical results provide insight into stresslet for relatively high $Ca_{eff}$ ($10^{-1} \leq Ca_{eff} \leq 1$), which correspond to venule and arteriole environments in humans. Although the particle pressure and normal stress differences are difficult to measure in experiments, we hope that our numerical results stimulate not only numerical but also experimental studies to clearly show the viscoelastic behavior of blood.

\subsection{Effective volume fraction}
Conventionally, the relative viscosity $\mu_{re}$ (= 1 + $\mu_{sp}$) of dilute and semi-dilute particulate suspensions can be described by a polynomial expression in the volume fraction $\phi$ \citep{Einstein1911, Taylor1932, Stickel2005}. For example, \cite{Einstein1911} proposed for dilute suspensions of rigid particles: $\mu_{re} = 1 + 2.5 \phi$, while \cite{Taylor1932} proposed a modified law for particles including an internal fluid: $\mu_{re} = 1 + 2.5 \tilde{\lambda} \phi$, where $\tilde{\lambda}$ is Taylor's factor defined as $\tilde{\lambda} = \left( \lambda + 0.4 \right)/\left( \lambda + 1 \right)$. However, our numerical results show that the intrinsic viscosity $\eta$ (= $\mu_{sp}/\phi$) of RBC suspensions is not constant but first decreases from dilute to semi-dilute suspensions because of the mode change of RBCs from rolling to tumbling. This suggests that a simple polynomial approach cannot be applied to RBC suspensions even for low volume fractions;
this issue cannot be solved by any higher-order expansions, since they necessarily involve particle-particle interactions and thus any higher-order coefficients would depend on the local flow and/or on the local microstructure.
For high volume fractions, an exponential expression may be applicable. 
\begin{figure}
  \centering 
  \includegraphics[height=5.5cm]{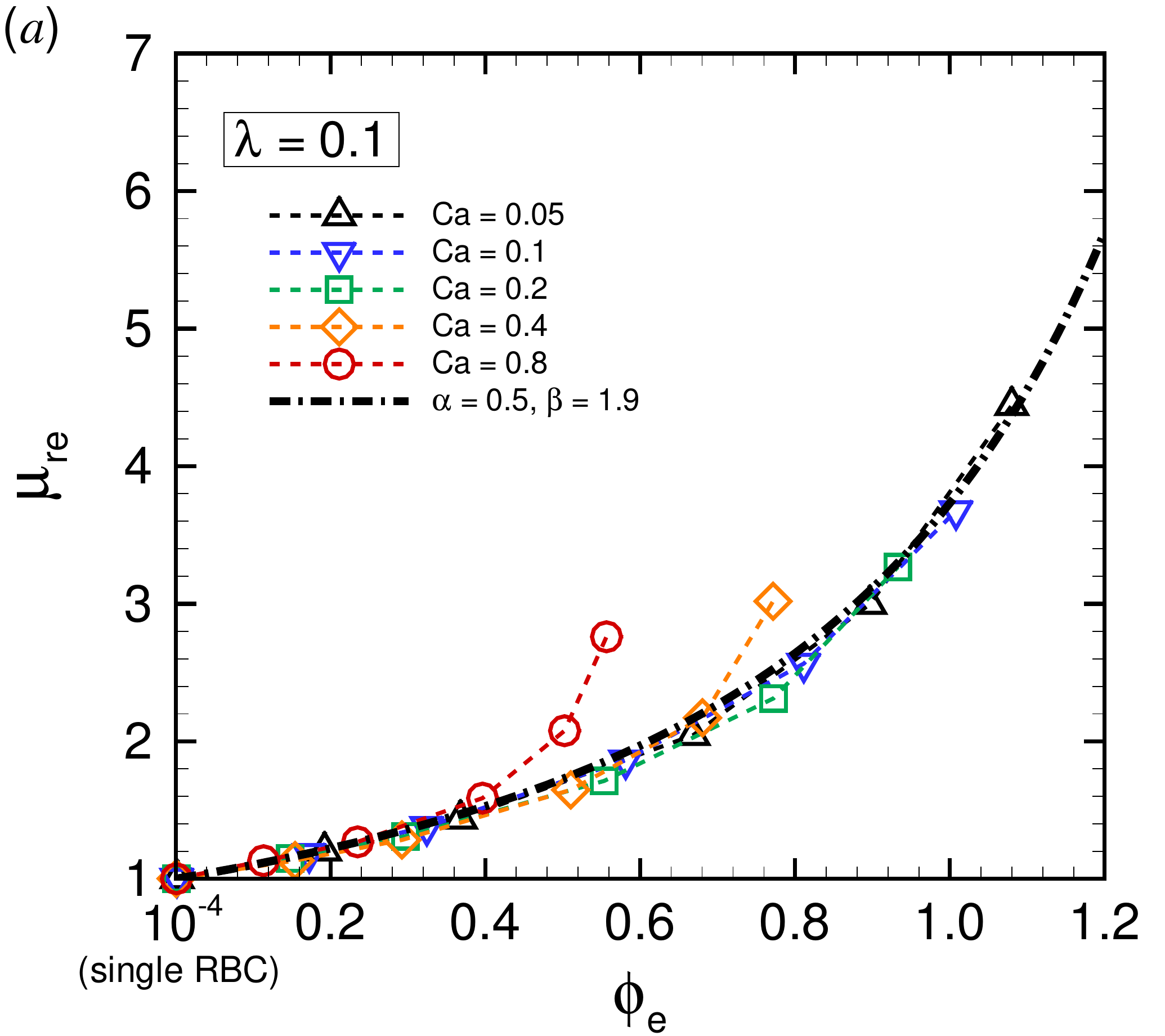}
  \includegraphics[height=5.5cm]{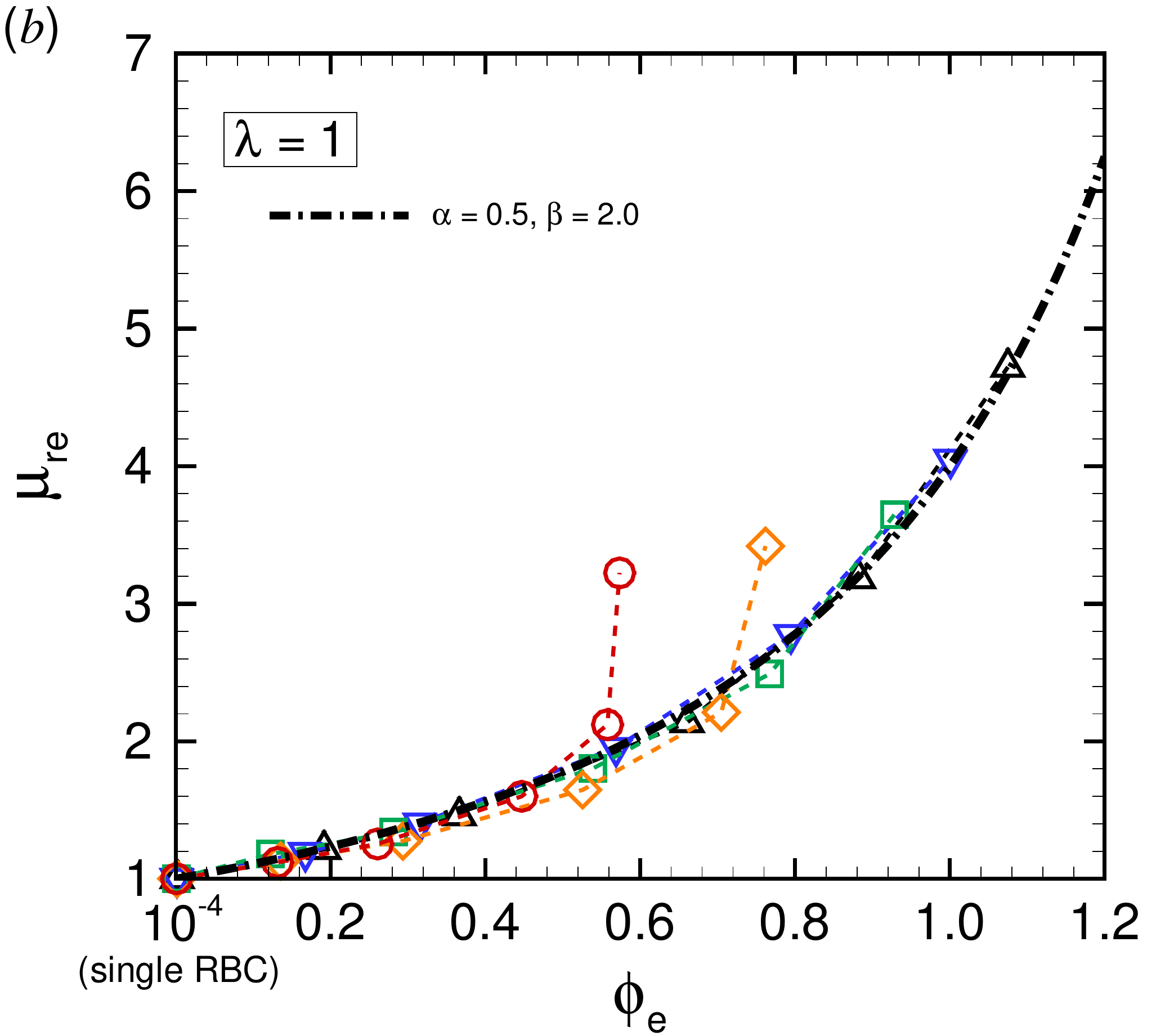}
  \includegraphics[height=5.5cm]{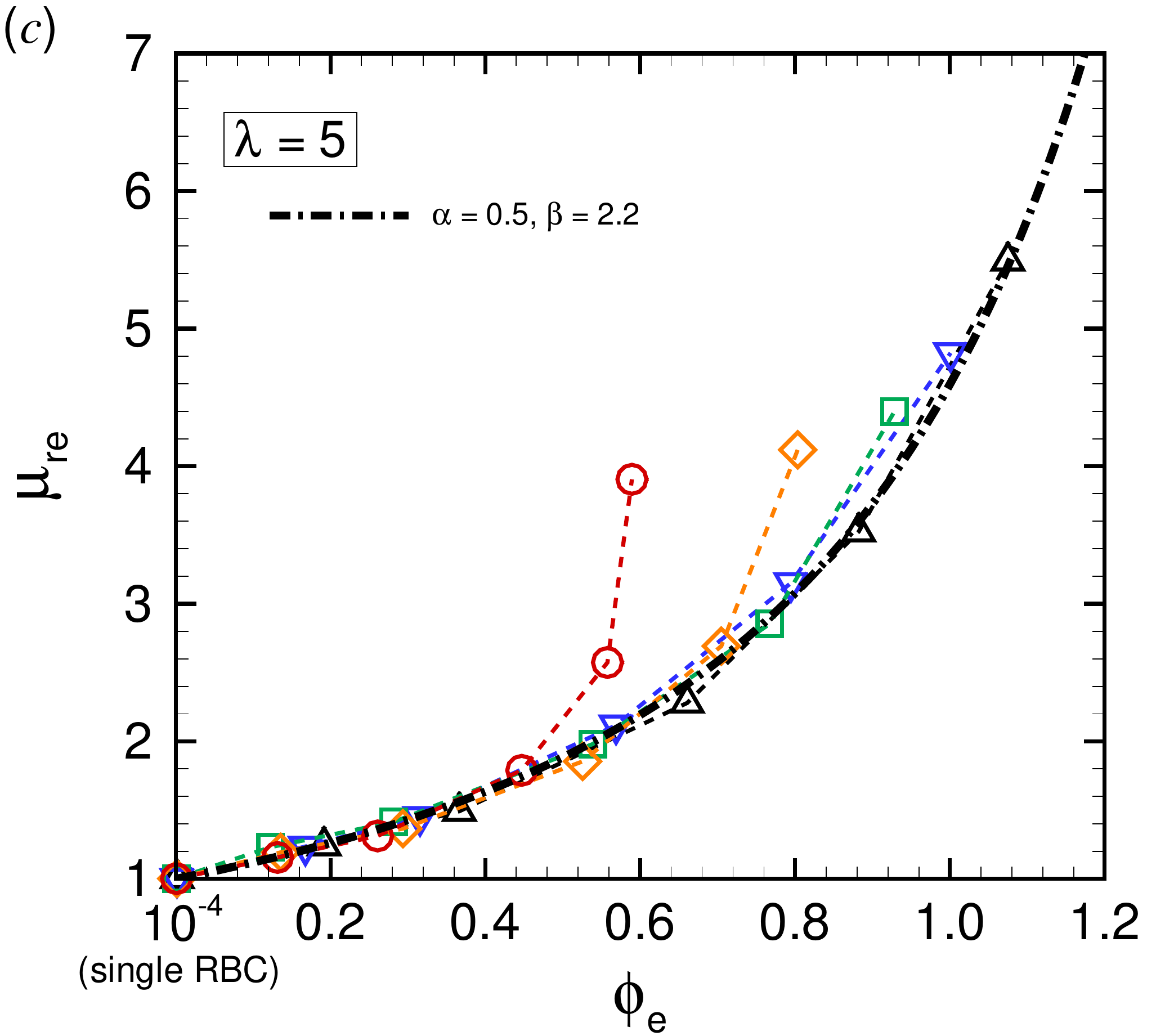}
  \includegraphics[height=5.5cm]{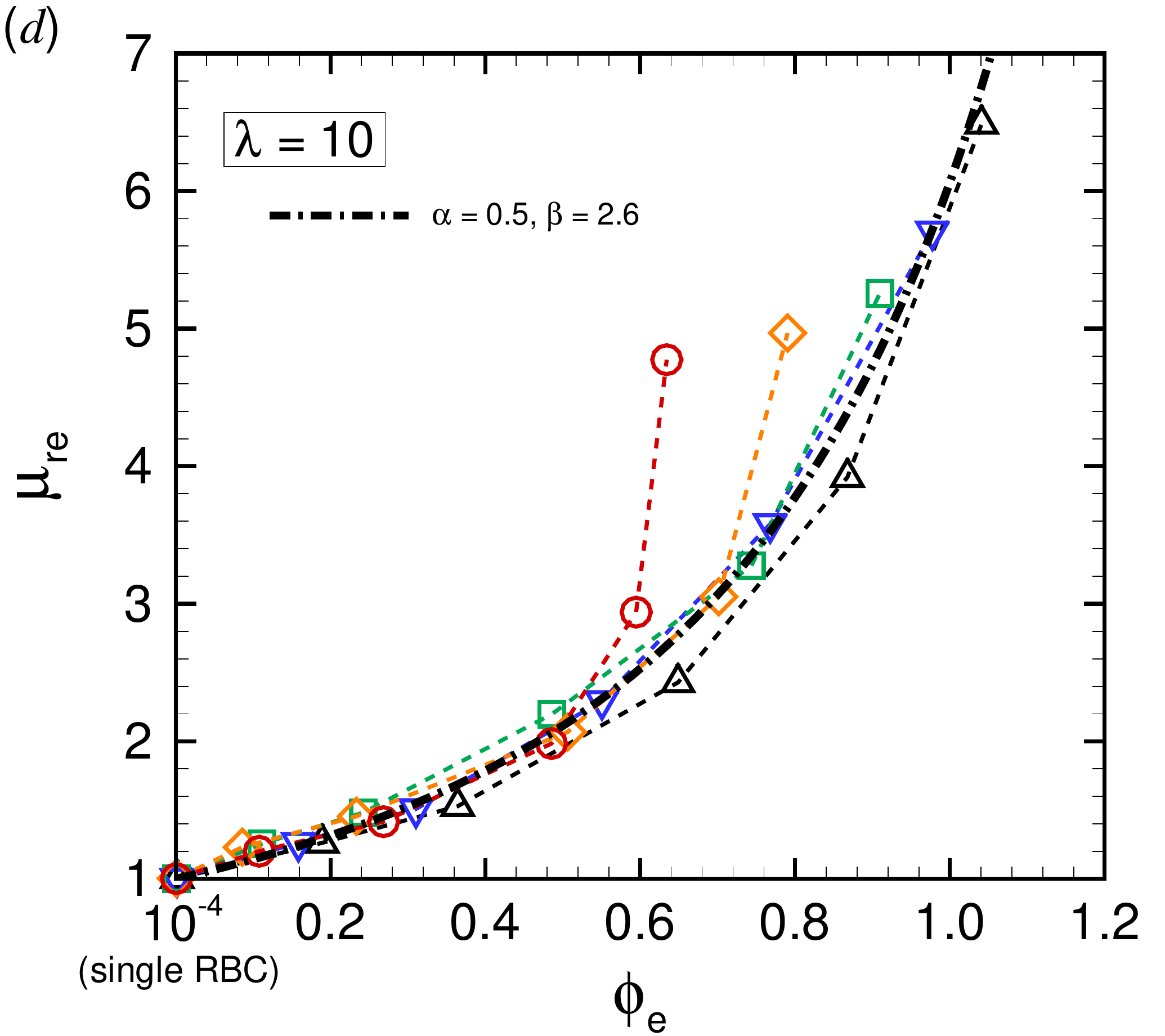}
  \caption{
  Relative viscosity $\mu_{re}$ as a function of the effective volume fraction $\phi_e$ for different $Ca$ at each specific viscosity ratio
  ($a$) $\lambda$ = 0.1,
  ($b$) $\lambda$ = 1,
  ($c$) $\lambda$ = 5 and
  ($d$) $\lambda$ = 10.
  The dashed-dot-lines are exponential curves defined as $\mu_{re} = \left( 1 - \alpha \phi_e \right)^{-\beta}$ with $\alpha$ = 0.5 and $\beta|_{\lambda=0.1}$ = 1.9, $\beta|_{\lambda=1}$ = 2.0, $\beta|_{\lambda=5}$ = 2.2 and $\beta|_{\lambda=10}$ = 2.6.
  }
  \label{fig:mure_ephi}
\end{figure}
\cite{Rosti2018a} proposed that the effective volume fraction $\phi_e$, which is a collective volume fraction of spheres whose radius is defined with the semi-minor axis $a_2$ (here, $a_1 \geq a_3 \geq a_2$) of deformed spherical particles, is able to describe the relative viscosity of suspensions of deformable particles. Here, we define the effective volume fraction $\phi_e$ with the semi-middle axis $a_3$ of the deformed RBC, i.e., $\phi_e  = N^R 4 \pi a_3^3/(3V)$, where $N^R$ is the number of RBC in the computational box of volume $V$. The length of the semi-middle axis $a_3$ of the deformed RBC is obtained from the eigenvalues of the inertia tensor of an equivalent ellipsoid approximating the deformed RBC \citep{Ramanujan1998}.
Figure \ref{fig:mure_ephi} shows the relative viscosity $\mu_{re}$ as a function of the effective volume fraction $\phi_e$: for each $\lambda$, the numerical results of $\mu_{re}$ successfully collapses on a single non-linear master curve, except for the case with high $Ca \geq$ 0.4, where the fit works only in the case of low/moderate volume fraction, and fails in the case of a fully dense suspension.
The fail of the fit for high $Ca$ and $\phi$ is limited to the cases of RBCs showing complex shapes, e.g. multilobes. Indeed, in these cases the shape is mostly asymmetric and its approximation with an equivalent ellipsoid not reliable.
The single non-linear curves are fitted by a general exponential expression:
\begin{equation}
  \mu_{re} = \left( 1 - \alpha \phi_e \right)^{-\beta},
  \label{exponential_curve}
\end{equation}
where, $\alpha$ = 0.5 and $\beta|_{\lambda=0.1}$ = 1.9, $\beta|_{\lambda=1}$ = 2.0, $\beta|_{\lambda=5}$ = 2.2 and $\beta|_{\lambda=10}$ = 2.6. The coefficients $\alpha$ and $\beta$ in equation (\ref{exponential_curve}) are related to those in the Krieger-Dougherty formula (\ref{Krieger1959}) by the following relations: $\alpha$ = 1/$\phi_m$ and $\beta$ = −$\eta \phi_m$; \cite{Tao2011} proposed the following values for the coefficients $\phi_m$ = 0.72 and $\eta$ = 2.3. \cite{Gross2014} proposed a toy model based on the effective medium theory by considering the effects of $Ca$ and higher volume fraction $\phi$, but up to now no model has been able to fully predict the behavior of the relative viscosity. The next challenge may be constructing a model, which is able to cover a wide range of viscosity ratios and large deformation of RBCs.

\section{Conclusion}
We numerically investigate the rheology of a suspension of red blood cells (RBCs) in a wall-bounded shear flow for a wide range of volume fractions $\phi$, viscosity ratios $\lambda$ and capillary number $Ca$ assuming the Stokes flow regime. The RBCs are modeled as biconcave capsules, whose membrane follows the Skalak constitutive law. The problem is solved numerically through  GPU computing, using the lattice-Boltzmann method for the inner and outer fluid and the finite element method to follow the deformation of the RBCs membrane. 

Single RBC subjected to low $Ca$ tends to orient to the shear plane and exhibits the rolling motion as a stable mode associated to higher intrinsic viscosity $\eta$ (= $\mu_{sp}/\phi$) than the tumbling motion. As $Ca$ increases, the mode shifts from the rolling to the swinging motion, and the intrinsic viscosity $\eta$ decreases. Hydrodynamic interactions (higher volume fraction) also allows RBCs to exhibit the tumbling or swinging motions resulting in a decrease of the intrinsic viscosity $\eta$ for dilute and semi-dilute suspensions. This suggests that a simple polynomial equation of the volume fraction $\phi$ for the relative viscosity $\mu_{re}$ (= 1 + $\mu_{sp}$) cannot be applied to RBC suspensions at low volume fractions. The relative viscosity $\mu_{re}$ for high volume fractions, however, can be well described as a function of an effective volume fraction $\phi_e$, defined by the volume of spheres of radius equal to the semi-middle axis of the deformed RBC. For all $\lambda$ considered, the relative viscosity $\mu_{re}$ successfully collapses on a single non-linear curve as a function of $\phi_e$ except for the case with $Ca \geq$ 0.4, where the fit works only in the case of low/moderate volume fractions.

We hope that our numerical results will stimulate the numerical and experimental study of hemorheology, aiming not only to gain insight into suspension rheology but also to the precise diagnosis of patients with hematologic disorders.

\section*{Acknowledgements}
This research was supported by JSPS KAKENHI Grant Numbers JP17K13015, JP18H04100, and by the Keihanshin Consortium for Fostering the Next Generation of Global Leaders in Research (K-CONNEX), established by Human Resource Development Program for Science and Technology, and also by MEXT as ``Priority Issue on Post-K computer'' (Integrated Computational Life Science to Support Personalized and Preventive Medicine)(Project ID:hp180202). M.E.R. and L.B. thank the financial support by the European Research Council Grant No. ERC-2013-CoG-616186, TRITOS, and from the Swedish Research Council (VR), through the Outstanding Young Researcher Award. Last but not least, N.T. thanks Dr. Daiki Matsunaga and also Dr. Toshihiro Omori for helpful discussions.

\section*{Supplementary movie}
Supplementary movies are available at \textit{http://xxx.yyy.zzz}.

\appendix
\section{Numerical setup}
\subsection{Behavior of a single capsule and RBC with different viscosity ratios}\label{appA1}
\begin{figure}
  \centering 
  \includegraphics[height=5.5cm]{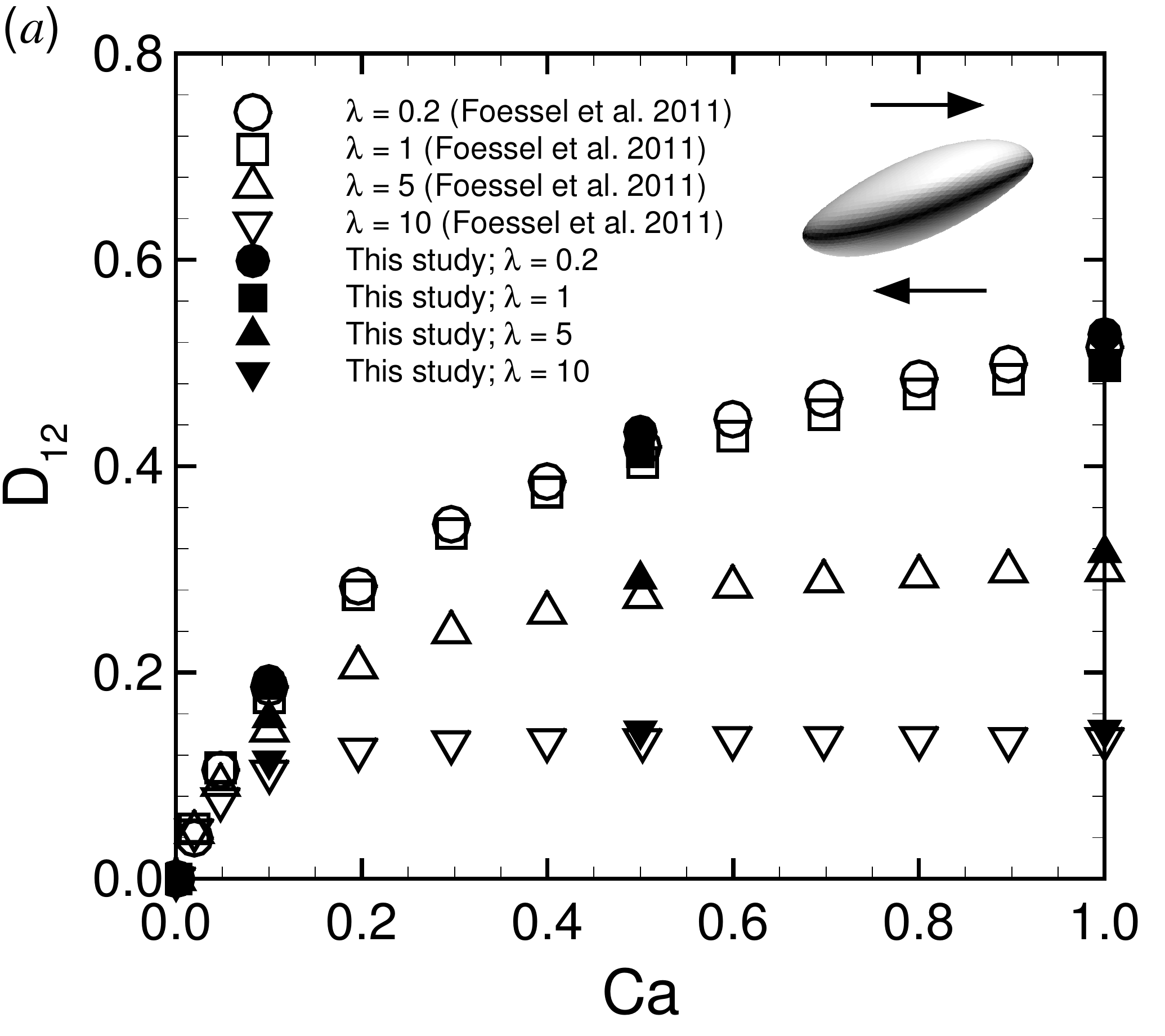}\\
  \includegraphics[height=5.5cm]{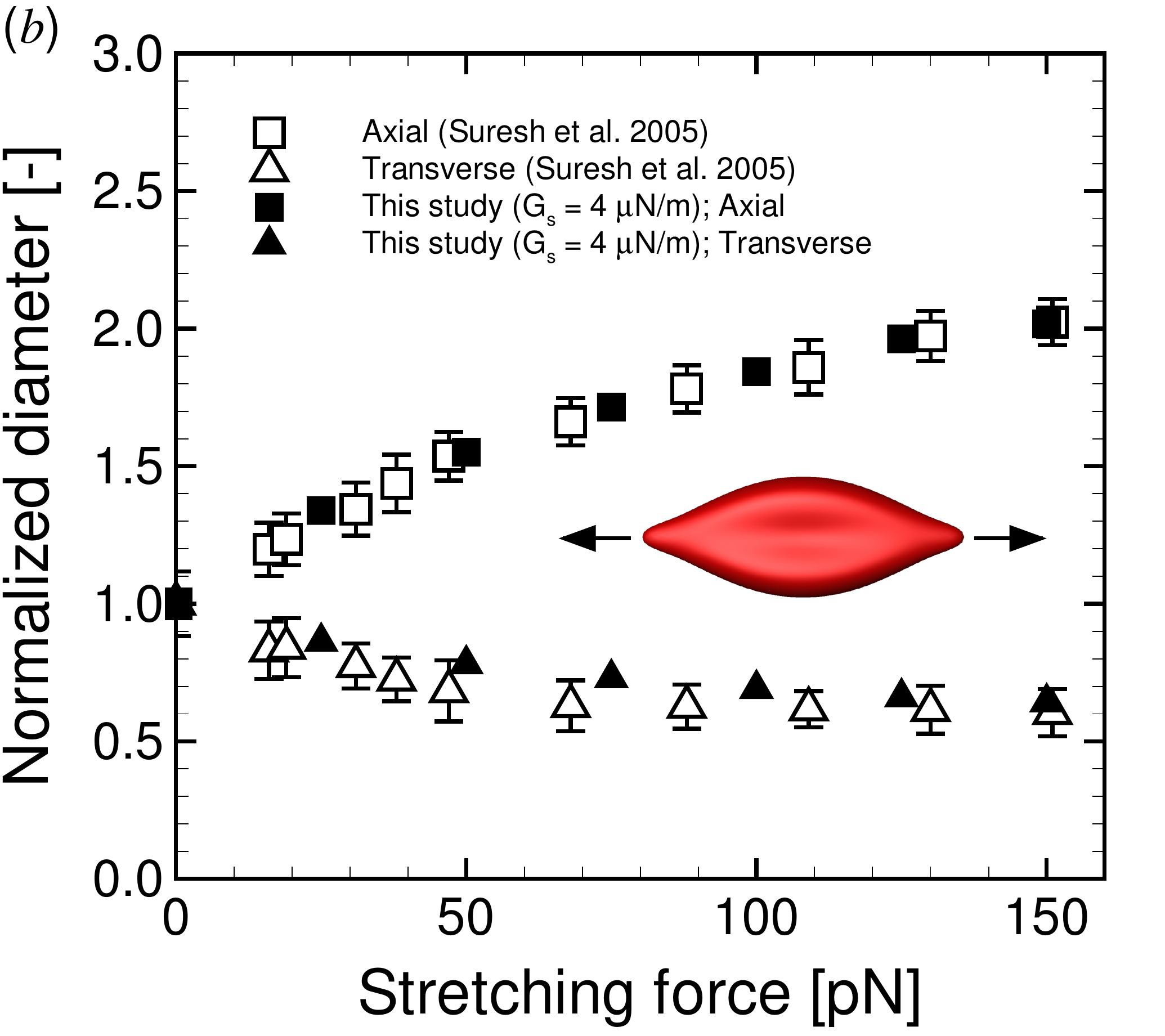}
  \includegraphics[height=5.5cm]{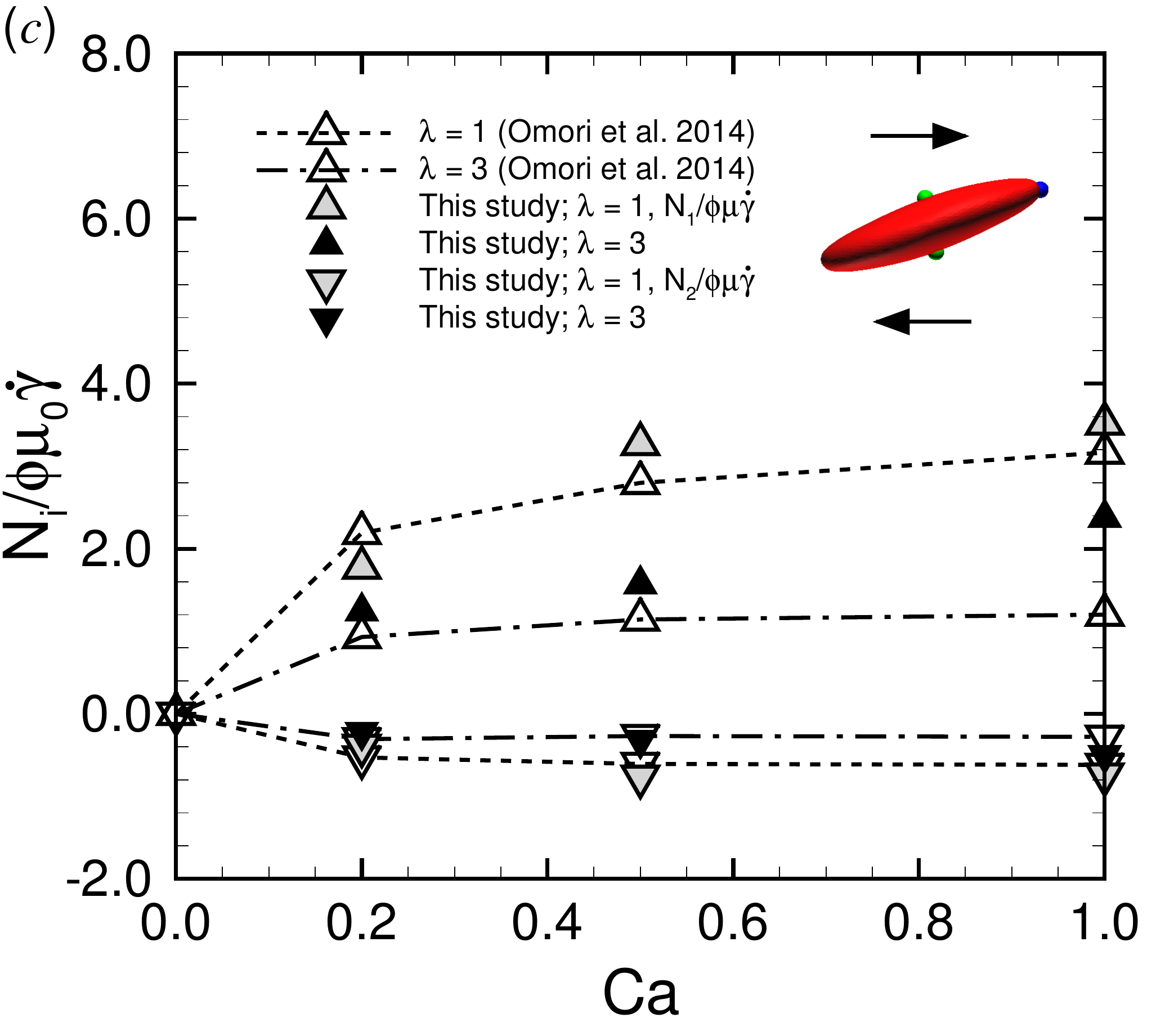}
  \caption{
  ($a$) Time averaged Taylor parameter $D_{12}$ of a SK-spherical capsule as a function of $Ca$ for different viscosity ratios $\lambda$ (= 0.2, 1, 5, and 10); previous numerical results by \cite{Foessel2011} are also displayed. The inset figure represents a tank-treading spherical capsule at $Ca$ = 1.0 and $\lambda$ = 1.
  ($b$) Comparison of the deformation of a stretched RBC in the experiment by optical tweezers \citep{Suresh2005} and in our numerical simulation: an RBC membrane with $G_s$ = 4 $\mu$N/m, $C$ = 10$^{2}$ and $k_b$ = 5$\times$10$^{-19}$ J is laterally stretched by applying constant forces to two points of the membrane surface of radius equal to 1 $\mu$m. The inset figure represents the stretched RBC by 100 pN forces.
  ($c$) Time averaged normal stress difference $N_i/\phi\mu_0\dot{\gamma}$ ($i$ = 1 and 2) for $\lambda$ = 1 and 3; previous numerical results by \cite{Omori2014} are also displayed. The inset figure represent a tank-treading RBC at $Ca$ = 1.0 and $\lambda$ = 1.
  }
  \label{fig:validation_single}
\end{figure}

To validate our numerical approach to update the viscosity in the fluid lattice, we tested the deformation of a single spherical capsule for different $Ca$ and different viscosity ratios $\lambda$ (= 0.2, 1, 5 and 10). The capsule deformation is quantified by the Taylor parameter $D_{12}$, which is defined as:
\begin{eqnarray}
   D_{12} = \frac{|a_1 - a_2|}{a_1 + a_2},
\end{eqnarray}
where $a_1$ and $a_2$ are the lengths of the semi-major and semi-minor axes of the deformed capsule (or RBC), and are obtained from the eigenvalues of the inertia tensor of an equivalent ellipsoid approximating the deformed capsule \citep{Ramanujan1998}. Time average starts after the non-dimensional time $\dot{\gamma}t$ = 40 to reduce the influence of the initial conditions, and continues to $\dot{\gamma}t$ = 100. Our numerical results are compared with previous numerical results obtained with the boundary integral method (BIM) \citep{Foessel2011}. The resolution of the fluid and membrane mesh are the same as in the analysis above. For reasonable comparison with previous numerical study \citep{Foessel2011}, the same parameter are considered and the membrane is modeled with the Skalak constitutive law (\ref{SK}) with the area dilation modulus $C$ = 1 and without bending resistance. Fig.\ref{fig:validation_single}($a$) shows that our numerical results are in good agreement with those by \cite{Foessel2011}.

To characterize the surface shear elastic modulus $G_s$ for RBCs, we performed a numerical simulation reproducing the stretching of RBCs by optical tweezers \citep{Suresh2005}, see Fig.\ref{fig:validation_single}($b$). An RBC membrane with $G_s$ = 4 $\mu$N/m, $C$ = 10$^{2}$ and $k_b$ = 5$\times$10$^{-19}$ J is laterally stretched by applying constant forces to two points of the membrane surface of radius equal to 1 $\mu$m. $G_s$ is thus obtained to capture the nonlinear deformation curve obtained from the experiment. Using these parameters, 
we also tested the behavior of single RBC, and compared the normal stress differences $N_i/\phi \mu_0 \dot{\gamma}$ ($i$ = 1 and 2) with those of previous numerical results obtained with the BIM \citep{Omori2014} in Fig.\ref{fig:validation_single}($c$). Again, our numerical results are in good agreement with those of the literature, although the value of $N_1/\phi\mu_0\dot{\gamma}$ obtained with $\lambda$ = 3 and $Ca$ (= 1) is slightly larger than that of the BIM.

\subsection{Effect of the domain size}\label{appA2}
We have tested the computational domain size, especially the wall-to-wall distance $H$, and investigated its effect on the suspension behavior. Although the influence of $H$ upon particle shear stress and relative viscosity was systematically investigated by \cite{Kruger2011}, and the same computational domain size as in our study has been successfully applied to previous numerical studies of particle suspensions \citep{Picano2013, Rosti2018a}, we also tested several parameters of deformed RBCs for different domain heights $H$ (= 7.5$a$, 12.5$a$ and 15$a$). The results of each parameter reported in Table \ref{tab:error} are compared with those of the reference domain height $H$ = 10$a$; in particular, we have analyzed the ensemble average of the Taylor parameter $\langle D_{12} \rangle$, the orientation angle $\langle \theta \rangle$, the specific viscosity $\mu_{sp}$, the particle pressure $\Pi_p/\mu_0 \dot{\gamma}$ and the normal stress difference $N_i/\mu_0 \dot{\gamma}$ ($i$ = 1 and 2). Here, the ensemble average of a parameter $\langle \chi \rangle$ is defined as
\begin{eqnarray}
   \langle \chi \rangle = \frac{1}{MN} \sum^M_m \sum^N_n \chi^{m,n},
\end{eqnarray}
where $M$, $N$ are the number of time steps and capsules respectively, The error for each observable are defined by
\begin{equation}
  \varepsilon_{\chi} = \left| \frac{\langle \chi \rangle - \langle \chi^{ref} \rangle}{\langle \chi^{ref} \rangle} \right|,
  \label{eq:error}
\end{equation}
where the superscript \emph{ref} indicates the reference values. The results of each parameter and the corresponding relative errors are listed in Table \ref{tab:error}. Since differences between the case with the largest height ($H$ = 12.5$a$) and our reference case are less than 5\% in the orientation angle and 3\% in the others, the results presented in this study are all obtained with the domain height of $H$ = 10$a$.

\begin{table}
  \begin{center}
\def~{\hphantom{0}}
  \begin{tabular}{lccc}  \\
       Height $H$ & 7.5$a$ &  10$a$ (\emph{ref}.) & 12.5a \\
       Number of RBCs & 258 & 344 & 442 \\  [5pt]
       $\langle D_{12} \rangle$ & 0.51130	& 0.50960 & 0.50700 \\
       $\varepsilon_{D_{12}}$ & 0.00334 & - & 0.00510  \\ [5pt]
       $\langle \theta \rangle/\pi $ & 0.08005 & 0.07835 & 0.07684 \\
       $\varepsilon_{\theta}$ & 0.04932 & - & 0.04949 \\ [5pt]
       $\mu_{sp}$ & 0.86880 & 0.85440 & 0.88010 \\
       $\varepsilon_{\mu_{sp}}$ & 0.01685 & - & 0.03008\\ [5pt]
       $\Pi_p/\mu_0 \dot{\gamma}$ & -0.58130 & -0.53580 & -0.52560 \\
       $\varepsilon_{\Pi_p}$ & 0.08492 & - & 0.01904  \\
       $N_1/\mu_0 \dot{\gamma}$ & 0.80310 & 0.75340 & 0.75790 \\
       $\varepsilon_{N_1}$ & 0.06597 & - & 0.00597 \\ [5pt]
       $N_2/\mu_0 \dot{\gamma}$ & -0.21390 & -0.19250 & -0.18940 \\
       $\varepsilon_{N_2}$ & 0.11117 & - & 0.01610  \\
  \end{tabular}
  \caption{Effect of domain height $H$ on the ensemble average of the Taylor parameter $\langle D_{12} \rangle$, the orientation angle $\langle \theta \rangle$, the specific viscosity $\mu_{sp}$, the particle pressure $\Pi_p/\mu_0 \dot{\gamma}$ and the normal stress difference $N_i/\mu_0 \dot{\gamma}$ ($i$ = 1 and 2).  The error of each parameter $\varepsilon_{\chi}$ is defined by (\ref{eq:error}). The simulation were performed at $\phi$ = 0.21, $Ca$ = 0.4 and $\lambda$ = 5.}
  \label{tab:error}
  \end{center}
\end{table}

\subsection{Suspension of spherical capsules}\label{appA3}
\begin{figure}
  \centering
  \includegraphics[height=3.0cm]{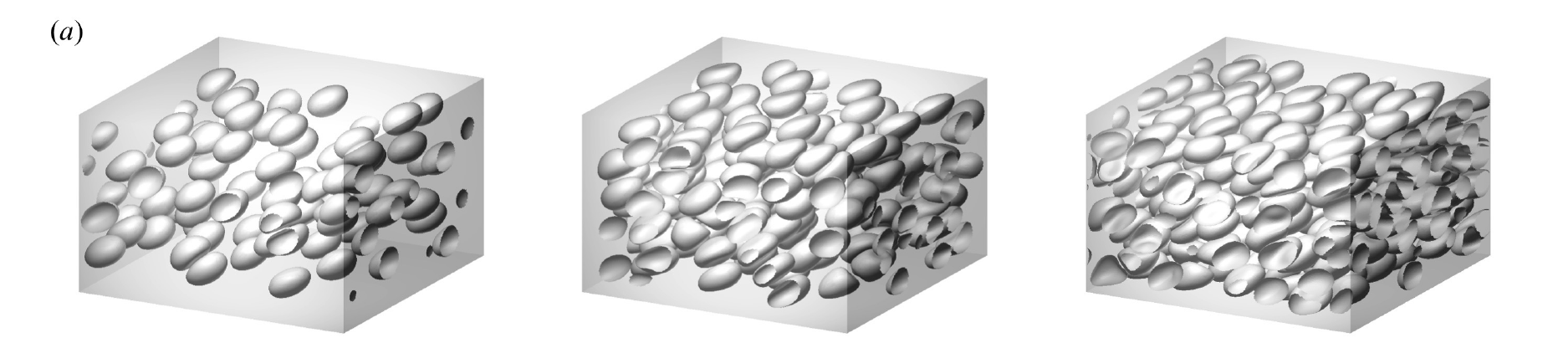}\\  \vspace{0.4cm}
  \includegraphics[height=5.5cm]{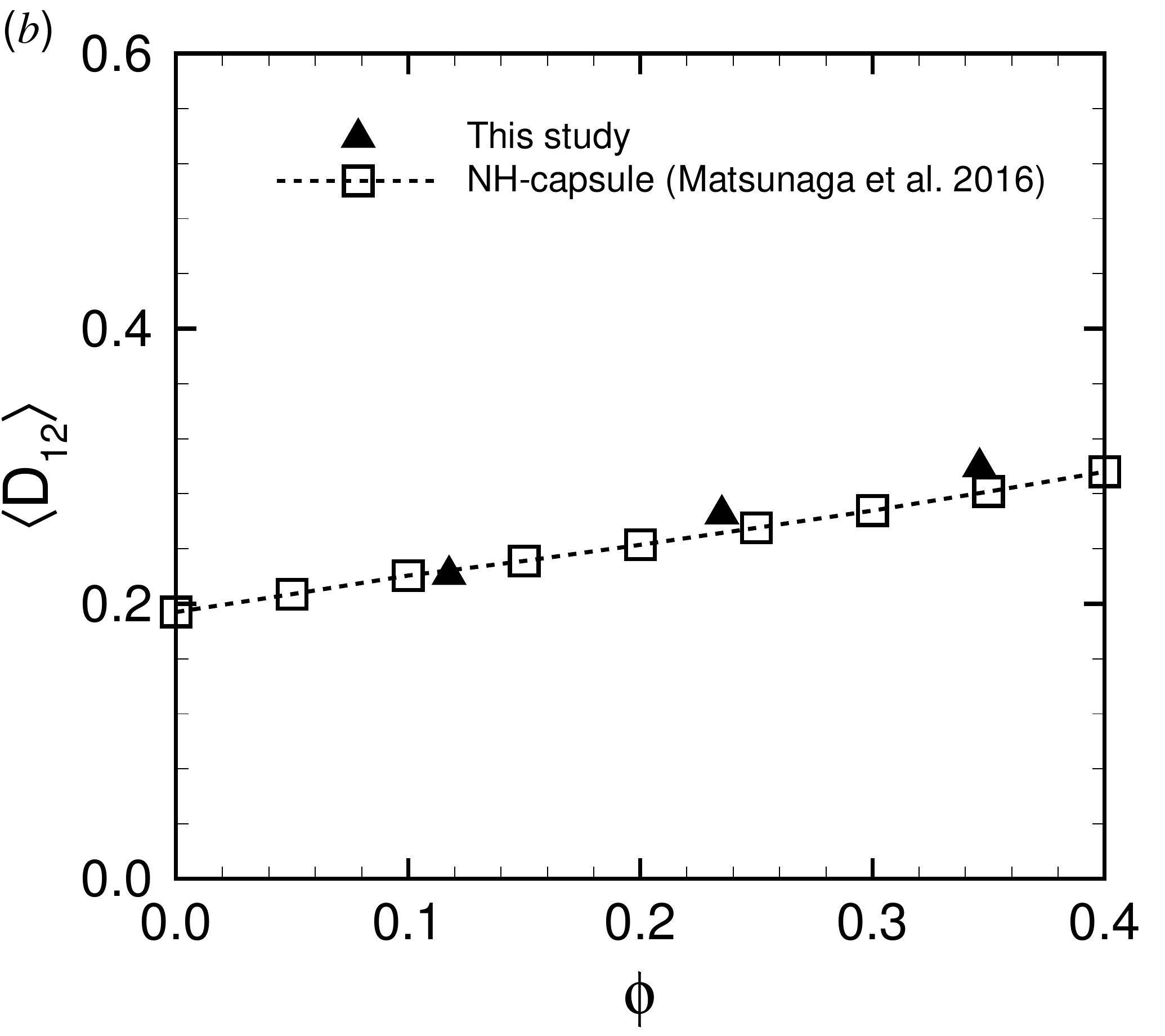}
  \includegraphics[height=5.5cm]{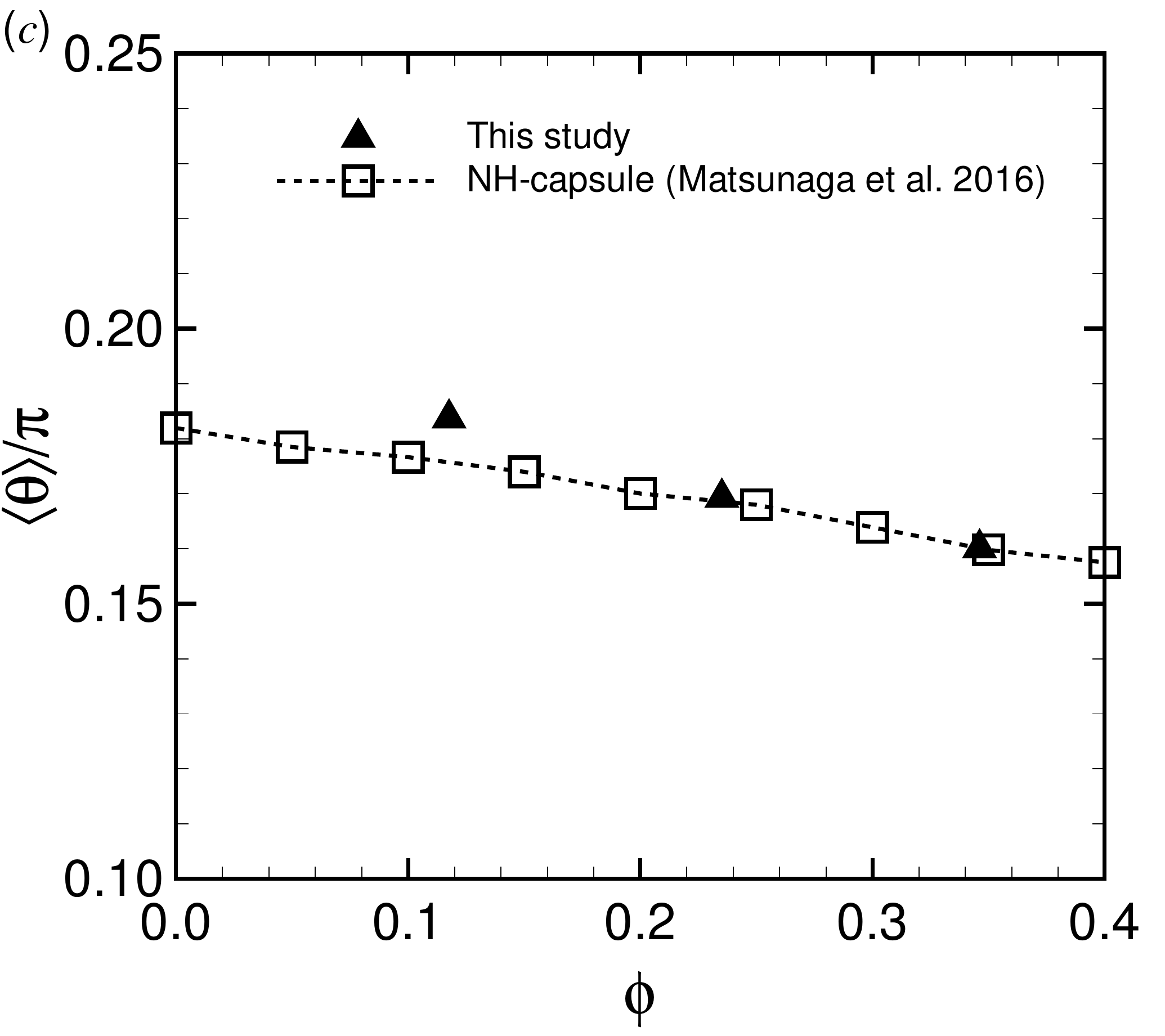}
  \includegraphics[height=5.5cm]{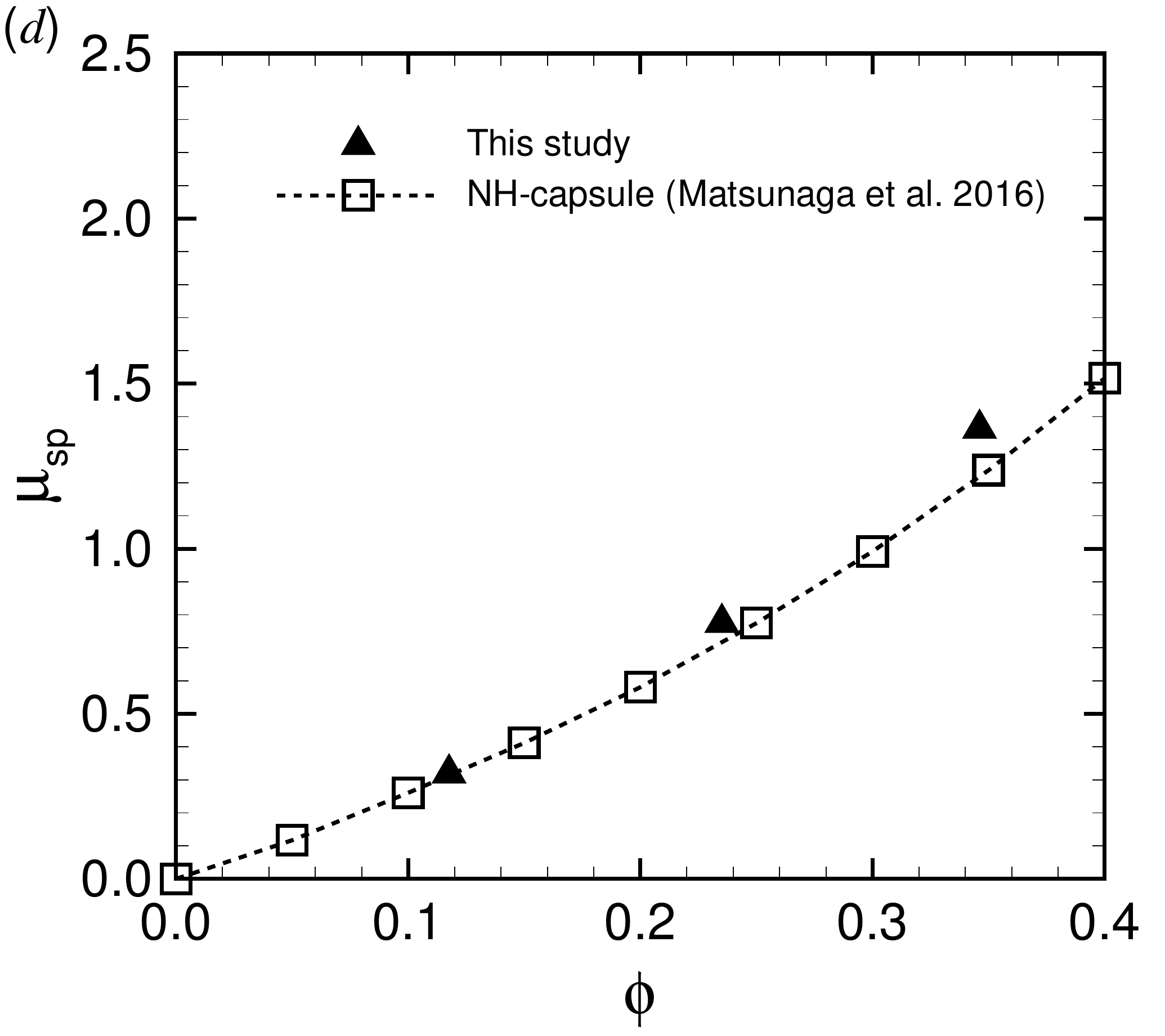}
  \includegraphics[height=5.5cm]{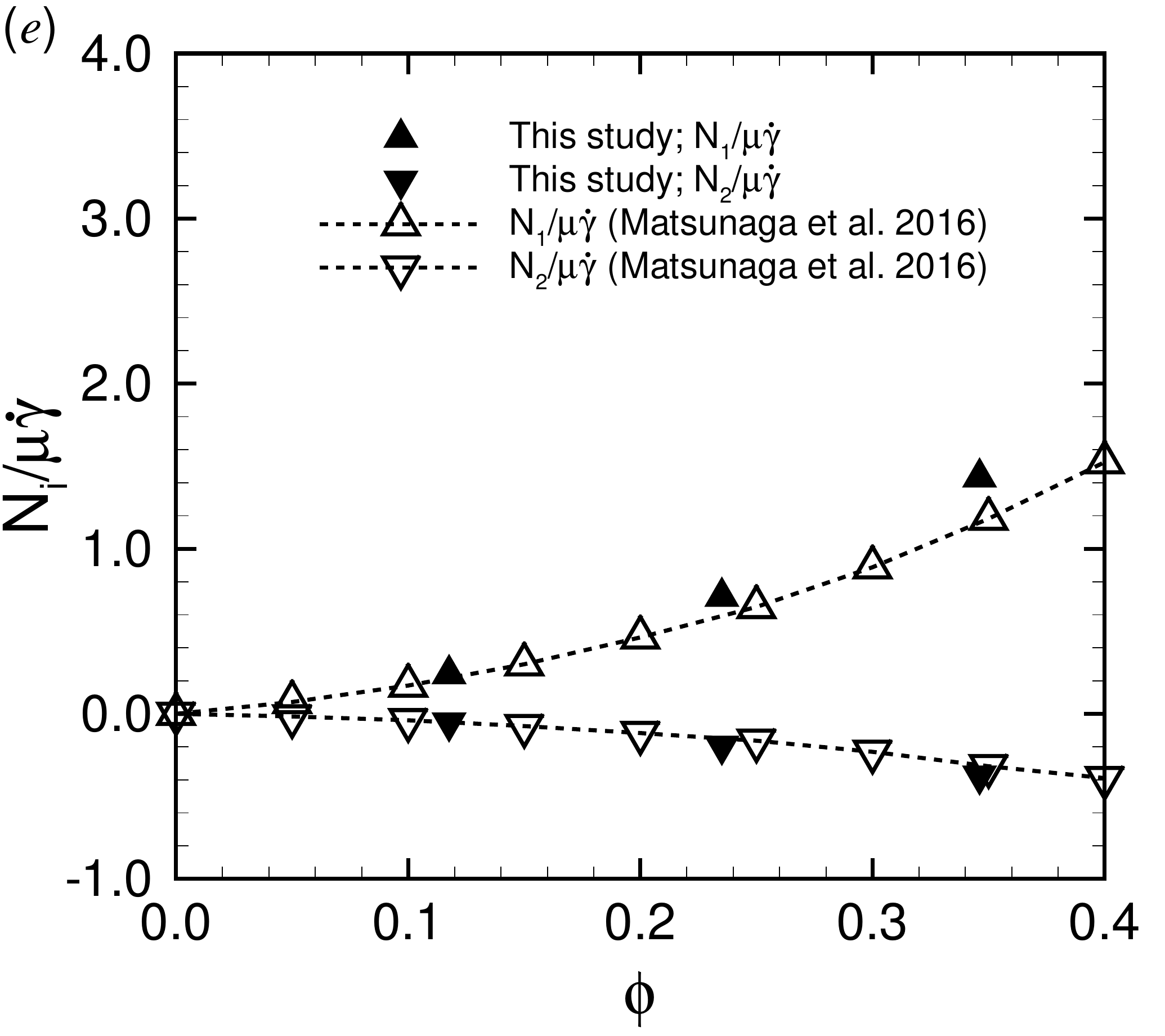}
  \caption{
  ($a$) Snapshots of numerical results of the suspension of NH-spherical capsules for different volume fractions $\phi$ = 0.12 ($left$), $\phi$ = 0.24 ($middle$) and $\phi$ = 0.35 ($right$).
  Ensemble average of
  ($b$) Taylor parameter $\langle D_{12} \rangle$,
  ($c$) orientation angle $\langle \theta \rangle/\pi$,
  ($d$) specific viscosity $\mu_{sp}$, and
  ($e$) first and second normal stress difference $N_i/\mu_0\dot{\gamma}$ ($i$ = 1 and 2) of NH-spherical capsules, with viscosity ratio $\lambda$ = 1, subjected to $Ca$ = 0.1 as a function of the volume fraction $\phi$.
  Our numerical results obtained with $\lambda$ = 1 are compared with those reported in the previous numerical study by \cite{Matsunaga2016}.
  }
\label{fig:nh-sphere_deformation}
\end{figure}
We simulate suspensions of neo-Hookean (NH) spherical capsules for different $Ca$ and $\phi$ = (0.12, 0.24 and 0.35), as reference for the results pertaining RBCs and to further validate our numerical model. The NH constitutive law is given as:
\begin{equation}
  w_s^{NH} = \frac{G_s}{2} \left( I_1 - 1 + \frac{1}{I_2 + 1} \right).
  \label{NH}
\end{equation}
The viscosity ratio is set to be $\lambda$ = 1, and the bending modulus is the same as the model of RBC in this study, i.e., $k_b$ = 5$\times$10$^{-19}$ J. Our numerical results are compared with previous numerical results of NH-spherical capsule simulated by the BIM in an unbounded domain by \cite{Matsunaga2016}.

Figure \ref{fig:nh-sphere_deformation}($a$) shows the snapshots of our numerical results for different $\phi$. The ensemble average of the Taylor parameter $\langle D_{12} \rangle$ and of the orientation angle $\langle \theta \rangle$ of the NH-spherical capsules as a function of the volume fraction $\phi$ are reported in the following panels: the values of $\langle D_{12} \rangle$ increase with the volume fraction (Fig.\ref{fig:nh-sphere_deformation}$b$), while the orientation angle decreases with it (Fig.\ref{fig:nh-sphere_deformation}$c$). Both these quantities are in good agreement with the results from the literature \citep{Matsunaga2016}. Time average of the specific viscosity $\mu_{sp}$ and of the normal stress differences $N_i$ ($i$ = 1 and 2) are also compared with those reported by \cite{Matsunaga2016}, and depicted in Fig.\ref{fig:nh-sphere_deformation}($d$) and \ref{fig:nh-sphere_deformation}($e$), respectively. The suspension viscosity increases with the particle volume fraction, as the absolute value of the normal stress difference, being the first positive and the second negative. In our simulations, a small bending resistance ($k_b$ = 5.0$\times$10$^{-19}$ J) is considered to avoid the membrane buckling.
Since our numerical results are in very good agreement with the previous study by \cite{Matsunaga2016}, and quantitatively similar to those obtained in a larger domain (Appendix \S\ref{appA2}), we will continue to use the rectangular box that is considered as reference in \S\ref{appA2} and include a weak bending stiffness.

\bibliographystyle{jfm}
\bibliography{jfm-instructions}

\end{document}